\documentclass[5p,twocolumn]{elsarticle}
\usepackage{lineno,hyperref}
\modulolinenumbers[5]
\usepackage[parfill]{parskip}
\usepackage[utf8]{inputenc}
\usepackage{multirow}
\usepackage{amsmath,amssymb,amsfonts,amsthm}
\usepackage{lineno,graphicx}
\usepackage{epstopdf}
\usepackage{siunitx}
\usepackage{color,soul,ulem}

\usepackage{colortbl}
\newcommand{\ccdue}{\cellcolor[cmyk]{0.43,0,0.78,0}} 
\newcommand{\cctre}{\cellcolor[cmyk]{1,0,1,0}}       
\newcommand{\ccqua}{\cellcolor[cmyk]{0,0,1,0}}       
\newcommand{\cccin}{\cellcolor[cmyk]{0,0.3,1,0}}     
\newcommand{\ccsei}{\cellcolor[cmyk]{0,0.65,1,0}}    
\newcommand{\ccbluno}{\cellcolor[cmyk]{0.21,0,0,0}}       
\newcommand{\ccbldue}{\cellcolor[cmyk]{0.70,0,0,0}}       
\newcommand{\ccbltre}{\cellcolor[cmyk]{0.88,0.34,0,0}}      
\newcommand{\ccblqua}{\cellcolor[cmyk]{0.76,0.64,0,0}}      
\newcommand{\funo}{\cellcolor[cmyk]{0.0049,0,0.2157,0.2}}
\newcommand{\fdue}{\cellcolor[cmyk]{0,0.23,0.74,0}}
\newcommand{\ftre}{\cellcolor[cmyk]{0,0.99,0.93,0}}
\newcommand{\fqua}{\cellcolor[cmyk]{0,1,0.33,0}}
\newcommand{\fcin}{\cellcolor[cmyk]{0.99,0.98,0,0}}
\newcommand{\fsei}{\cellcolor[cmyk]{0.99,0,0.02,0}}
\newcommand{\fset}{\cellcolor[cmyk]{0.87,0,0.88,0.31}}
\usepackage{hhline}
\newcommand{\STAB}[1]{\begin{tabular}{@{}c@{}}#1\end{tabular}}
\usepackage{array}
\newcolumntype{L}[1]{>{\raggedright\let\newline\\\arraybackslash\hspace{0pt}}m{#1}}
\newcolumntype{C}[1]{>{\centering\let\newline\\\arraybackslash\hspace{0pt}}m{#1}}
\newcolumntype{R}[1]{>{\raggedleft\let\newline\\\arraybackslash\hspace{0pt}}m{#1}}

\def\beq{\begin{equation}}
\def\eeq{\end{equation}}

\def\beqy{\begin{eqnarray}}
\def\eeqy{\end{eqnarray}}
\journal{ENGEO}
\biboptions{authoryear}
\begin{document}
\begin{frontmatter}
\title{Rockfall susceptibility and network--ranked susceptibility along the Italian railway}
\author{Massimiliano Alvioli\texorpdfstring{\corref{corrauthor}}{}}
\cortext[corrauthor]{Corresponding author}
\ead{massimiliano[dot]alvioli[at]irpi[dot]cnr[dot]it}
\author{Michele Santangelo}
\author{Federica Fiorucci}
\author{Mauro Cardinali}
\author{Ivan Marchesini}
\author{Paola Reichenbach}
\author{Mauro Rossi}
\author{Fausto Guzzetti}
\author{Silvia Peruccacci}
\address{Consiglio Nazionale delle Ricerche, Istituto di Ricerca per la Protezione Idrogeologica,
  via Madonna Alta 126, I-06128, Perugia, Italy}
%
\begin{abstract}
  Rockfalls pose a substantial threat to ground transportation, due to their rapidity, destructive
  potential and high probability of occurrence on steep topographies, often found along roads and
  railway routes. Approaches for the assessment of rockfall susceptibility range from purely
  phenomenological methods and statistical methods, suitable for modeling large areas, to purely
  deterministic ones, usually easier to use in local analyses.
  A common requirement is the need to locate potential detachment points, often found uphill
  on cliffs, and the subsequent assessment of the runout areas of rockfalls stemming from such
  points.\\
  Here, we apply a physically based model for the calculation of rockfall trajectories along the
  whole Italian railway network, within a corridor of total length of about 17,000 km and varying
  width.
  We propose a data--driven method for the location of rockfall source points based on expert
  mapping of potential source areas on sample representative locations.
  Using empirical distributions of gridded slope angle values in source areas mapped by
  experts, we derived probabilistic maps of rockfall sources in the proximity of the railway network,
  regardless of a particular trigger.\\
  Source areas act as starting points of simulated trajectories, within the three--dimensional
  model STONE. The program provides a pixel--by--pixel trajectory count, covering 24,500 km$^2$
  and representing the largest homogeneous application of the model to date. We classified the
  raster map into a vector susceptibility map, analyzing the railway track as a collection of
  segments, for which we provide segment--wise rockfall susceptibility.\\
  Eventually, we  considered an equivalent graph representation of the network, which
  helps classifying the segments both on the basis of rockfall susceptibility
  and the role of each segment in the network, resulting in a network--ranked susceptibility.
  Both maps are useful for subsequent hazard assessment, and to prioritize safety improvements
  along the railway, at national scale.
\end{abstract}
\end{frontmatter}
\tableofcontents
\newpage
{\large
\section*{Highlights}
\begin{itemize}
\item[$\bullet$] We performed physically based rockfall susceptibility analysis along the Italian railway, over an area of size 25,400 km$^2$ 
\item[$\bullet$] Sample rockfall sources mapped by experts helped devising a probabilistic identification of potential sources in the whole area
\item[$\bullet$] Locations of sources was a key input for three--dimensional simulation of rockfall runout areas within the STONE model
\item[$\bullet$] We performed classification of railway segments into rockfall susceptibility classes and with a new joint classification
  index including network properties
\end{itemize}}
\section{Introduction}\label{sec:intro}
Rockfall susceptibility along a transportation corridor is a recurring topic in landslide research.
Conclusions based on quantitative studies are relevant for practical purposes, providing key information
for risk mitigation. They may allow planning of protection barriers in key spots to optimize
available resources and, as in the case motivating this work, they provide a quick estimate of the
possible critical locations along a transportation network after a triggering event.

A susceptibility map does not contain information about the magnitude of possible events, nor their
expected frequency over a period of time. Aiming to such a complete description (\textit{i.e.}, a hazard
map) would require knowledge of the processes responsible for the mobilization of individual blocks or
rock masses at the local scale. The input to prepare a susceptibility map is less demanding, though it
would still not be trivial to gather the required data homogeneously over a large area, as we aim to in
this work. The minimum requirement is knowledge of the point of origin of the falling mass, the
underlying topography, and a model for the description of how the mass would reach a rest state,
downhill.
The majority of existing rockfall studies made use of physically based models, for
they give a more accurate description of the whole process, and only a few examples exist of
statistical models for rockfall susceptibility \citep{Frattini2008}.

Phenomenological (easier) ways to relate rockfall sources to their arrival point exist,
for example using an empirical shadow angle (\textit{i.e.}, the  angle between the distal limit
of the shadow and the top of  the  talus slope) \citep{Evans1993}. Here we committed to using
the model STONE by \cite{Guzzetti2002}. The model requires a digital elevation model (DEM) of
the area,  predetermined source locations, and numerical parameters controlling the energy restitution
at the location of impacts of boulders with the topography.
We used the trajectory count per grid cell produced by the model to calculate rockfall
susceptibility.

To this end, in this paper we (i) applied a novel, data--driven method to specify source areas for
rockfall trajectories \citep{Alvioli2020b}, based on expert--mapping of source areas in sample
locations and statistical generalization across the whole study area; (ii) performed trajectory
count simulation in a probabilistic way, using the model STONE, which we eventually classified as
a susceptibility map along the national railway; (iii) performed a network analysis, within graph
theory, trying to infer the impact of possible events on the functionality of the network as a whole.
A national rockfall inventory consisting of more then 70,000 rockfall polygons helped validating
the results.

The paper is organized as follows.
\textbf{Section \ref{sec:background}} gives some context for this study, both for physically based
rockfall modeling and for network analysis within graph theory. \textbf{Section \ref{sec:data}}
describes the data used in this study. \textbf{Section \ref{sec:methods}}
describes the core method, and it is split into four separate paragraphs describing
data preparation, how we singled out potential rockfall source locations, physically based rockfall
susceptibility and analysis of the railway network within graph theory; a short paragraph
describes the computational demand of this work. Results and discussion are split in a similar way, in
\textbf{Sections \ref{sec:results}} and \textbf{\ref{sec:discussion}}, respectively. Conclusions are drawn
in \textbf{Section \ref{sec:conclusions}}.
%
\section{Background}\label{sec:background}
\subsection{Physically based rockfall modeling}\label{sec:background_phys_based}
Relevant stages of rockfalls  are detachment of a rocky mass from a slope (usually a cliff face),
typically by sliding, toppling or falling, and subsequent movement down slope, through a process which can
assume different dynamics. These phenomena can vary among a rather broad spectrum, with two
limiting cases represented
by independent falls of individual fragments with no interactions among each other, and flow of granular
masses with or without interaction with the substrate \citep{Bourrier2013}.
Phenomena which do not fall strictly under the two limiting cases contain elements of both, and are the
most difficult to model.

In this work we did not adopt a physical model for the detachment process. To our knowledge, no model
exists that would cover the multifaceted mechanism of initiation of a rockfall \citep{Collins2016,Matasci2018}.
One possibility to avoid a detailed description of this step would be direct mapping through observation
of the rocky slopes, for example by laser scanners \citep{Jaboyedoff2012,Riquelme2014,Li2019}, photogrammetry
\citep{Buyer2020}, and similar approaches combined with remotely piloted aircrafts
\citep{PerezRey2019,Santangelo2019,Giordan2020}. Such measurements and algorithms may provide detailed
information about the fractured status and/or discontinuities of rocks, block orientations and distribution of
joints, giving insight to infer locations of possible detachment areas \citep{Turanboy2018,Menegoni2021,Mavrouli2020}.

The methods outlined above to identify potential rockfall sources are only applicable on small areas,
and probably on individual slopes.
Expert--mapping of possible source areas by photo--interpretation \citep{Santangelo2019,Santangelo2020}
is another approach which can be very effective, and can be applied to relatively larger areas.
In this work, we aim to a very large area assessment,
so none of these methods could be applied. A promising method to this end could be a multivariate
statistical regression technique as the one outlined by \cite{Rossi2021}, which was not implemented
here, since we selected a data--driven method \citep{Alvioli2020b}, described in the following sections.

Here we only deal with
individual masses following independent trajectories. These can be described by ballistic dynamics,
during free fall, and bouncing or rolling on the ground. In particular, the model STONE
adopted here assumes point--like boulders under the only action of gravity and the constraints due to
topography and geology. The computer program STONE models the motion of three possible states in a
number of time steps in a three--dimensional fashion along the trajectory, until the motion comes to
an end due to friction exhausting the initial kinetic energy.

Types of movements of rocky masses other than ballistic trajectories interrupted by bouncing or rolling
are not included in this work, as they require
models suited to describe avalanches and granular fluxes, while STONE does not include interactions between
boulders, neither it accounts for their fragmentation. Recently, \cite{Matas2020} and \cite{RuizCarulla2020}
studied in detail the fragmentation mechanism of large boulders by directly observing blocks debris after
they impacted the slope. Data about the angular distribution of the fragments helped calibrating numerical
parameters of the model RockGIS \citep{Matas2017}, which implements a fractal model for the mass of the fragments,
which is absent in STONE.

To go beyond a spatial susceptibility assessment, one should consider the magnitude of possible rockfalls
\citep{Guzzetti2003,Guzzetti2004},
in a given area, and their temporal recurrence. Magnitude may vary in a wide range of volumes, from cubic
decimeters to millions of cubic meters \citep{Hungr1999}.
\cite{Melzner2020} also investigated frequency--size distributions, finding a dependence on
the method used to map rockfalls and, similarly to \cite{Corominas2018}, recommended caution when considering
the large--size region of the distributions. We did not consider the magnitude of rockfalls, here, as the
model STONE does not directly account for that.

We also did not account for a specific trigger: as a matter
of fact, rockfalls may be triggered by different phenomena, mostly by intense rainfall and seismic
activity, with their occurrence influenced by a variety of factors \citep{Macciotta2015}.
Static factors are steep terrain gradient and structural setting, and dynamic factors are
thermal deformations and or freezing--thawing cycles of the rocks.
Studies exist about correlations between rockfalls and rainfall, for example by means
of probabilistic rainfall thresholds \citep{Melillo2020}, or studies of the relation between rockfalls
and rain, freezing periods, and strong temperature variations \citep{Delonca2014}. Inclusion in our
method of a trigger-- or time--dependent description is also beyond the model STONE, thus it falls out
of the scope of this paper and needs to be considered elsewhere.
\subsection{Analysis of the railway network within graph theory}\label{sec:background_graph}
Graph theory is the mathematical representation of pairwise relations between a collection of objects,
and a whole science of graphs exists with examples in a broad spectrum of theoretical and applied research
\citep{Chartrand2015}. One of major field of application of graph theory is the representation of a
physical network. Literature on network and/or graph science is extensive, and it deals with a variety
of phenomena. Examples are biological and social networks, power grids, telecommunications and, obviously,
the world wide web and transportation networks \citep{Barthelemy2011}. Studies related to transportation
networks often involve urban settings, including interdependent and/or multi modal networks of different
means of transportation, and communication networks also influence the structure of cities \citep{Alvioli2020c}.
  
The effect of disruption of a number of vertices or edges was also investigated in the literature;
\cite{Mattson2015} contains a review on the resilience of transportation networks. A few general
results are that redundancy in a complex network is not sufficient for tolerance upon failure, while
tolerance is much higher for networks exhibiting scale invariance \citep{Albert2000}
as for the world wide web \citep{Cohen2000}. On the other hand, scale--free networks
are subject to disruption upon removal of a few of the relevant vertices.
Features of networks relevant for their resilience are the presence of loops \citep{Katifori2010}
and vulnerable spots \citep{Bassolas2020}.

In this work, we use basic notions of graph theory, useful to make sense of the relevance of a
particular edge of the graph within the entire graph. We refer to ``network'' as the collection
of physical links (track segments) and of nodes (stations or intersections), and to ``graph'' as
the equivalent representation of edges and vertices. The elements of the physical network are
associated with specific
%
\begin{figure*}[!ht]
  \leftline{\hspace{0.5cm}\includegraphics[width=0.45\textwidth]{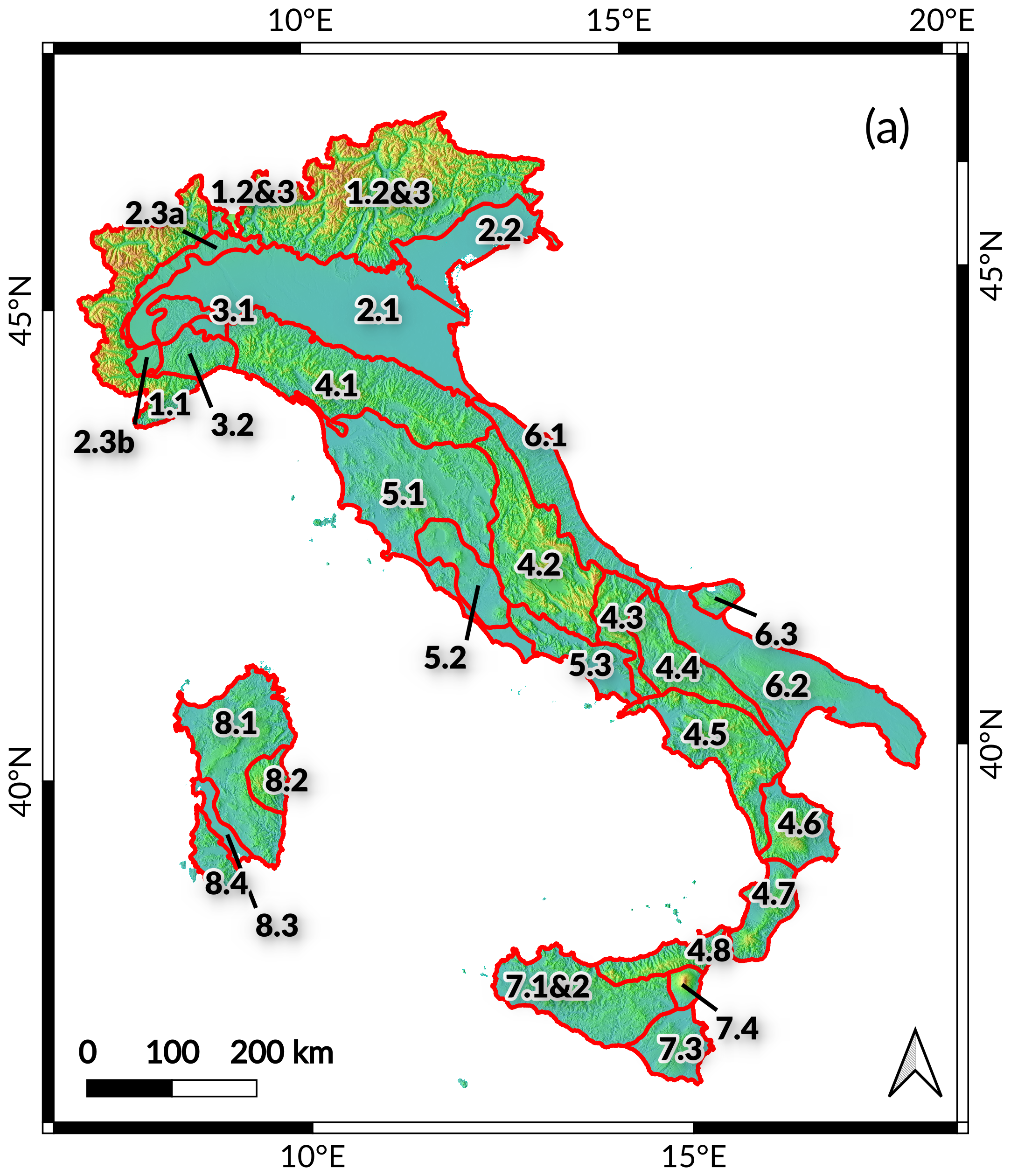}}
  \vskip -9.56cm
  \rightline{\includegraphics[width=0.45\textwidth]{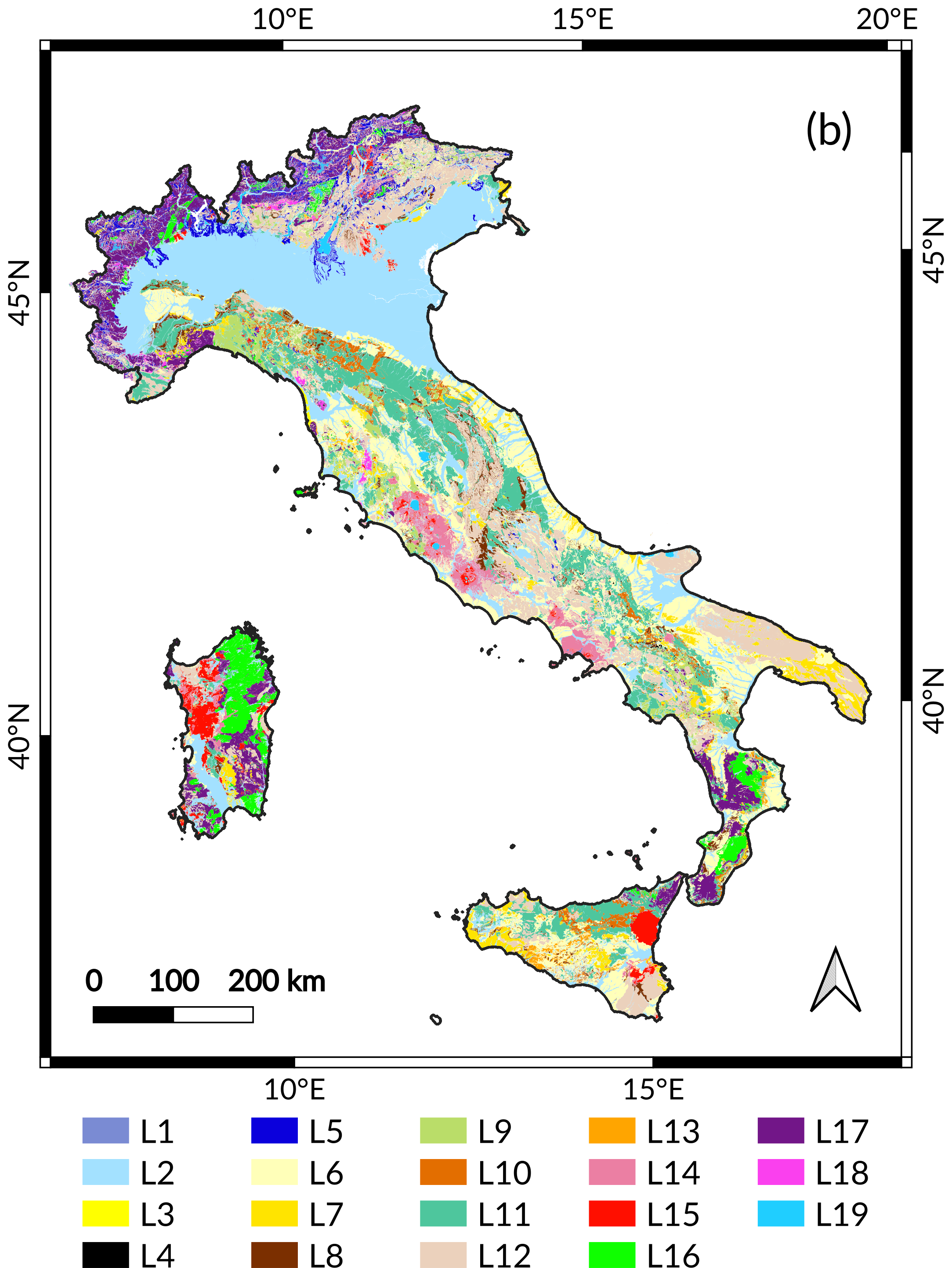}\hspace{0.5cm}}
  \vskip -1cm
  \leftline{\hspace{0.5cm}\includegraphics[width=0.45\textwidth]{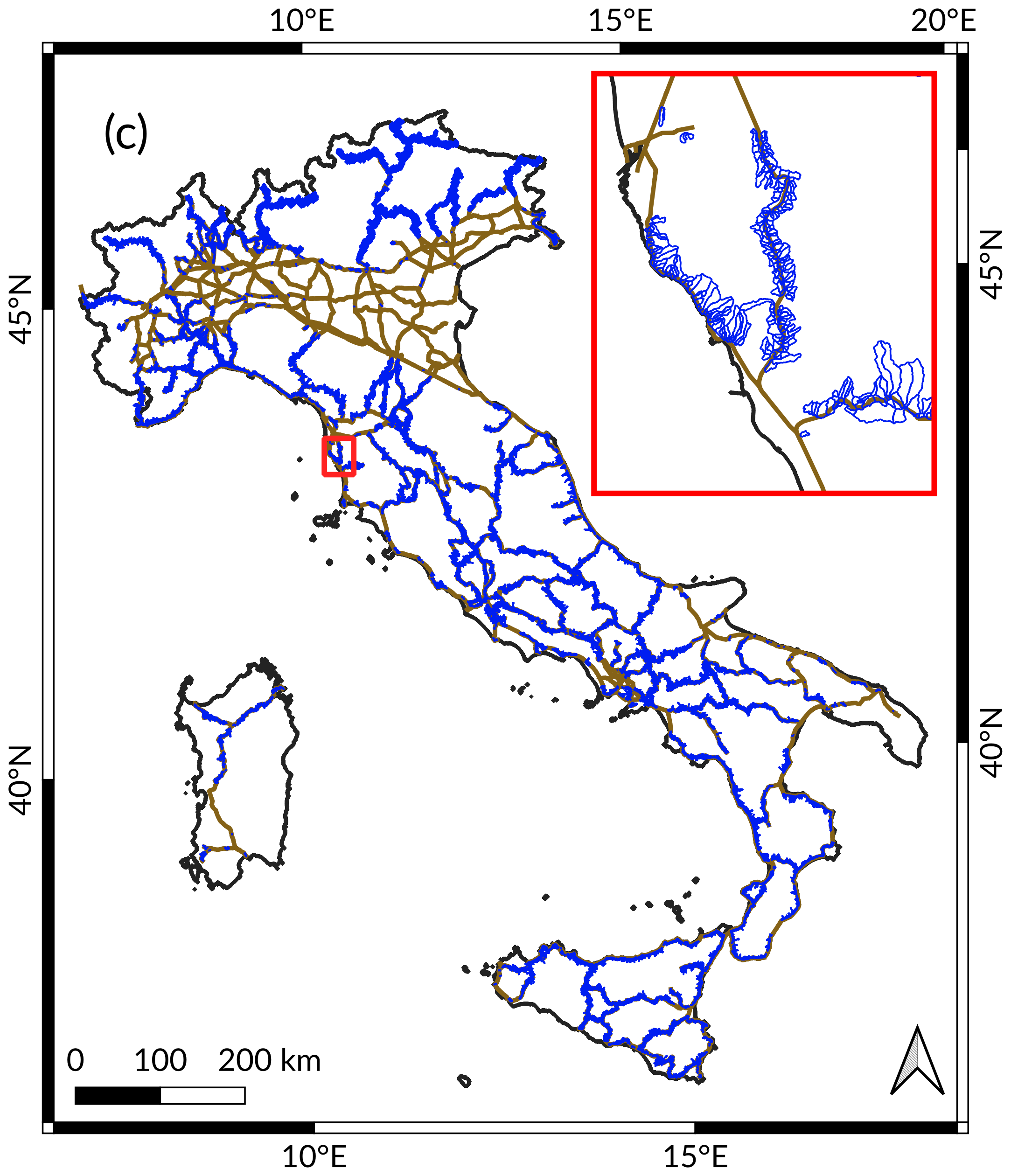}}
  \vskip -8.cm
  \hspace{0.54\textwidth}\begin{minipage}[t]{0.42\textwidth} 
  \caption{(a) Map show boundaries of 29 topographic units in Italy, first defined by Guzzetti and Reichenbach (1994),
    with red lines, superimposed to a shaded relief calculated from the EU-DEM, 25 m resolution (after \cite{Alvioli2020a}).
    Code IDs of topographic units match the codes in \textbf{Tables \ref{tab03}} and \textbf{\ref{tab04}}, while the units'
    names are listed in Section \ref{subs:results_sources}. Map is in EPSG:4326, datum in EPSG:3035.
    (b) Lithological map. ID codes for the different lithologies match the codes in
    \textbf{Table \ref{tab01}}. After \citep{Fongo2018,Bucci2021}.
    (c) The spatial buffer along the railway network adopted in this work. The buffer
    consists of the subset of slope units from \cite{Alvioli2020a} intersecting a buffer
    of 1 km around the whole railway track, depicted in brown. Slope units are in blue.
    The total number of selected slope units is 32,200, out of the over 330,000 polygons,
    for a planimetric area of 25,400 km$^2$ (\textit{cf.} \textbf{Table \ref{tab03}}).
  }\label{fig01}\end{minipage}
  \vskip 1.9cm
\end{figure*}
%
geographical locations, while the associated graph is an abstraction
that only retains the topology of the physical network and, optionally, a few other quantities.

We further refer to betweenness centrality of vertices \citep{Freeman1978} and edges \citep{Lu2013},
a measure of centrality of each such element in the graph. Betweenness centrality (in the following,
betweenness) of a vertex is given by the number of shortest paths going through the vertex, out of
the total set of shortest paths connecting any two pair of vertices in the graph (all--to--all routes).
In a similar way, we can define betweenness of edges. Values of betweenness of vertices and edges depend
on the weights of edges in the graph. As the graph considered here is associated to a physical network,
we weight edges with the actual length of railway links. Centrality of a vertex or an edge in the network
gives a measure of its relevance for functioning of the whole network.

Eventually, we devised a new (simple) centrality index intended to quantify the impact on the network
when one edge is removed, which mimics an interruption due to an external event. The new index helps
performing a combined classification of the railway segments, based simultaneously on centrality of
the segments and their rockfall susceptibility. We call the new classification network--ranked
susceptibility. A similar approach, namely calculation of network risk assessment combining exposure
to link failures and identification of link importance, was proposed by \cite{Cats2016} for a multi-modal
public transport network in the Netherlands.
%
\section{Available Data}\label{sec:data}
This work involved both material which was already available to us and newly developed maps.
Among the existing material, this study required (i) a digital elevation model, needed for the
whole processing chain; (ii) a nation--wide inventory of rockfalls; (iii) a small--scale
lithological map; (iv) a vector map of the railway network. Newly developed maps were a
nation--wide slope unit and sample rockfall source area maps. Here follows a short description
of such pieces of information.
\begin{itemize}
\item[$\bullet$] Digital elevation model. The highest resolution DEM freely available for the whole
  of Italy is TINITALY \citep{Tarquini2007}. The grid is 10 m x 10 m, in the reference system with
  EPSG:32633. \textbf{Figure \ref{fig01}(a)} shows a shaded relief map based on the 25 m-resolution
  EU-DEM; its use in this work is limited to this figure.
\item[$\bullet$] Inventory of landslide phenomena in Italy, known as IFFI \citep{Trigila2010,IFFI2018}.
  The IFFI inventory contains over 620,000 landslide polygons, of which 70,576 are classified as
  rockfalls, and 4,051 are located in the vicinities of the railway track and were used in this work.
  Rockfall polygons source and runout areas.
\item[$\bullet$] Lithological map of Italy, scale 1:100,000 \citep{Fongo2018,Bucci2021}. \textbf{Figure
  \ref{fig01}(b)} shows the map, with 19 lithological classes \textbf{(Table \ref{tab01})}.
\item[$\bullet$] Map of subdivision in topographic units (TUs) of Italy \citep{Guzzetti1994}.
\item[$\bullet$] Nation-wide slope unit map \citep{Alvioli2020a}, also published in the research report
  ``Unit\`a territoriali di riferimento'' P-05-1 of ``Progetto RFI-SERVICE'', CNR IRPI, August 21, 2018.
  The map contains more than 300,000 slope unit polygons, of variable shape and size.
\item[$\bullet$] Vector layer containing the railway network track. \textbf{Table \ref{tab02}} shows the
  distribution in elevation of the various links of the network, and \textbf{Fig. \ref{fig01}(c)} shows
  the slope units selected from this work along the railway track. The figure also shows
  the track on plain areas, where slope units are undefined.
\item[$\bullet$] A map of sample potential source area for rockfalls in Italy, from expert interpretation
  of Google Earth\texttrademark{ }images. Section \ref{subs:methods_sources} contains a description
    of the rationale and the criteria used during expert mapping.
\end{itemize}
\section{Methods}\label{sec:methods}
Performing nation-wide rockfall susceptibility assessment, along the railway track, required careful
planning. The long chains of actions, from data selection to the preparation of the final map, is in
fact prone to propagating early--stage mistakes in unpredictable ways. Selection of STONE as a rockfall
runout modeling software \citep{Guzzetti2002,Agliardi2003} was dictated by the need to compromise between
a well--defined relationship between sources and runout, and overall computational demand.

\begin{figure}[!ht]
  \centerline{
    \includegraphics[width=0.45\textwidth]{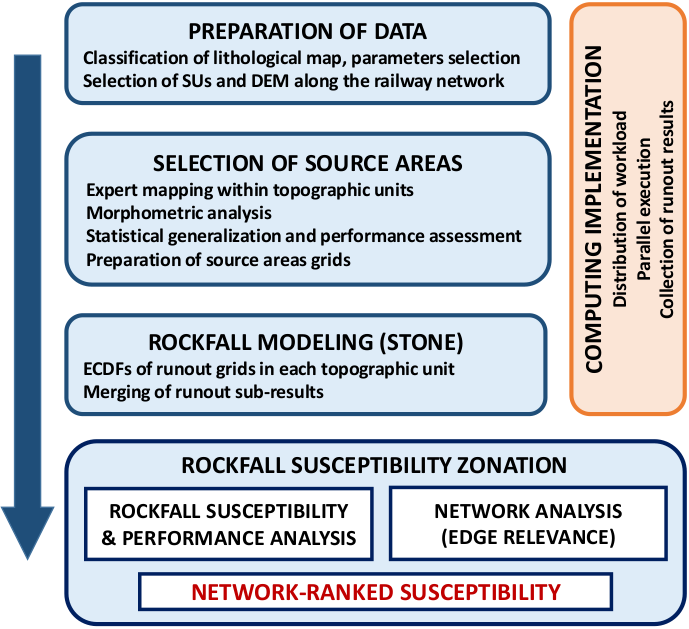}}
  \caption{A flowchart of the procedure followed in this work to obtain maps of rockfall susceptibility
    and network--ranked susceptibility. The different steps involved in the procedure are described in
    detail in Section 3 (Available data), Section 4 (Methods) and Section 5 (Results).
  }
  \label{fig02}
\end{figure}

To prepare a susceptibility map based on the results of the model STONE we adopted the following steps:
(i) preparation of the data, (ii) selection of possible source areas, which is probably the most important
input of the model, (iii) execution of the program, (iv) collection and classification of the results
analyzing the susceptibility results from the perspective of the properties of the network.
The flowchart in \textbf{Figure \ref{fig02}} illustrates the procedure; the following
paragraphs describe in detail all the steps performed in the analysis.
%
\subsection{Preparation of data}\label{sec:methods_data}
We adopted the topographic subdivision, previously defined by \cite{Guzzetti1994} and slightly modified
for slope units delineation \citep{Alvioli2020a}.
The original subdivision contained 30 sections, obtained by quantitative analysis and visual interpretation
of morphometric variables obtained from a digital elevation model. In the modified version, we neglected
topographic units much smaller than the surrounding ones, and split one into two different ones, ending
up with 29 topographic units, shown in \textbf{Fig. \ref{fig01}(a)}. Use of topographic units was dictated
by the need of similar terrain types, allowing to use the same soil parameters in each unit, for numerical
simulations.

Utilization of topographic domains also allowed us to define potential source areas of rockfalls using the
statistical generalization proposed in this work separately within the 29 sections, corresponding to 29
different terrain types. A lithological map of Italy at the 1:100,000 scale provided the necessary insight
for the determination of different sets of input parameters for the software STONE in different lithological
classes. These parameters control the behaviour of boulders during rolling or bouncing on the ground, along
the simulated trajectories. Their final values were inferred from previous studies.
%
\begin{table*}[ht!]
\footnotesize
\begin{center}
  \caption{Numerical values of the parameters used in STONE. \textbf{Figure \ref{fig01}(b)} shows the
    corresponding lithological map \citep{Fongo2018,Bucci2021}. STONE performs random
    sampling of values of the parameters in a $\pm$ 5\% range around the nominal values
    listed here.
  }
    \vskip 0.5cm
  \label{tab01}{\renewcommand{\arraystretch}{1.2}
  \begin{tabular}{c|c|c|c|c}  
\hline
\multirow{2}{*}{\textbf{Class ID}} & \textbf{Lithological} & \textbf{Dynamic} & \textbf{Normal} & \textbf{Tangential}\\
& \textbf{Class} & \textbf{Friction} & \textbf{Restitution} & \textbf{Restitution}\\ \hline
L1  & Anthropic deposits                                            & 0.65 & 35 & 55\\
L2  & Alluvial, lacustrine, marine, eluvial and colluvial deposits  & 0.80 & 15 & 40\\
L3  & Coastal deposits, not related to fluvial processes            & 0.65 & 35 & 55\\
L4  & Landslides                                                    & 0.65 & 35 & 55\\
L5  & Glacial deposits                                              & 0.65 & 35 & 55\\
L6  & Loosely packed clastic deposits                               & 0.35 & 45 & 55\\
L7  & Consolidated clastic deposits                                 & 0.40 & 55 & 65\\
L8  & Marl                                                          & 0.40 & 55 & 65\\
L9  & Carbonates-siliciclastic and marl sequence                    & 0.35 & 60 & 70\\
L10 & Chaotic rocks, m\'elange                                      & 0.35 & 45 & 55\\
L11 & Flysch                                                        & 0.40 & 55 & 65\\
L12 & Carbonate Rocks                                               & 0.30 & 65 & 75\\
L13 & Evaporites                                                    & 0.35 & 45 & 55\\
L14 & Pyroclastic rocks and ignimbrites                            & 0.40 & 55 & 65\\
L15 & Lava and basalts                                              & 0.30 & 65 & 75\\
L16 & Intrusive igneous rocks                                       & 0.30 & 65 & 75\\
L17 & Schists                                                       & 0.35 & 60 & 70\\
L18 & Non--schists                                                  & 0.30 & 65 & 75\\
L19 & Lakes, glaciers                                               & 0.95 & 10 & 10\\\hline
\end{tabular}}
\end{center}
\end{table*}
%

A second, much finer subdivision of the territory was adopted here, consisting in groups of slope units
intersecting the railway. Use of slope units was both a natural choice, given that slope units are mapping units
well suited for landslide studies, and a solution to the technical problem of defining meaningful sub--areas
for execution of the simulations in parallel. We favor SU as suitable mapping units for the
description of landslide phenomena \citep{Alvioli2016,Camilo2017,Schloegel2018,Bornaetxea2018,Tanyas2019a,
  Tanyas2019b,Jacobs2020,Amato2020,Chen2020,Li2020}. More specifically, we adopted SUs instead of a geometric
buffer around the track, because rockfall trajectories initiated in a given SU will be bounded within the SU,
with reasonable confidence. The set of SU used here were specifically calculated and optimized over the whole
of Italy \citep{Alvioli2020a} for this work, and for a series of additional works regarding selection of source
areas of debris flows \citep{Marchesini2020} and soil slides susceptibility (yet unpublished), along the railway
network. The subset of SUs selected for this work is shown in \textbf{Fig. \ref{fig01}(c)}. The
selected polygons (\textit{cf.} point 2a) cover an overall area of 25,400 km$^2$.
%
%
%
\begin{table}[ht!]
\footnotesize
\begin{center}
  \caption{Distribution of the railway network in intervals of elevation. Total length is 17,727 km,
    a portion of which is outside of TINITALY DEM (denoted as N/A in the table), and the total number
    of segments is 3,419. Number of segments overlapping with at least one slope unit is 2,427, with
    total length 13,746 km.}
  \vskip 0.5cm
  \label{tab02}{\renewcommand{\arraystretch}{1.5}
  \begin{tabular}{c|r|r|r}
\hline
 \textbf{Elevation}  & \multirow{2}{*}{\textbf{\# Segments}} & \textbf{Length} & \multirow{2}{*}{\textbf{Percentage}}\\
 {[m]}               &       & {[km]\hspace{0.3cm}} &       \\\hline
 {[0,100)}           & 1,746 &  8,671 & 45.0\%\\
 {[100,500)}         & 1,447 &  7,814 & 46.5\%\\
 {[500,1000)}        &   188 &  1,034 &  7.5\%\\
 {[1000,2000)}       &    24 &    141 &  0.8\%\\
 {[2000,3000]}       &     2 &      7 &  0.0\%\\\hline
 N/A                 &    12 &     59 &  0.2\%\\\hline
  \end{tabular}}
\end{center}
\end{table}
%

We prepared all of the data splitting the study area into sub--areas and working within sub--areas
in parallel, computing--wise. In particular, for each group of SUs singled out as described above, we
prepared an independent simulation with STONE. This step requires selecting the relevant portion of DEM,
of sources areas and of grids containing the numerical values of parameters needed by STONE (\textbf{Table \ref{tab01}}).
\subsection{Identification of rockfall source areas}\label{subs:methods_sources}
Identification of potential rockfall source areas is a key step of physically based simulation of rockfall
runout, because they substantially influence the extent of runout. A detailed investigation
of the location of potential sources of rockfalls requires expert analysis of the cliffs in the study area
\citep{Guzzetti2004,Santangelo2019,Santangelo2020}, which is typically a time consuming and expensive procedure.
This makes identification of potential sources a limiting factor for systematic rockfall studies over large
areas. Visual mapping of all of the potential sources by photointerpretation over the whole of the railway
network in Italy would be impossible, in a reasonable time. Moreover, once a potential source is located on
a DEM, it is desirable to assign a probability for the likelihood of that location to evolve into a rockfall.

A straightforward way of selecting source areas for STONE is to establish a slope threshold above which any
grid cell acts as a potential rockfall source \citep{Guzzetti2003}. Other criteria are based on a combination
of geomorphological analysis and mapped sources \citep{Agliardi2003}, a combination of a slope threshold and
geomorphological analysis \citep{Sarro2020}, or a multivariate statistical analysis \citep{Rossi2021}. In
this work, we adopted a method to both locate sources and assign a probability of failure in a homogeneous
way over a large area, on the 10 m-resolution DEM of Italy (TINITALY, \cite{Tarquini2007}), according to the
following steps:
\begin{itemize}
\item[1a.] selection of a 1 km-wide buffer region around the railway track;
\item[2a.] selection of the subset of slope units (SU) intersecting the buffer, out of the whole
  set of about 330,000 SUs calculated for Italy by \cite{Alvioli2020a}; the subset of selected SUs
  is the final working buffer used here;
\item[3a.] expert mapping of potential rockfall source areas within the selected
  SUs, in sample locations we considered representative of the conditions that could trigger rockfalls
  in the specific topographic unit \citep{Guzzetti1994} under investigation;
\item[4a.] development of a data--driven procedure to identify source areas in different grid cells,
  with respect to the areas mapped as in 3a within the same topographic unit, using statistical
  generalization;
\item[5a.] visual analysis of the source areas map obtained from point 4a and assimilation in the final source
  areas map of additional sources in locations chosen in expert way.
\end{itemize}
Selection of the SU overlapping with a 1--km buffer around the railway track (\textit{cf.} point 1a)
ensures that the simulations will include all of the potential sources that could develop into rockfalls
and whose runout could concern the railway network (this work does not attempt to account for any anthropogenic
mass displacements in the vicinities of the railway track).

Expert mapping was performed on Google Earth\texttrademark{ }images. Image interpretation to detect and map potential
rockfall source areas can be carried out using different data and tools. We chose Google Earth\texttrademark{ }images
because they provide simultaneously morphological (elevation) and optical information, resulting in 2.5D images. They
are less accurate than stereoscopic optical images, but the advantage is completeness and free availability at the national
scale. This is far better than using 2D images in GIS, which have the same image availability as Google Earth\texttrademark{ },
but come with no morphological information. The vector format editing tool allows mapping directly on the interpreted images,
and exporting the polygons.

To map potential rockfall source areas, the geomorphologists considered that they correspond to sub--vertical rocky slopes
showing talus or rockfall deposit underneath. They usually appear not vegetated, even if in some locations tall vegetation
can hide potential detachment areas. For each topographic unit, the geomorphologists mapped all of the potential rockfall
source areas within each SU in a significant subset of SUs, which were considered representative of the main litho--structural
and morphological settings.

\begin{figure}[!ht]
  \centerline{
    \includegraphics[width=0.45\textwidth]{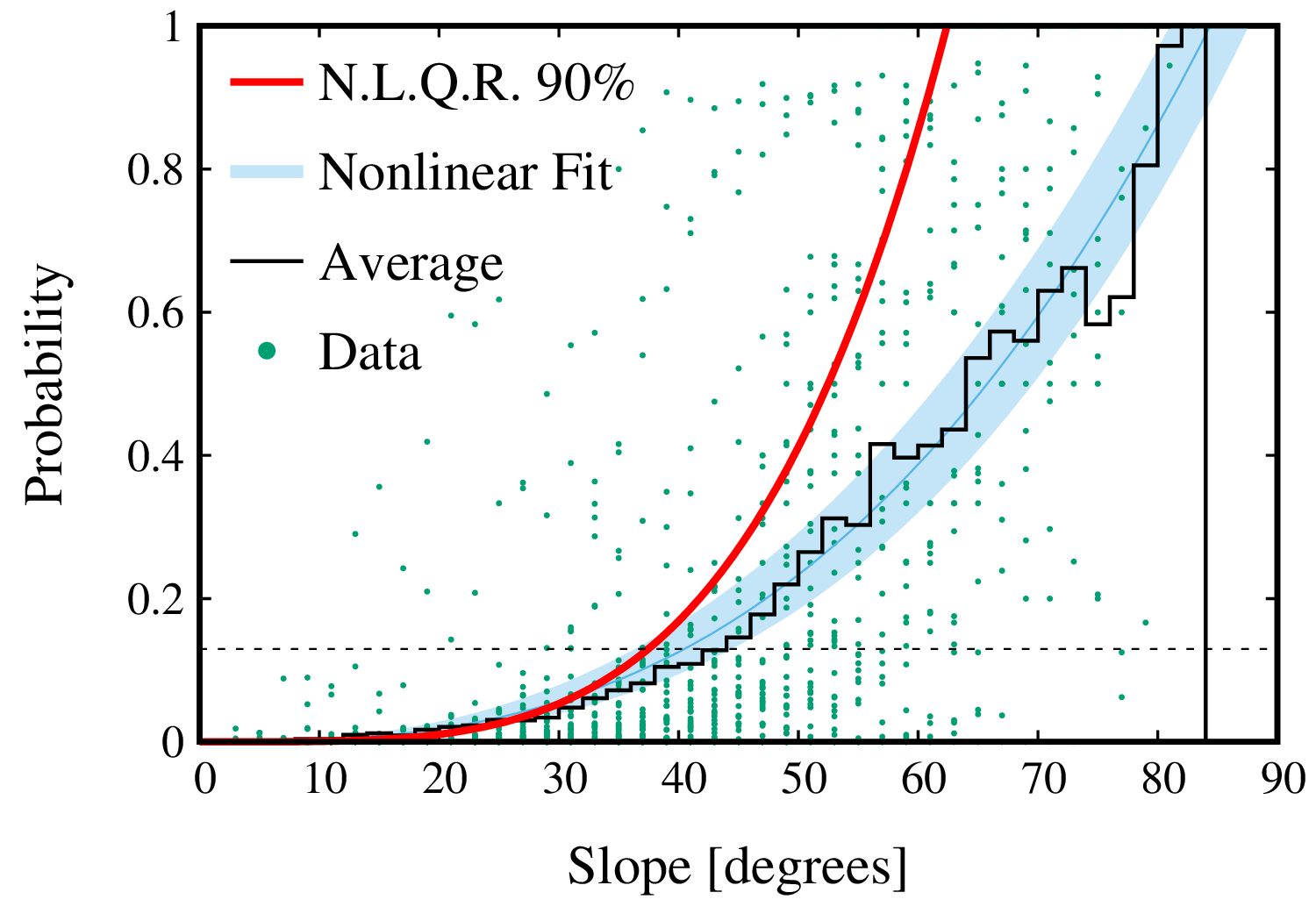}}
  \caption{Example result of statistical assessment of the probability of a grid cell with given slope
    to represent a source area for rockfalls. Data (green dots) and numerical models (curves) correspond
    to topographic unit 1.2 \& 1.3 in \textbf{Figs. \ref{fig01}(a)} and \textbf{Table \ref{tab01}}.
    The red curve, a non--linear 90\% quantile regression corresponding to \textbf{Eq. (\ref{eq2})}, is the
    adopted model. The horizontal line represents the limit under which probability is set to null.}
  \label{fig03}
\end{figure}

We used the information of expert--mapped sources to calculate the probability of any grid cell to be a source
as follows. For a given topographic unit, we selected SUs containing mapped polygons. We calculated
histograms of the distribution of values of slope angle in the cells within the mapped polygons,
and within the whole slope unit. For each bin of \ang{2} width, we took the ratio of the two histograms:
values of slope within the sources over values within the whole slope units. The values of the numerator
are smaller than at the values of the denominator, in each bin, by construction (the number of cells belonging
to source areas falling in each bin are less or equal to the number of cells, in the same bin, from the whole
slope unit). Since we took care of delineating all of the possible rockfall
sources within each of the considered slope units, we are confident that the ratio represents the probability
for a cell with slope in the \ang{2} bin range of being a rockfall source, in that specific slope unit. This
way, we have a probability point for each \ang{2} slope bin. Considering all of the sampled SUs in the given
topographic unit, we end up with a collection of probability values as the green dots in \textbf{Fig. \ref{fig03}},
%
%
for TU 1.2 \& 1.3, which we chose as an illustrative example.

The purpose of the procedure is to obtain probability--slope relations which can be extended to the entire set
of selected SU, in each topographic unit. Green dots in \textbf{Fig. \ref{fig03}} represent the probability
values from expert mapping, and curves show three possible approaches to define a functional dependence between
probability and slope. The black histogram is the simple average, in each slope bin, of the observed probability
values. The blue curve is a non--linear fit of the values estimated from observed data, with a function of the
following form:
\beq
\label{eq1}
P_{FIT}(S)\,=\,a\,\left(\frac{S}{90}\right)^b\,,
\eeq
where $S$ is the slope in degrees, and $a$, $b$ are parameters of the fit. The band around the fit is the
area between the minimum and maximum curves obtained using $a \pm \sigma_a$ and $b \pm \sigma_b$ (with
$\sigma_{a,b}$ the standard deviations provided by least squares minimization), shown to illustrate goodness
of fit. The red curve is a non--linear quantile regression with the 90$^{th}$ percentile, with a function
containing a single parameter, $c$, as follows:
\beq
\label{eq2}
P_{QR}(S)\,=\,c\,\left(\frac{S}{90}\right)^4\,.
\eeq
Thus, \textbf{Eq. (\ref{eq2})} is the curve dividing the observed sample in a way that 10\% of the green
dots in \textbf{Fig. \ref{fig03}} fall above the curve. We decided to use the method of quantile regression,
\textbf{Eq. (\ref{eq2})}, to assign the probability of a grid cell to represent a source area (\textit{cf.}
point 4a). We believe that it maximizes the information extracted from observed data, with respect to the
average and the fit of \textbf{Eq. \ref{eq1}}. Moreover, we used a function with a fixed exponent (the fourth
power of $S$/90) because using a parametric exponent as in \textbf{Eq. (\ref{eq1})} the resulting curves
systematically underestimated the probability of source presence for large values of slope. The procedure
provides an individual curve $P_{QR}(S)$ for each topographic unit, obtained form expert mapping of potential
rockfall source areas and from non--linear quantile regression of probability values observed in the same
topographic unit.

In each topographic unit, cell--by--cell values of probability obtained from \textbf{Eq. (\ref{eq2})} were
directly related with the number of simulated trajectories considering the cell as the starting point of
a falling boulder. We imposed a lower threshold probability under which the cell is not considered as a source.
The threshold is the minimum between the value of $P_{QR}$ providing a source map covering at least 80\% of
the sources mapped by visual interpretation, and $P_{QR}$ = 10\%. We simulated for each source cell a number
of trajectories corresponding to the numerical value of $P_{QR}$ in percent; for example, a cell with probability
X\% would trigger X simulated trajectories. This allows to exploit the possibility in STONE of simulating a
different number of falling boulders based on the probability of a given cell to be a source area.
This method increases the degree of randomization of the simulations (together with the random selection of
the value of input parameters within a given percent around the mean value provided as an input, built--in
in STONE).

Eventually, we evaluated the degree of match between model and ground truth using hit rate, defined
as $HR$ = $TP/(TP+FN)$, where $TP$ stands for true positives and $FN$ for false negatives. Given that
$TP$ + $FN$ = $P$, the total attempts of making a prediction (we do not consider negatives, here), the
index gives a measure of the percentage of times a prediction was a hit, out of the total attempts.
%
\subsection{Simulation of rockfall trajectories with STONE}\label{subs:methods_stone}
After the selection of source areas, we performed the following steps to calculate rockfall susceptibility
within a polygonal buffer around the whole Italian railway network, using the program STONE and the 10 m--resolution
DEM used to identify sources of rockfalls:
\begin{itemize}
\item[1b.] attribution of dynamic friction, normal and tangential restitution coefficients values, necessary
  for execution of the program STONE. This step was performed by expert assignment of values of coefficients
  available from the literature to each lithological class;
\item[2b.] preparation of data grids necessary for execution of STONE in each topographic unit, further
  split in rectangular domains containing contiguous SU polygons, as selected at point 2a in
  \textbf{Section \ref{subs:methods_sources}};
\item[3b.] execution of actual simulations with STONE. We performed distributed (parallel) runs within the
  921 sub--areas described in point 2b;
\item[4b.] collection on the various raster maps containing the trajectory count per grid cell (``counter''),
  in each of the sub--areas of each of the 29 topographic units;
\item[5b.] transformation (classification) of the counter map into a probabilistic map.
\end{itemize}
Numerical values of dynamic friction, tangential and normal restitution are necessary for running the program
STONE (\textit{cf.} point 1b). We selected them as follows. We considered a lithological map at 1:100,000
scale, obtained from a digital map in vector format. The original map, from Italian Geological Service (ISPRA \citep{Tacchia2004,Battaglini2012}),
contained more than 5,000 unique descriptions of geological formations \citep{Fongo2018,Bucci2021}. Using expert criteria,
with the help of about 400 geological sheets (digital photographs from the same source) at higher scale, the original
unique description were classified into 19 lithological categories. The numerical values of coefficients to run
the program STONE were obtained using previous estimations for similar lithologies available in the literature.
\citep{Guzzetti2002,Guzzetti2003,Guzzetti2004,Agliardi2003,Santangelo2019}. The selected values are listed in
\textbf{Table \ref{tab01}}.

The actions of points 2b, 3b, 4b were performed in 921 sub--areas singled out by considering separately groups of
adjacent SUs: (i) a DEM of the sub--area under investigation; (ii) a grid containing the positions of the rockfall
source areas, specifying the number of trajectories to be simulated; (iii) three grids, containing the values of
dynamic friction and of normal and tangential restitution coefficients, for each grid cell of the DEM.

The software STONE produces as an output three grids containing: (a) a count of the number of trajectories that
%
\begin{figure}[!ht]
  \centerline{\includegraphics[width=0.45\textwidth]{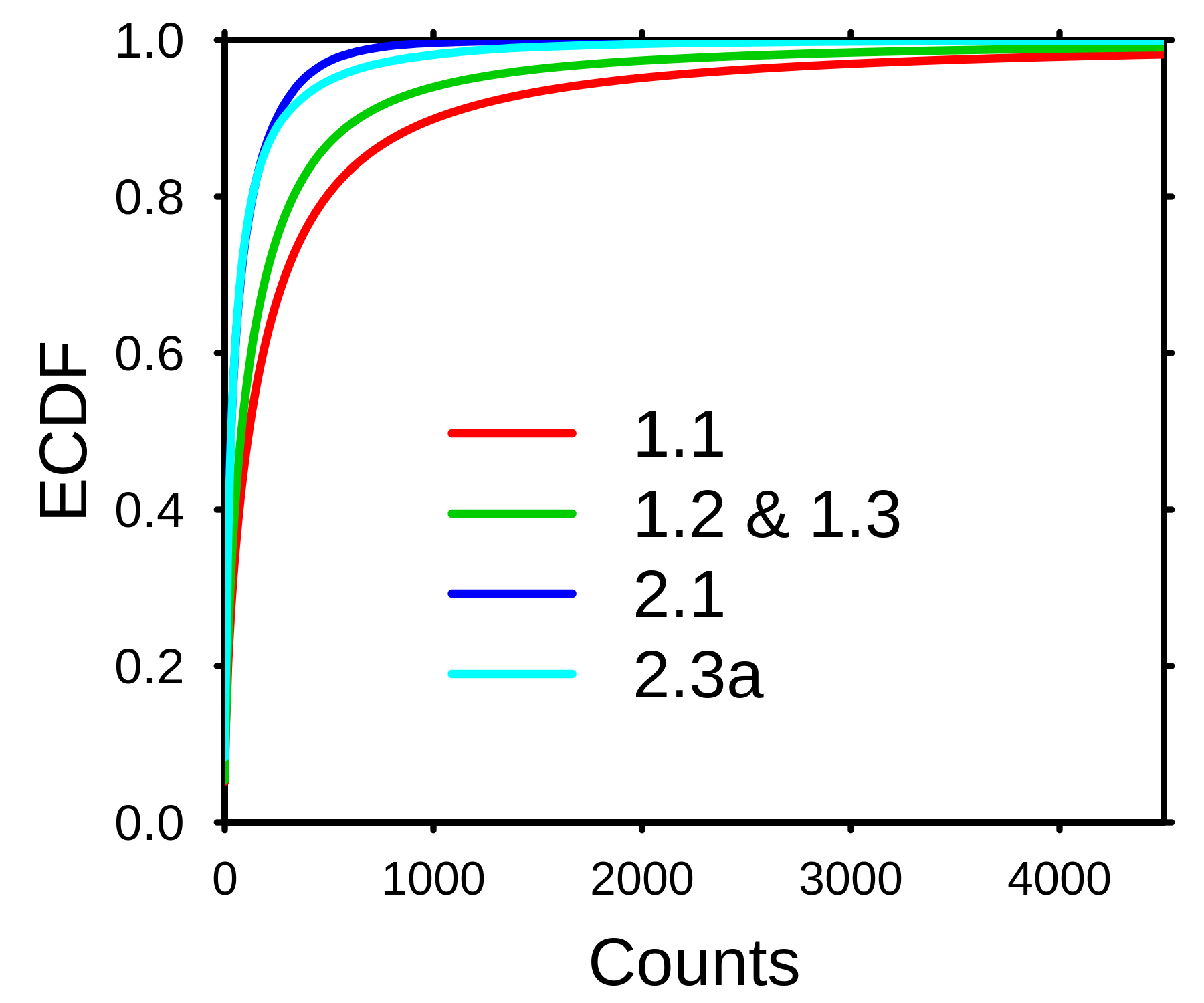}}
  \caption{The figure shows sample ECDFs, obtained in four topographic units;
    (\textit{cf.} \textbf{Fig. \ref{fig01}{(a)}} and \textbf{Table \ref{tab01}}).
    We used an approach based on ECDF to map the output of STONE, consisting in a cell--by-cell
    trajectory count (abscissa), into a relative probability map (ordinate).
  }
  \label{fig04}
\end{figure}
%
crossed the cell during the simulations; (b) the maximum height of the trajectories, measured from the local elevation;
(c) the maximum velocity of the simulated boulders in each cell. The three grids can be used jointly, in principle,
to estimate rockfall hazard in each cell. As a matter of fact, for given source areas, (a) can be interpreted as
the relative probability of having a trajectory crossing each grid cell, while (b) and (c) may be jointly considered
as a proxy of the magnitude of the expected rockfalls in each cell. Estimating temporal frequency of rockfall events,
instead, requires additional information \citep{Guzzetti2003,Guzzetti2004}. In this work, we only used output (a),
as an estimate of the relative spatial probability of rockfall events.

To convert the trajectory count per cell into a probabilistic index (\textit{cf.} point 5b) we used the following
rationale. We considered the national inventory of landslide phenomena in Italy (IFFI - \cite{IFFI2018} and \cite{Trigila2010}).
We  extracted polygons labeled in the inventory as ``falls'' or as ``extended areas containing falls''.
We calculated empirical cumulative distribution functions (ECDFs) of the trajectory count, limited to the grid cells
underlying the polygons extracted from IFFI inventory. In this way, we considered as relevant only the cells known to
actually contain a rockfall. We used the ECDF to map all of the cells in the trajectory count grid with non--null value
onto the [0,1] interval, which allows a probabilistic interpretation required for a susceptibility map.

Using an ECDF, instead of a simple normalization of the trajectory count values, has two advantages. First, it
allows to attribute less relevance to cells with a small number counts, given the huge variance of the counts
in the output grids. Second, using ECDFs makes results independent from the absolute values of the number of
simulated trajectories per source grid cell and makes only relative values relevant, consistently with our method
of assigning these numbers (\textit{cf.} Section \ref{subs:methods_sources}). We performed the counter--to--relative
probability conversion operation separately in each of the 29 topographic units shown in \textbf{Fig. \ref{fig01}(a)}.
\textbf{Figure \ref{fig04}} shows sample ECDFs obtained in four topographic units.
\subsection{Analysis of the railway network within graph theory}\label{subs:methods_graph}
The railway track running through Italy can be viewed as a collection of links, \textit{i.e.} portions of
track running without intersections from a starting point to an arrival point, and of nodes, \textit{i.e.}
the intersections. Nodes can be connected to  multiple links. This collection of links and nodes represents
a geographical network.

The links--nodes description can be further assimilated with a graph, whose vertices are the network nodes
and whose edges are the network links. We performed a simple preliminary analysis of the graph corresponding
%
\begin{figure*}[!ht]
  \centerline{
    \includegraphics[width=0.7\textwidth]{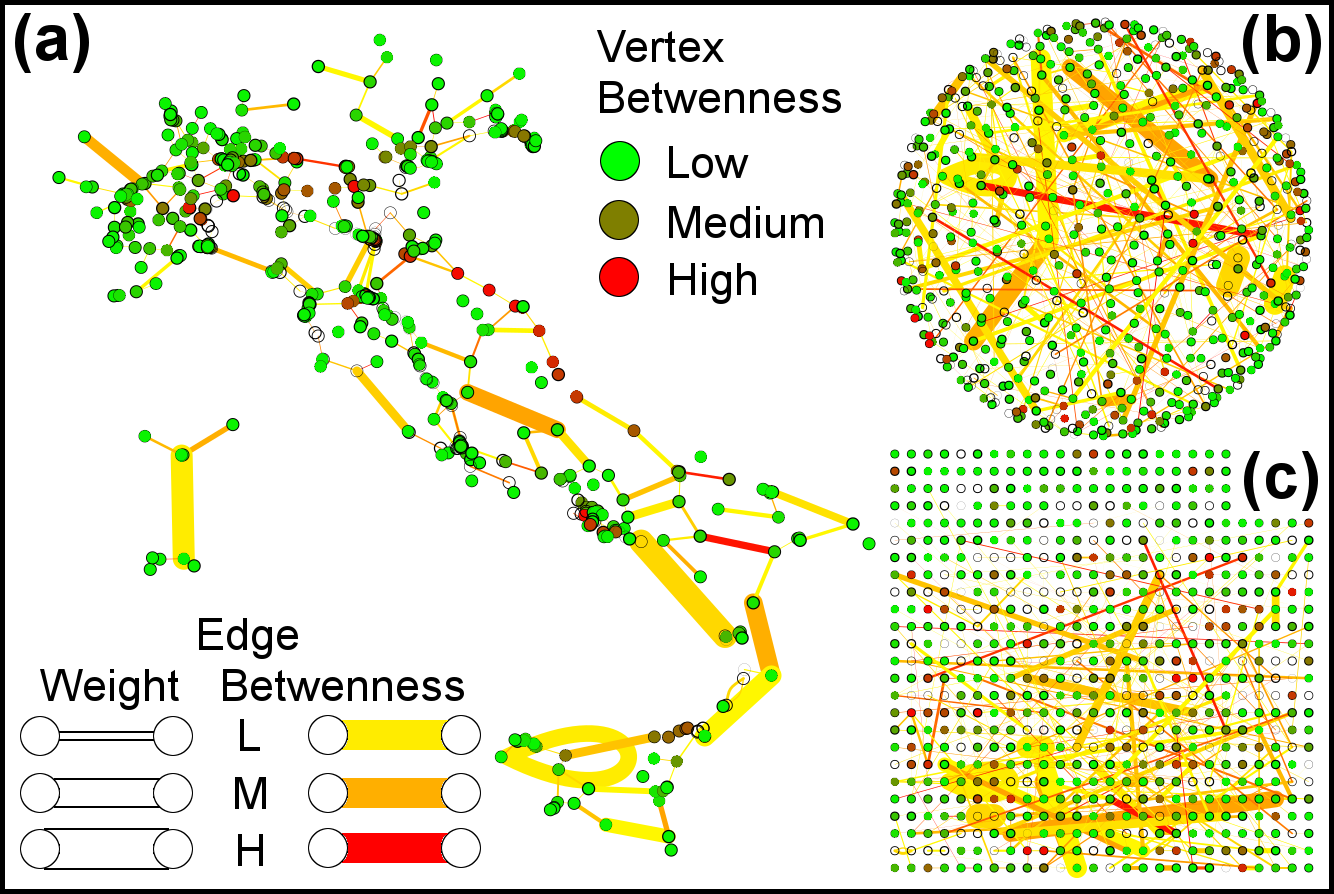}}
  \caption{Graphs associated to the railway network. (a) Spatial distribution of vertices    
    (circles) is consistent with their geographical location, and links of the network are
    represented as straight lines (edges) connecting vertices. (b)-(c) Two alternative
    representations of the same graph; graphical representations of a graph are infinite,
    in principle, and a specific one may help highlighting peculiar aspects. In (b), nodes
    are uniformly distributed on the surface of a sphere; in (c), nodes are on a square grid,
    showing that most of the nodes are rather peripheral.
    All of the graphs contains three quantitative measures: (i) color of the vertices,
    displayed with a green--to--red palette, represents betweenness; (ii) color of the
    edges, displayed with a yellow--to--red palette, represents the edge betweenness; (iii)
    thickness of each edge is proportional to the actual (spatial) distance between the two
    vertices (nodes) it connects. Information content is the same in the three representations.
  }
  \label{fig05}
\end{figure*}
%
to the railway network, calculating betweenness of vertices and edges. This analysis describes the topology
of the graph, and it is a starting point for discussing rockfall susceptibility in relation to the railway
network operativity. \textbf{Figure \ref{fig05}} is a graphical representations of such analysis.

\textbf{Figure \ref{fig05}} is a graph representation of the geographical network, where vertices are now
connected by straight lines -- graph edges. In \textbf{Fig. \ref{fig05}}, the dots colored with a green--to--red
palette represent betweenness of vertices (\textit{cf.} \textbf{Section \ref{sec:background_graph}}), with
red corresponding to higher values. The representation of edges is twofold. The color of the edges, quantified
with a yellow--to--red palette, represents betweenness of the edges: more central edges (\textit{i.e.} edges
crossed by a larger number of all--to--all shortest paths) have larger betweenness. Thickness of the edges,
instead, is proportional to the physical distance between the nodes they connect. Edge betweenness in the
(weighted) graph, is known to be one of the most important variables in networks \citep{Barthelemy2011},
and we used it in the subsequent analysis.

To understand the impact of possible interruptions due to rockfalls on the railway network,
from graph theory point of view, we performed two separate analyses. First, we considered the impact
on the graph (and, correspondingly, on the real network) of removing edges (\textit{i.e.} railway links,
consisting of a variable number of 1--km segments) from the network, one by one. Second, we considered
jointly the classification of the railway segments into susceptibility classes and their role in the
equivalent graph as described above.

Removing one edge from the graph, all of the remaining edges (as associated to the residual network)
change their values of betweenness by a small amount. For each edge in the graph, we calculated total
variation betweenness from the original graph to the graph obtained removing the one edge, and assigned
this value as an attribute to the removed edge in the original graph. We looped over the whole set of
edges, ending up with a new ranking of edges according to the newly calculated attribute. We used the
newly defined quantity as an additional index for a joint classification of railway segments. The joint
classification is based both on rockfall susceptibility and on the variation of betweenness index,
and we named it network--ranked susceptibility.
%

\subsection{Computing implementation}\label{subs:methods_computing}
Simulations over a large area with the software STONE has the limitation that the software calculates
the user-defined number of trajectories in a serial way. Despite
each calculation is relatively fast, the large number of trajectories needed for this study would
make it impossible to run in a single instance. We adopted a simple strategy to overcome this problem,
splitting the study areas in many sub--areas (here, groups of contiguous SUs) and running several
instances of STONE in parallel (\textit{cf.} Section \ref{subs:methods_stone}).

Raster grids were prepared for each sub--area in automatic way within GRASS GIS \citep{GRASS,GRASS_ivan}.
We used a multi--core machine, equipped with 48 computing cores and 330 GB RAM memory, to perform in parallel
the GIS operations necessary to prepare the grids (\textit{cf.} point 2b, \textbf{Section \ref{subs:methods_stone}}),
execution of the program itself (\textit{cf.} point 3b) and collection of the 921 sub--results (\textit{cf.}
point 4b). The total running time for each complete run was about two days, using a master--slave strategy,
which best balanced workload.
%
\section{Results}\label{sec:results}
This section describes the results of the physically based model STONE \citep{Guzzetti2002} along the
national railway network in Italy. We performed simulations in areas encompassed by all of the slope
units \citep{Alvioli2016,Alvioli2020a} overlapping with a 1 km buffer delineated around the railway
track.

Results are described in three different steps:
\begin{itemize}
\item[(i)] agreement between modeled and observed rockfall source areas mapped by experts, and with
  sub--areas of the polygons in the IFFI inventory (\textit{cf.} \textbf{Section \ref{subs:methods_sources}});
\item[(ii)] assessment of the trajectory count output of the program STONE, performed calculating the agreement
  between runout areas and polygons in the IFFI inventory; classification of railway segments into rockfall
  susceptibility classes (\textit{cf.} \textbf{Section \ref{subs:methods_stone}}); 
\item[(iii)] analysis of the impact of the network classification for the railway network within the newly
  defined network--ranked susceptibility (\textit{cf.} \textbf{Section \ref{subs:methods_graph}}).
\end{itemize}

We believe it interesting to show and discuss separately the results of source area modeling, point (i),
since they represent a relevant input to the physically based model used in this
work and, probably, to similar approaches. The procedure we developed for the purpose was calibrated
on a sample of expert--mapped source areas.
The output of the statistical generalization is a grid with with probabilistic interpretation;
comparison between modeled probability and mapped source areas used for calibration, thus, is meaningful.
To this end, we calculated the number of grid cells encompassed by polygons representing
mapped source areas in which the statistical procedure assigned non--null probability, and the number of
cells with values of probability larger than 0.8, to check if larger probabilities are found within
the mapped polygons.

Relevance of point (ii) is straightforward, as rockfall susceptibility along the railway track is the
primary goal of this work. Point (iii) gives insight into the mutual relationship between susceptibility
classes and the role of each railway segment within the network as a whole.
\subsection{Results of the procedure to identify source areas}\label{subs:results_sources}
\begin{figure*}[!ht]
  \centerline{
    \includegraphics[width=0.28\textwidth]{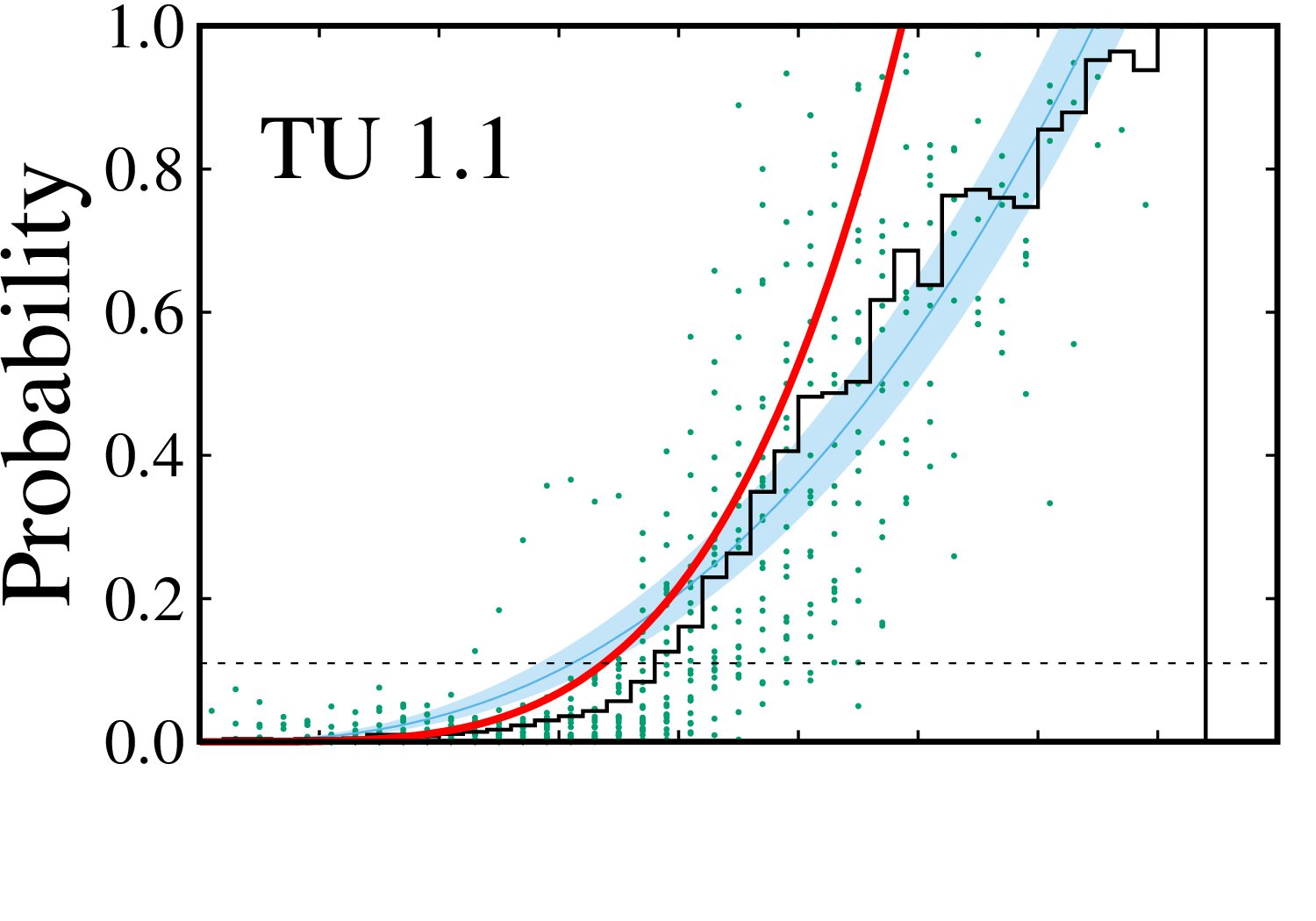}\hspace{-0.95cm}
    \includegraphics[width=0.28\textwidth]{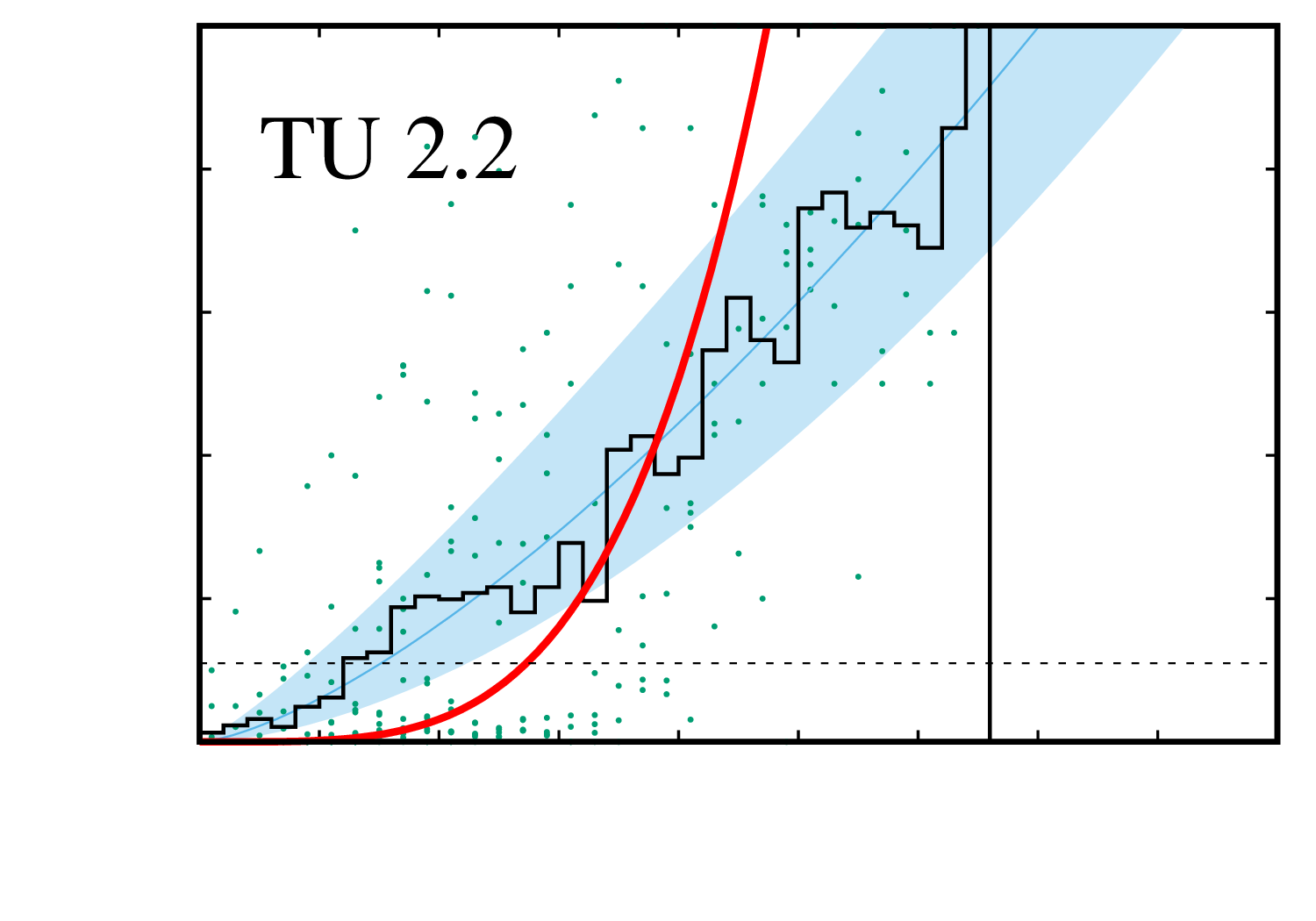}\hspace{-0.95cm}
    \includegraphics[width=0.28\textwidth]{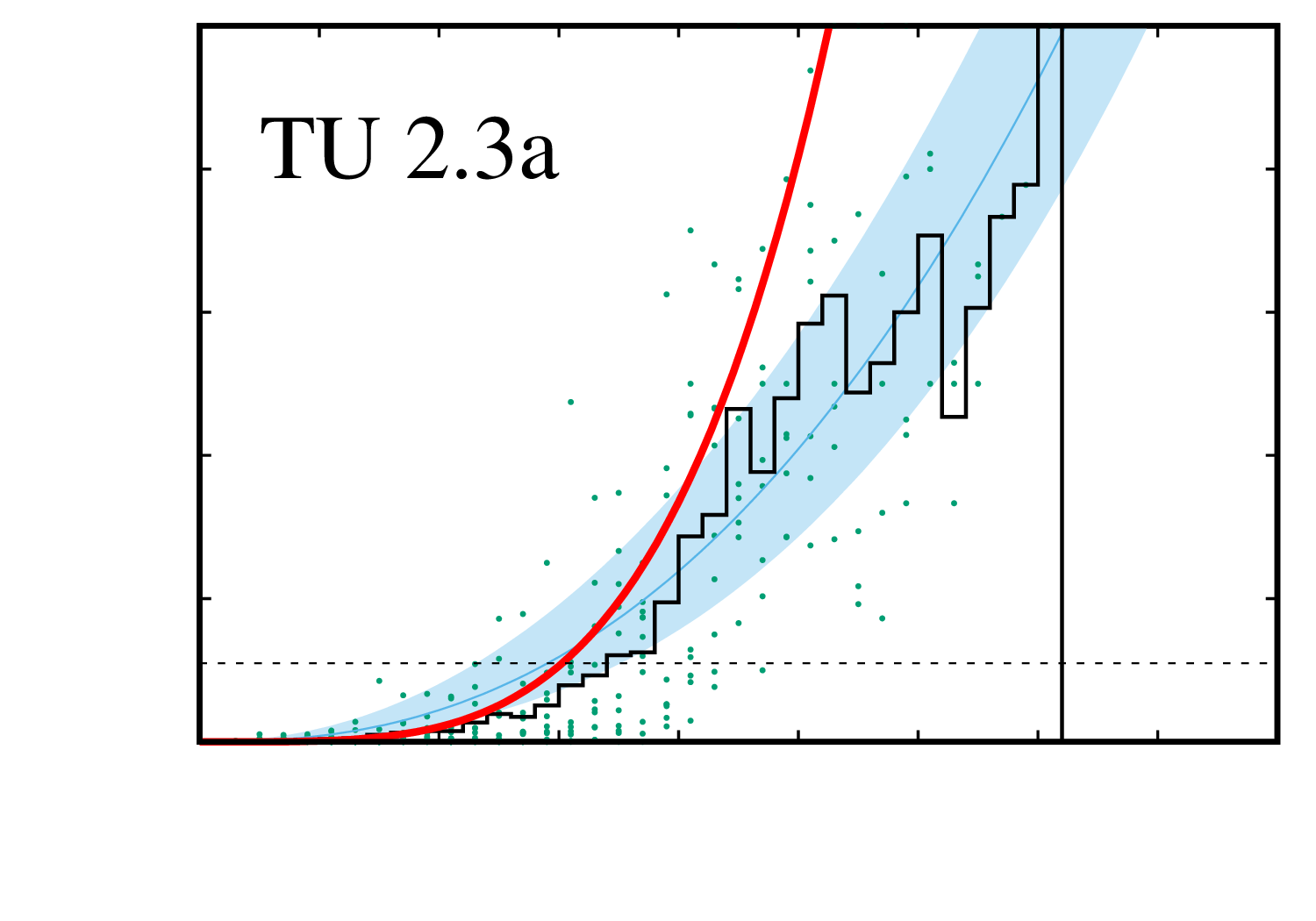}\hspace{-0.95cm}
    \includegraphics[width=0.28\textwidth]{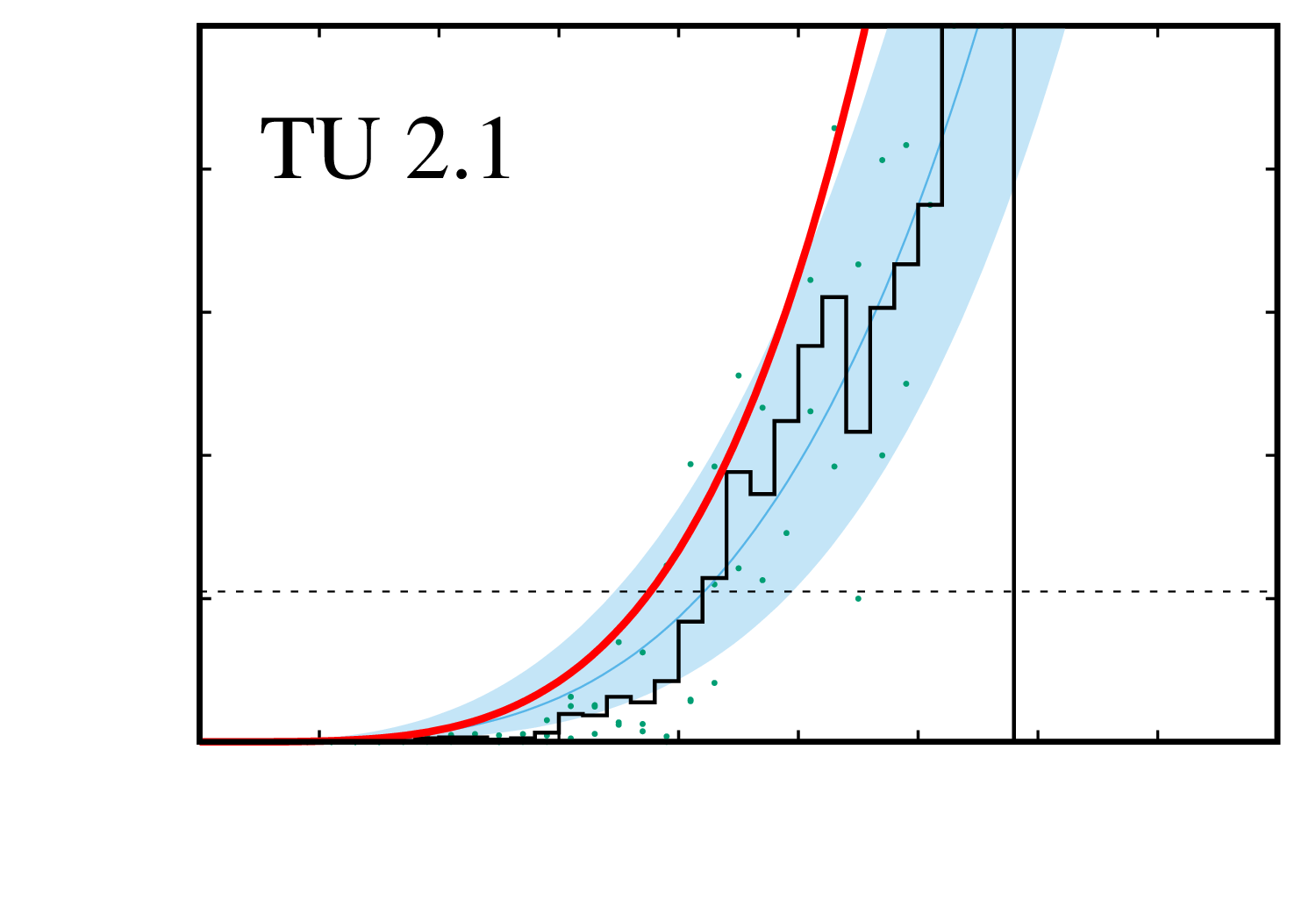}}
  \vskip -0.7cm
  \centerline{          
    \includegraphics[width=0.28\textwidth]{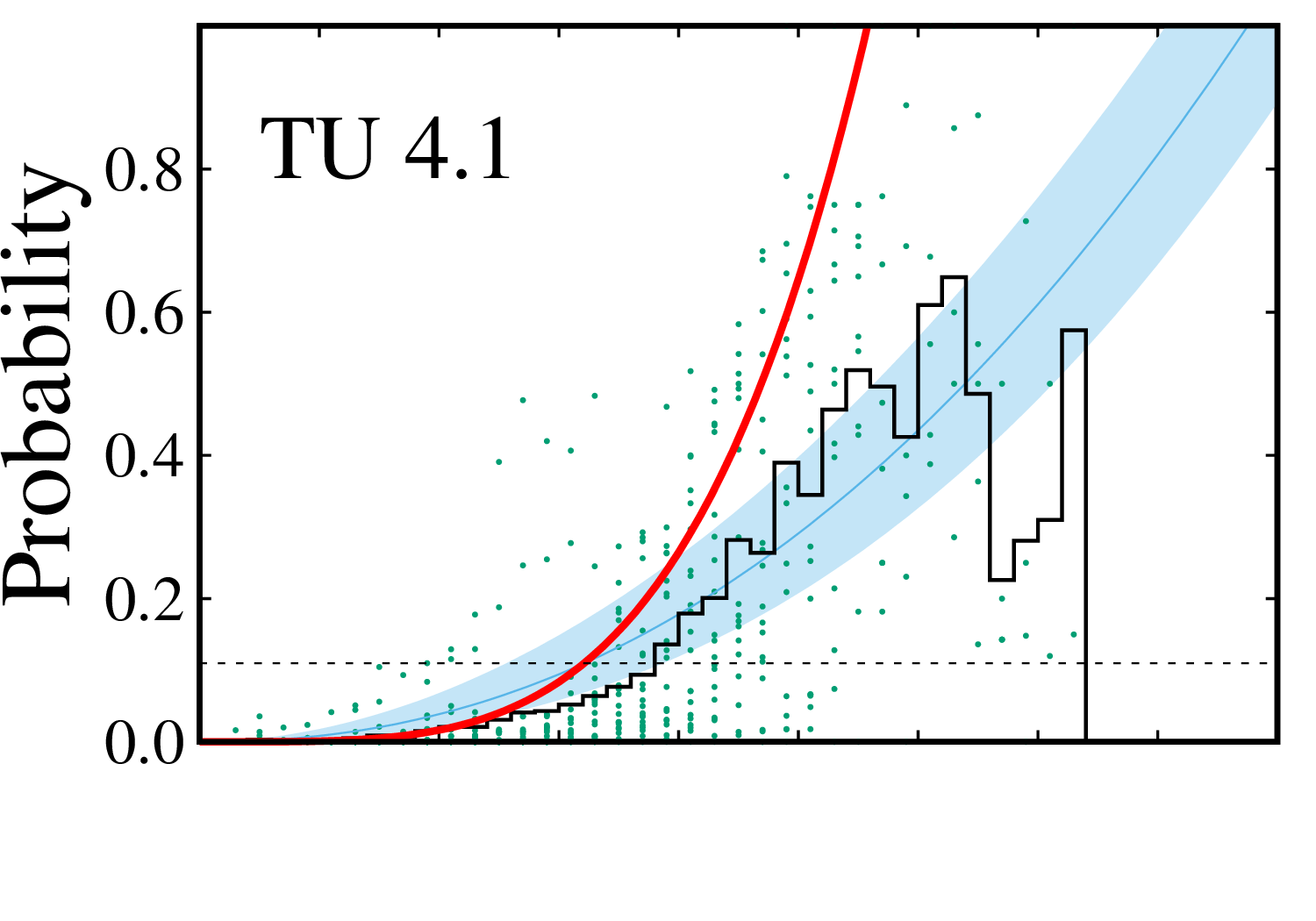}\hspace{-0.95cm}
    \includegraphics[width=0.28\textwidth]{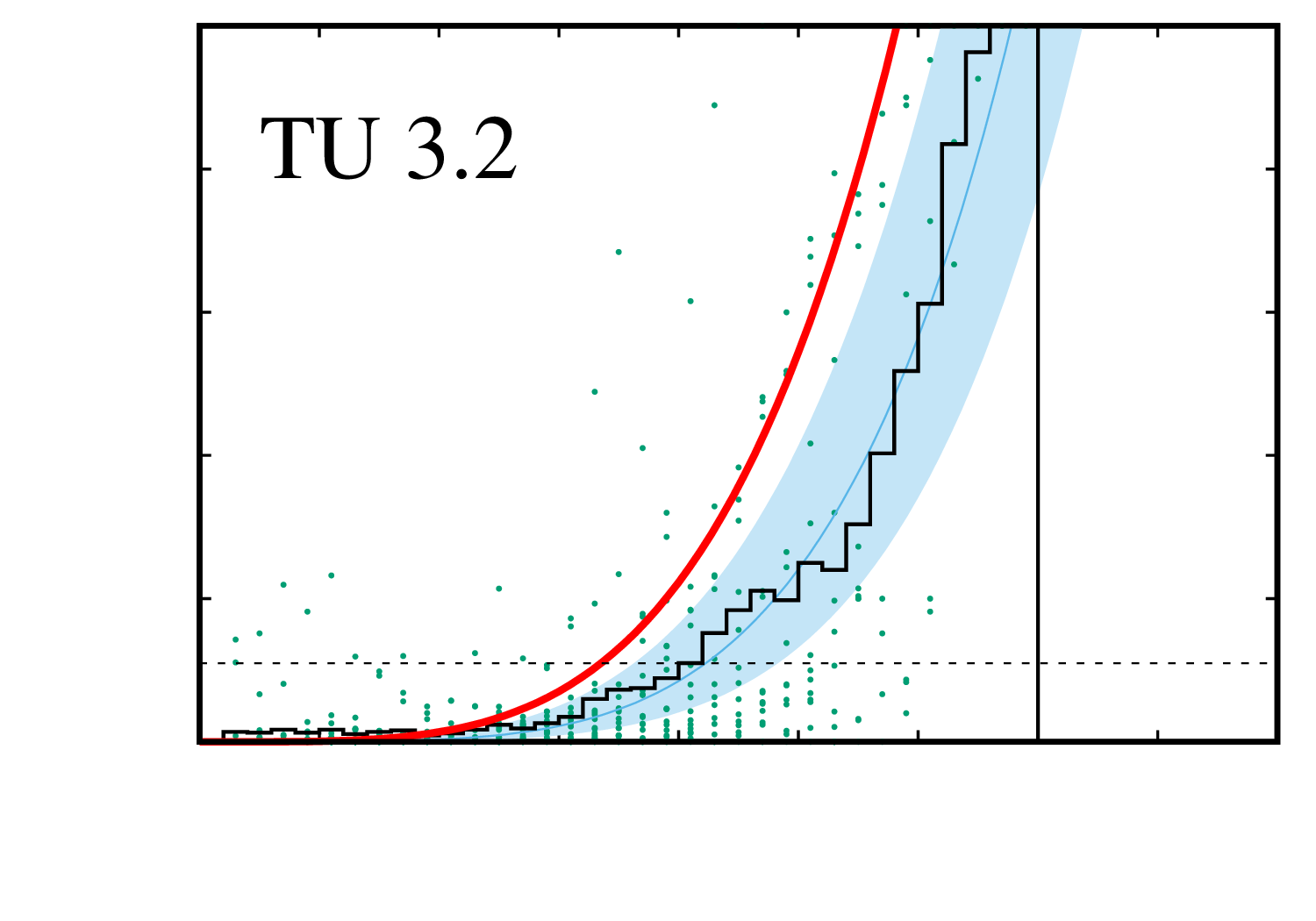}\hspace{-0.95cm}
    \includegraphics[width=0.28\textwidth]{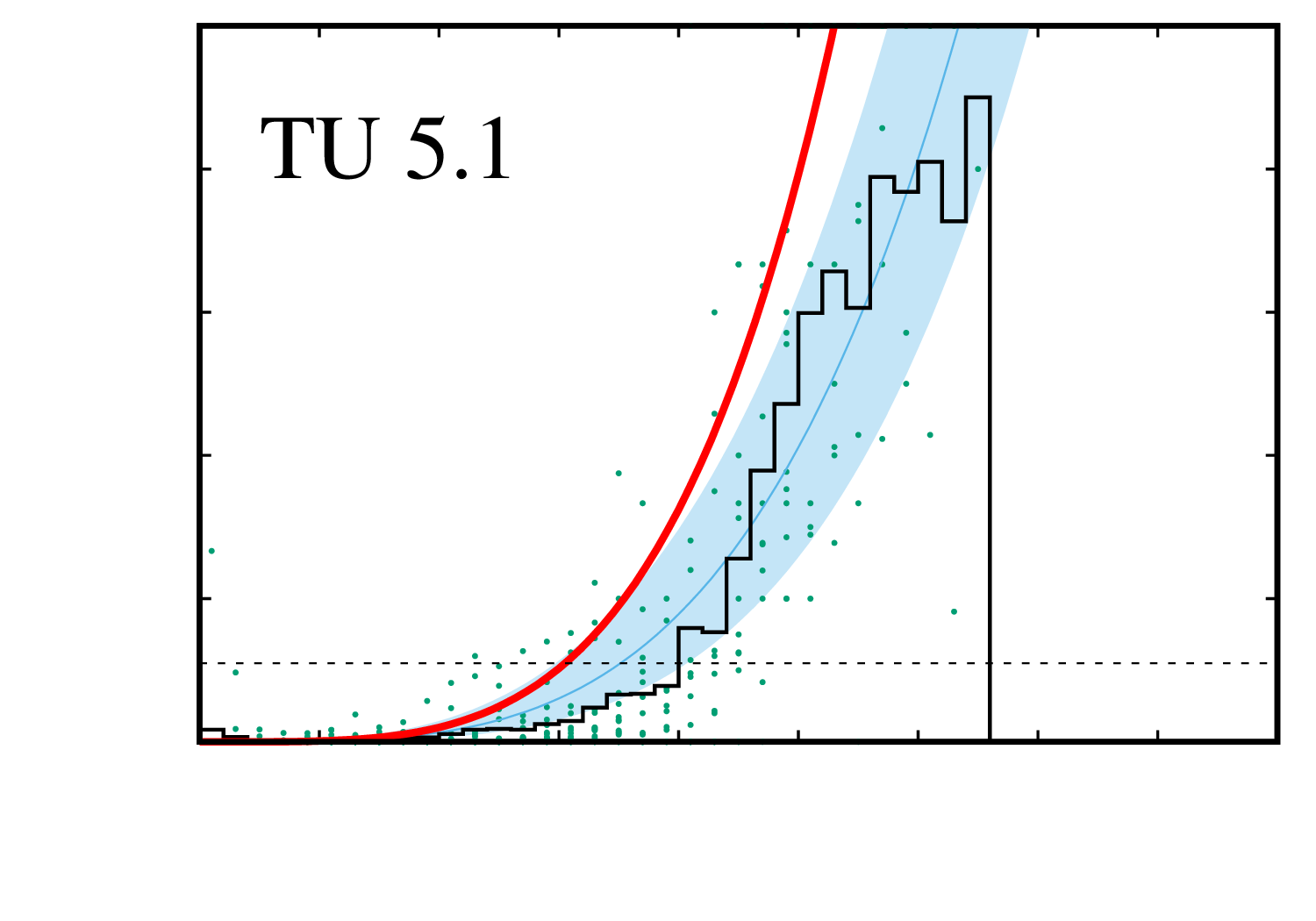}\hspace{-0.95cm}
    \includegraphics[width=0.28\textwidth]{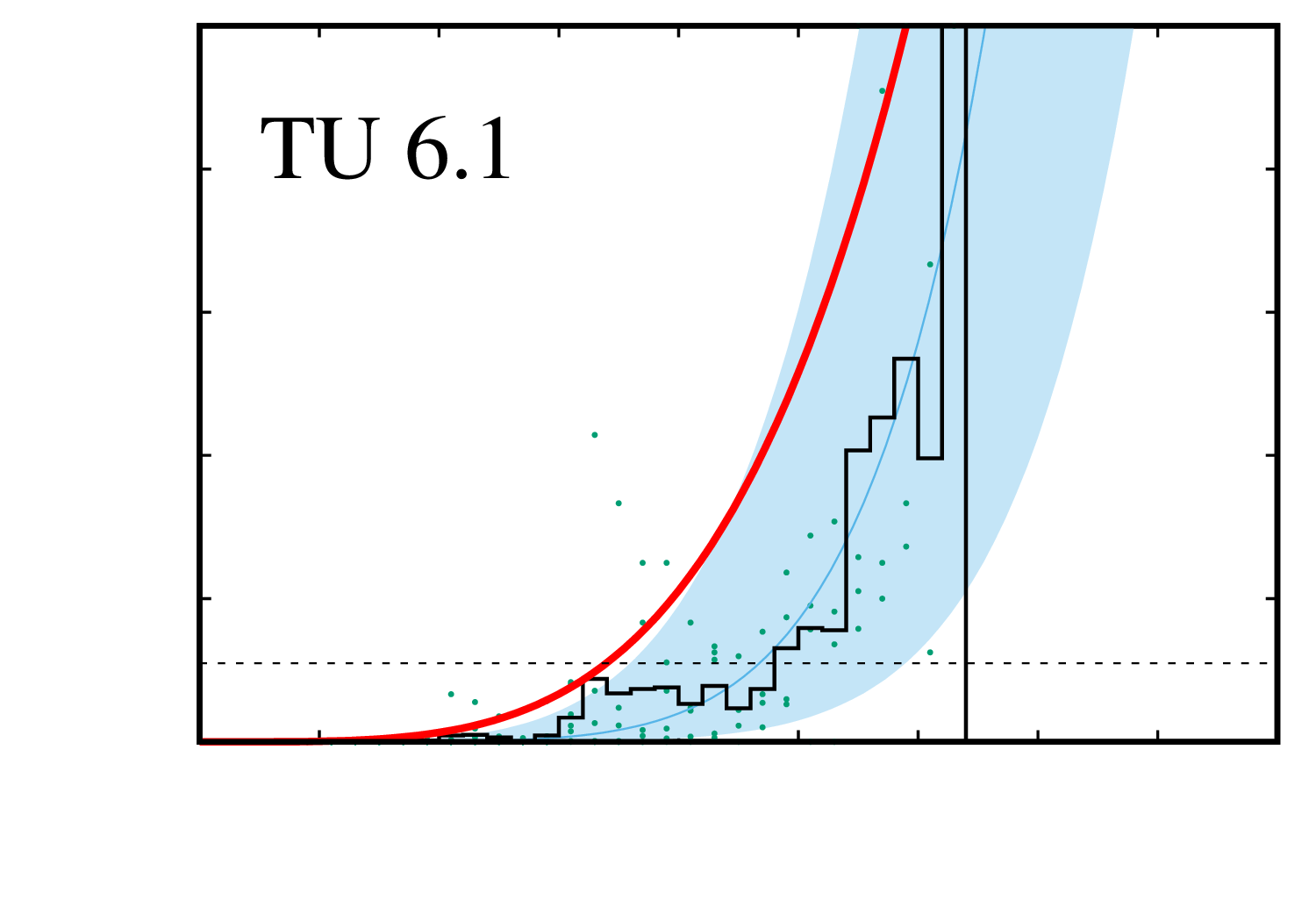}}
  \vskip -0.7cm
  \centerline{            
    \includegraphics[width=0.28\textwidth]{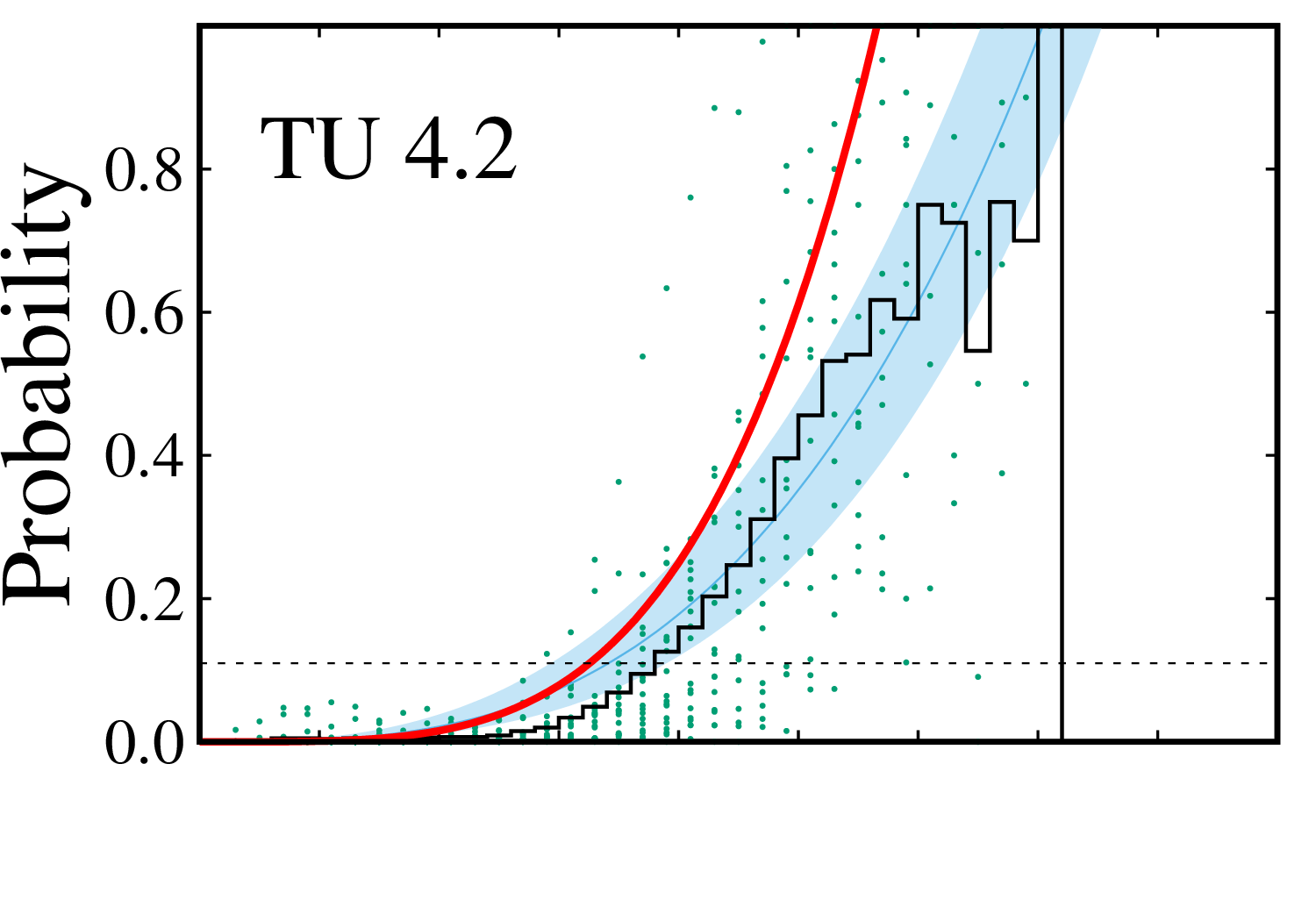}\hspace{-0.95cm}
    \includegraphics[width=0.28\textwidth]{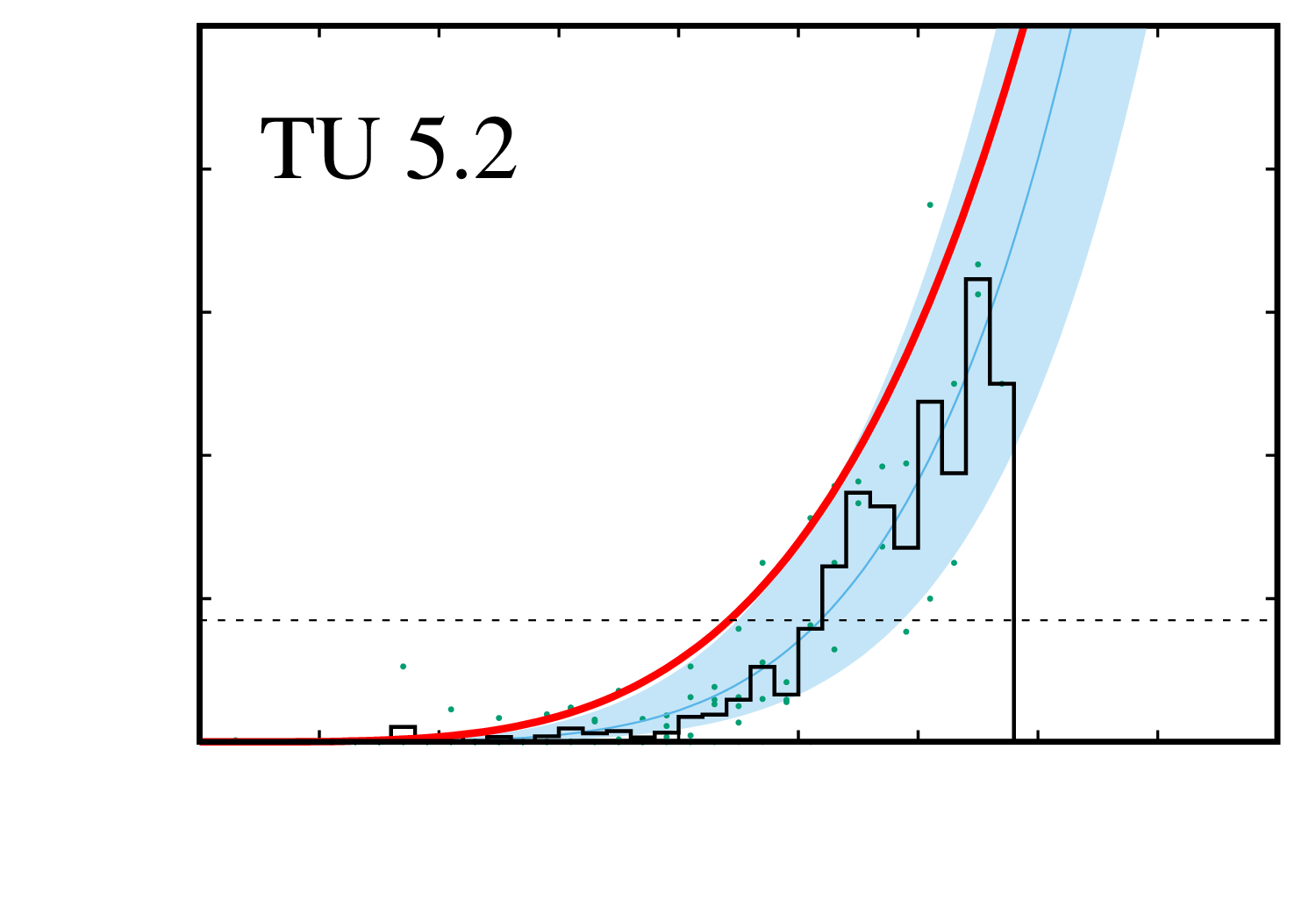}\hspace{-0.95cm}
    \includegraphics[width=0.28\textwidth]{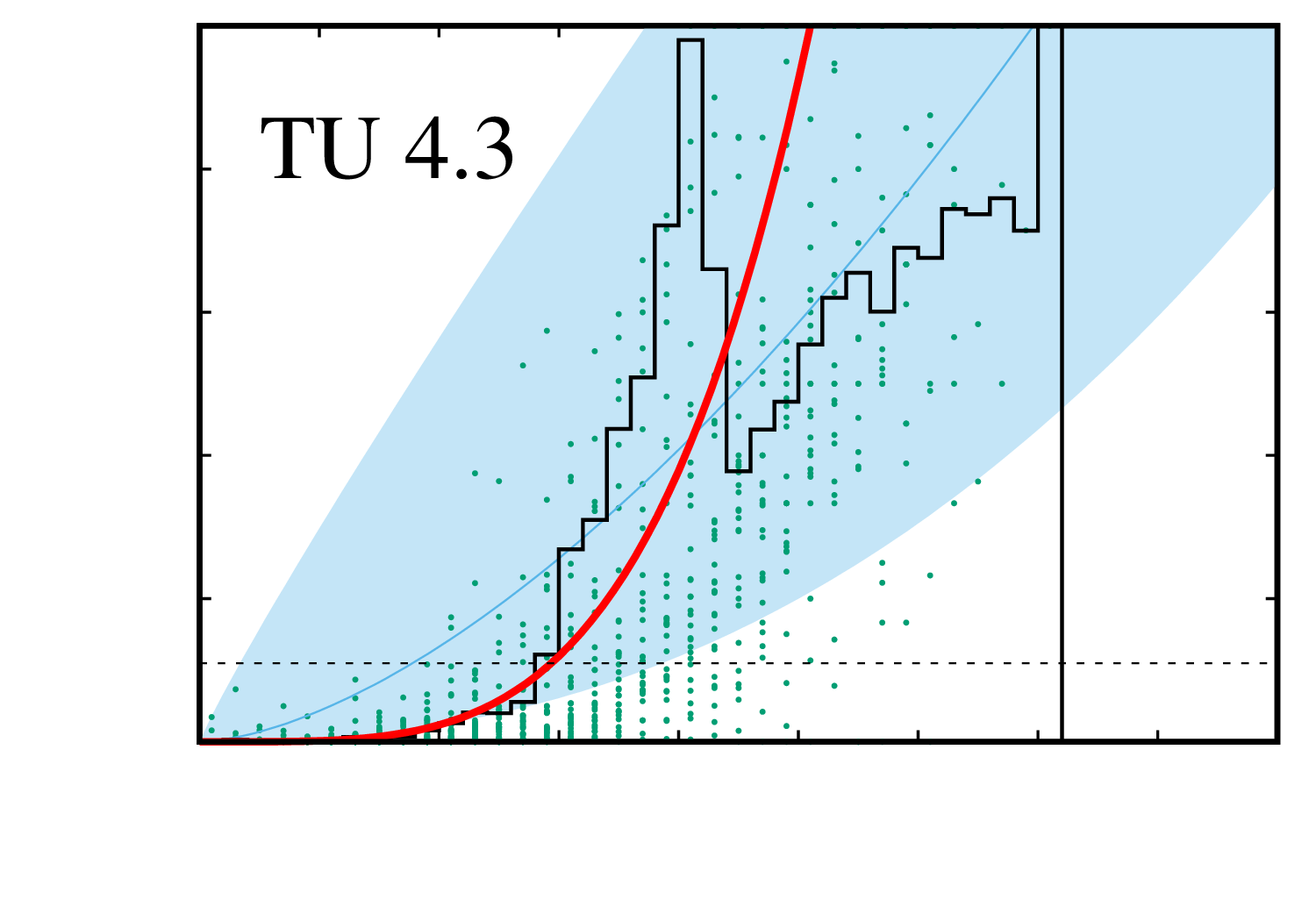}\hspace{-0.95cm}
    \includegraphics[width=0.28\textwidth]{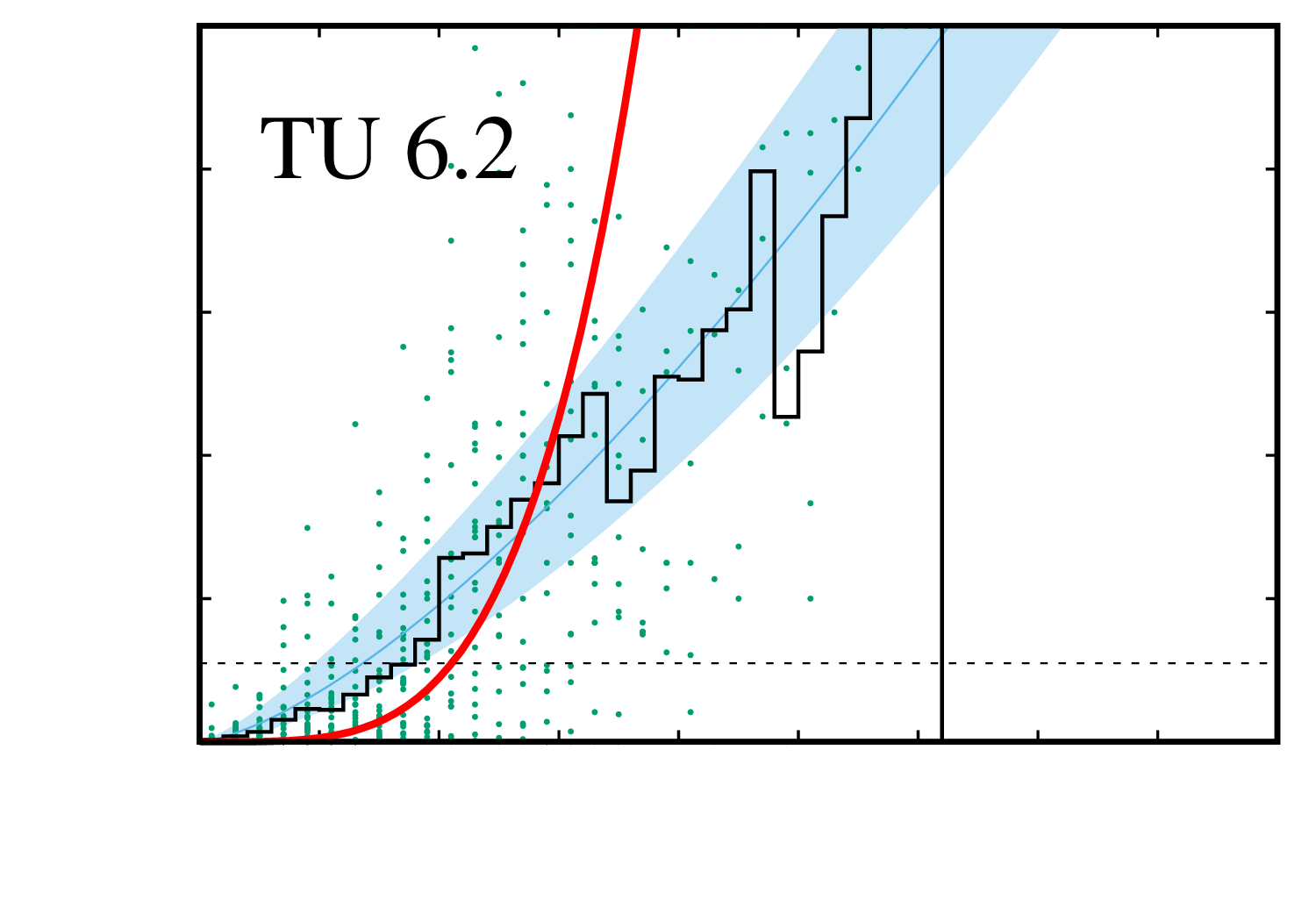}}
  \vskip -0.7cm
  \centerline{              
    \includegraphics[width=0.28\textwidth]{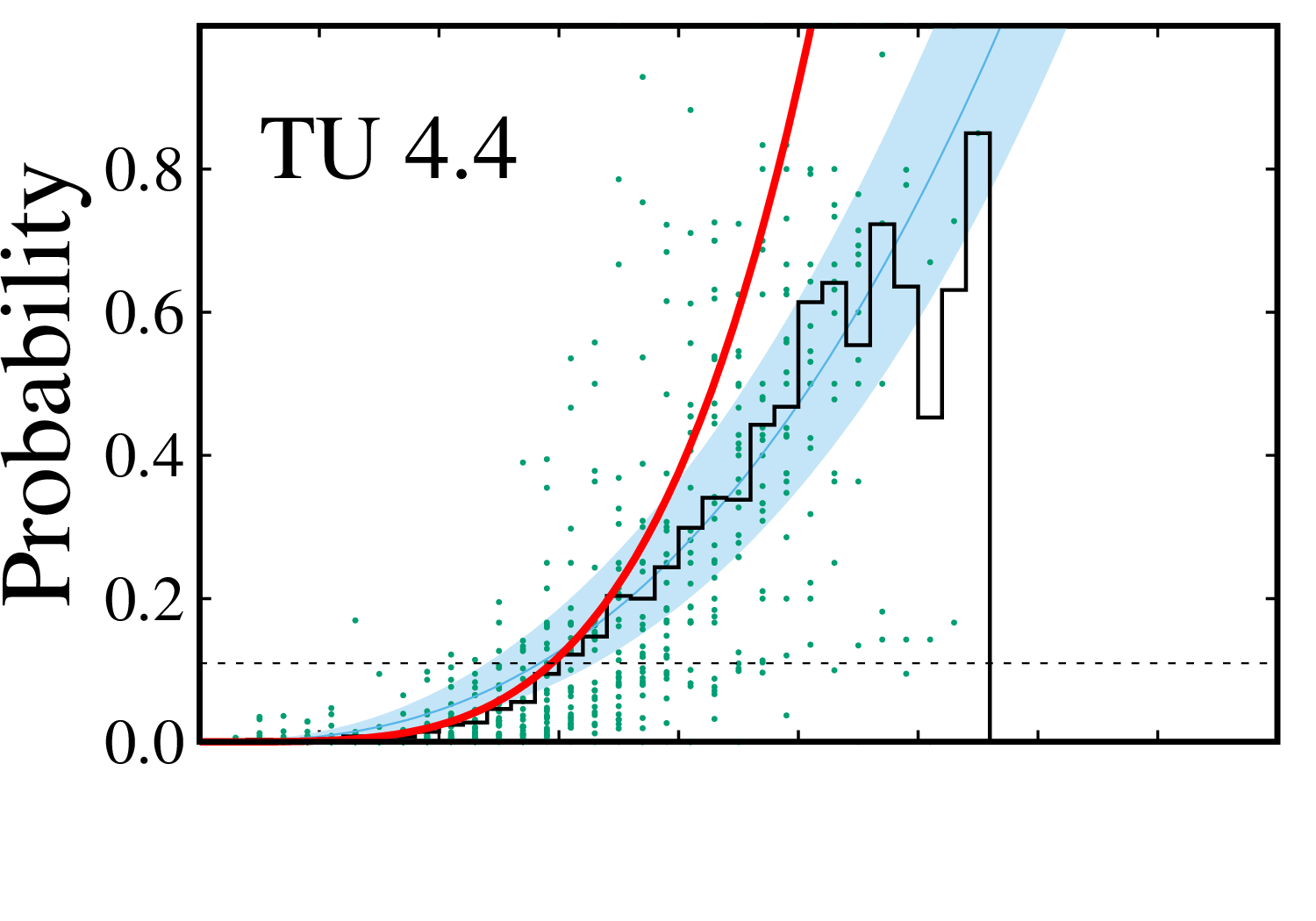}\hspace{-0.95cm}
    \includegraphics[width=0.28\textwidth]{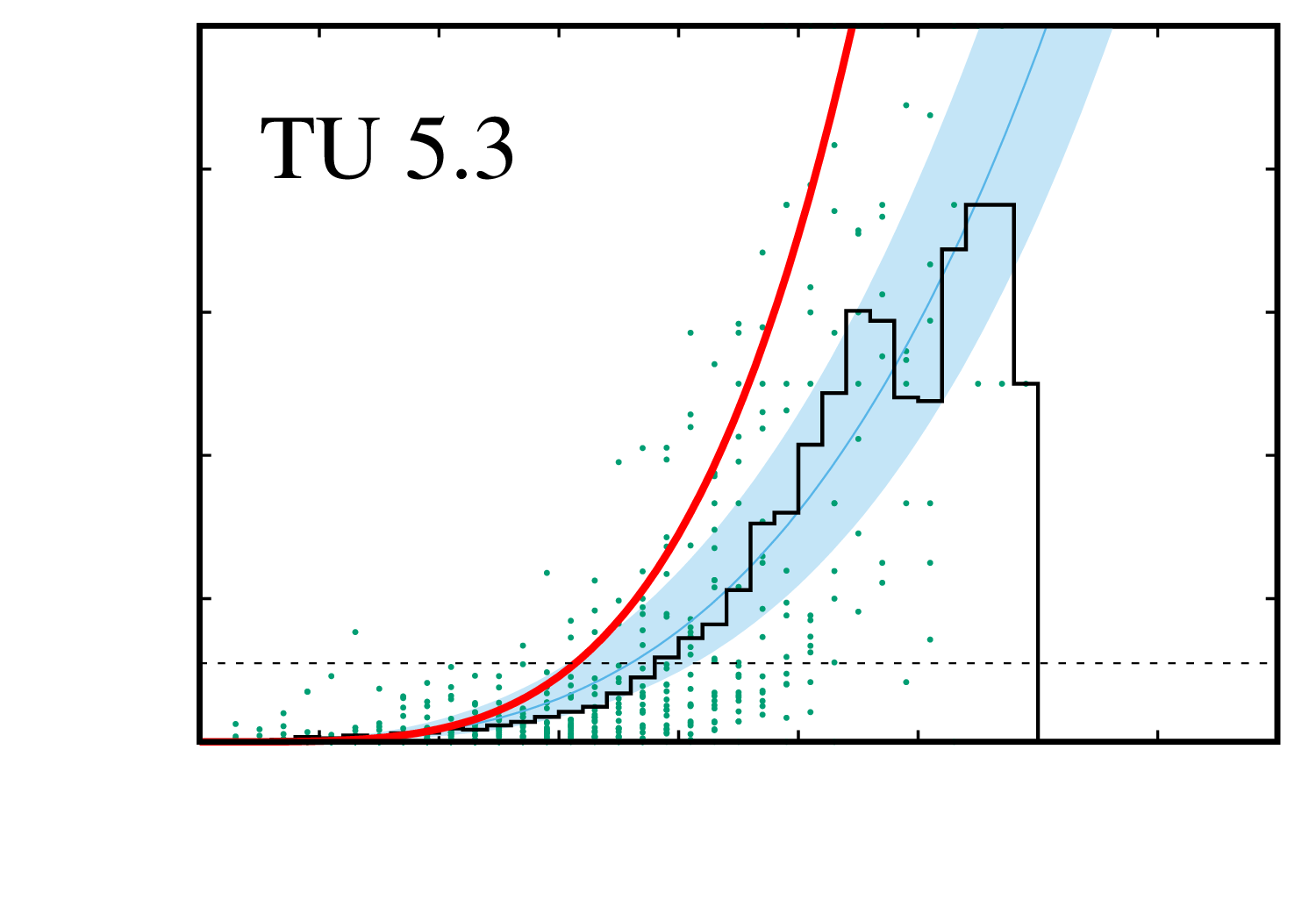}\hspace{-0.95cm}
    \includegraphics[width=0.28\textwidth]{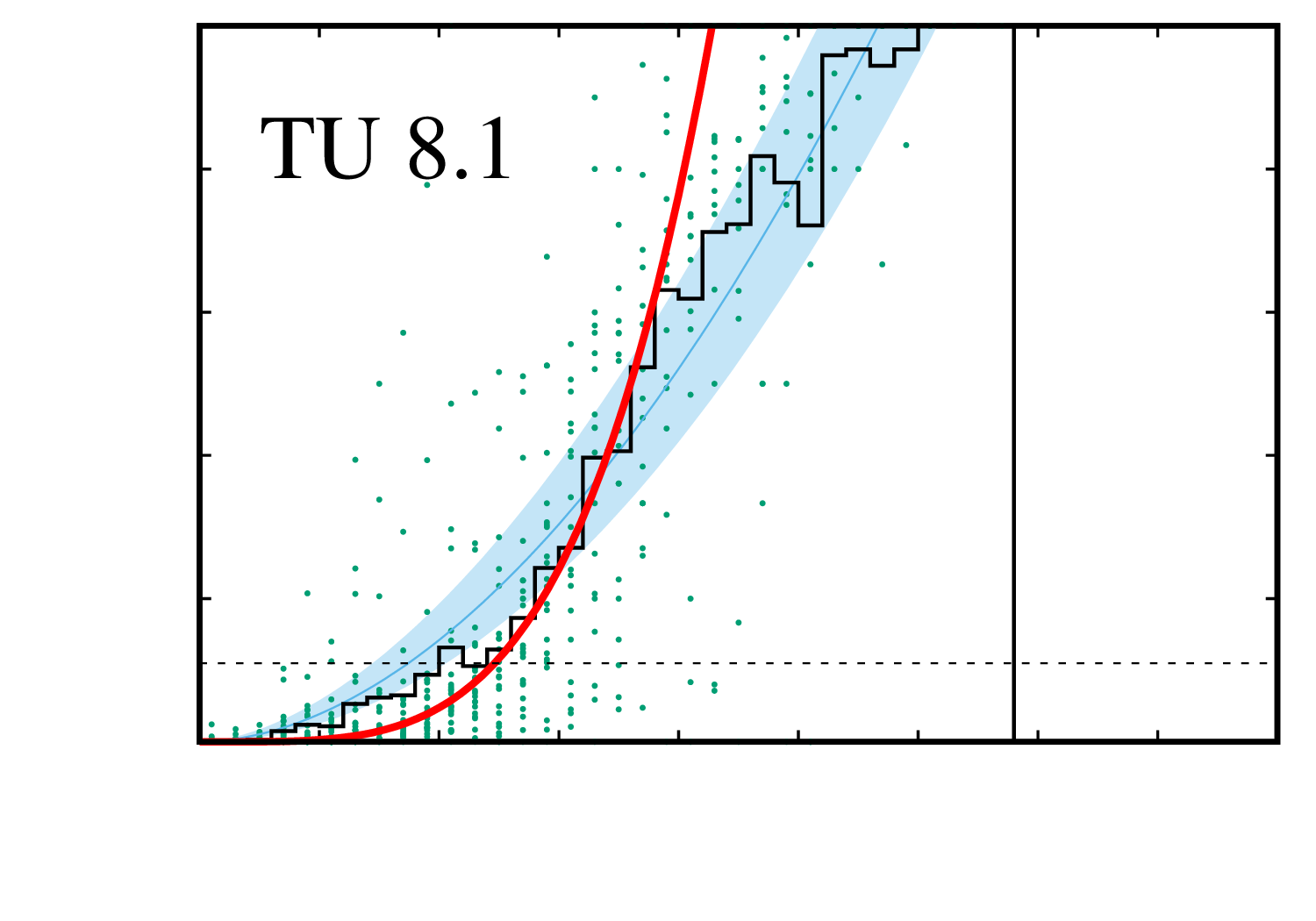}\hspace{-0.95cm}
    \includegraphics[width=0.28\textwidth]{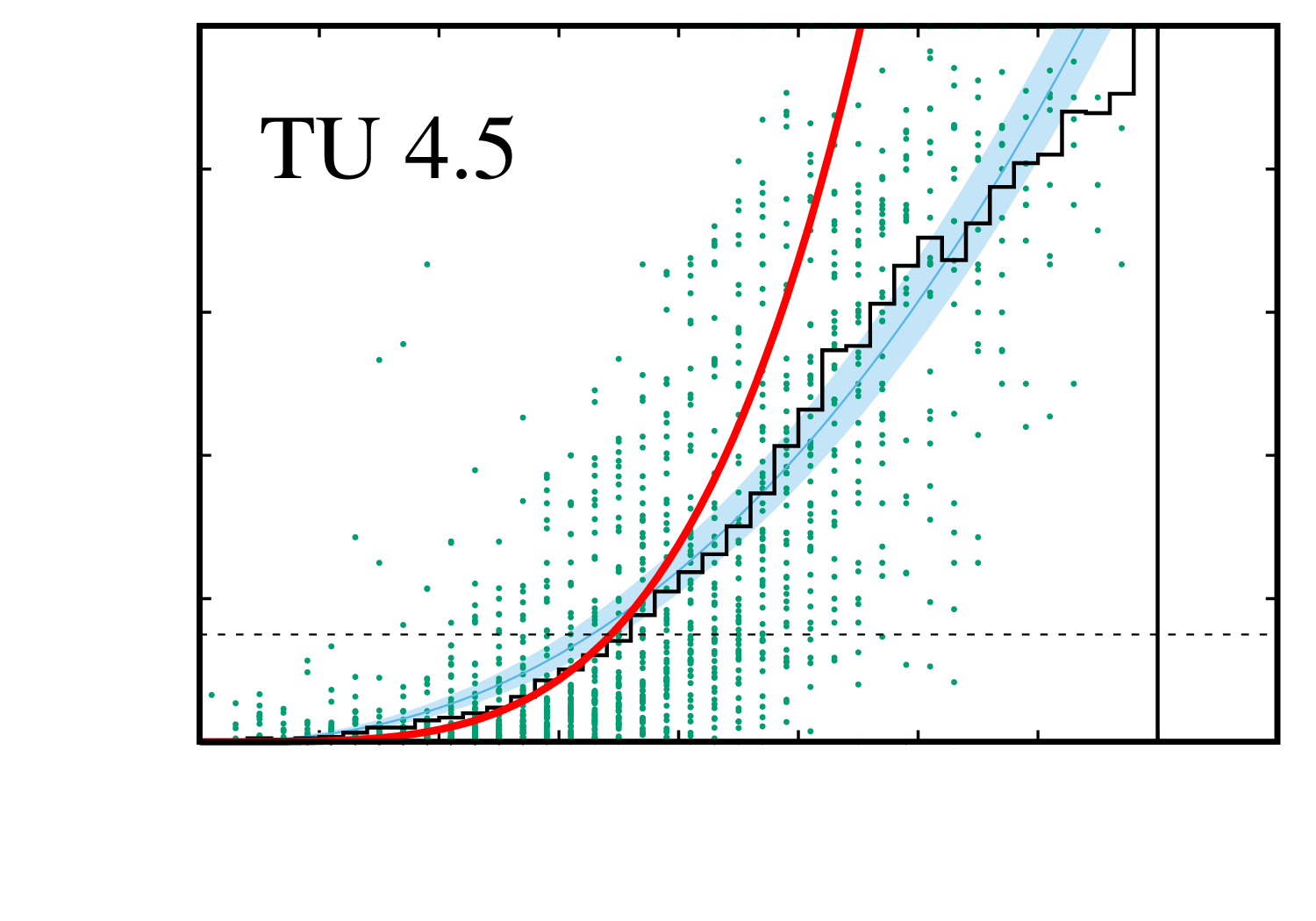}}
  \vskip -0.7cm
  \centerline{              
    \includegraphics[width=0.28\textwidth]{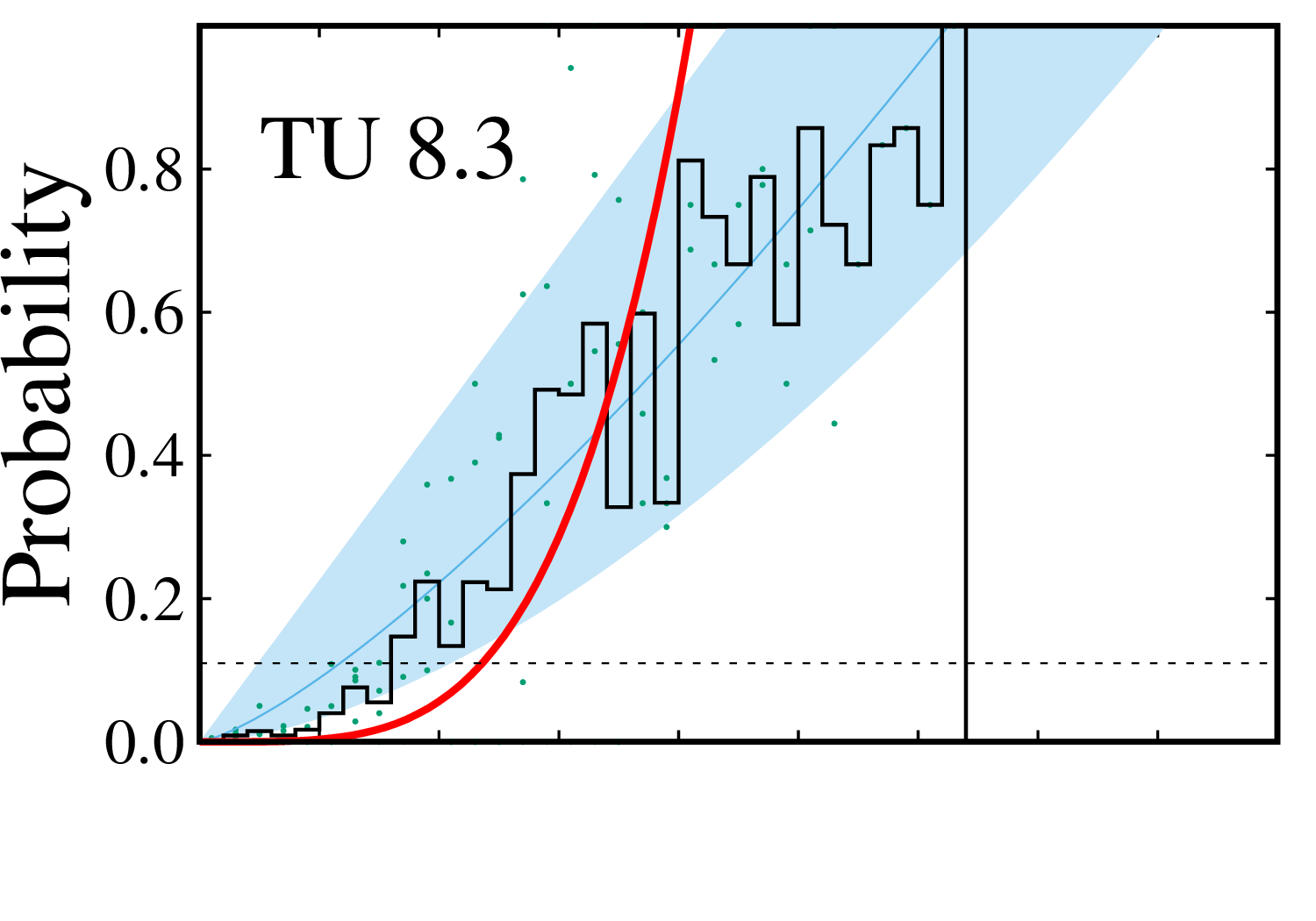}\hspace{-0.95cm}
    \includegraphics[width=0.28\textwidth]{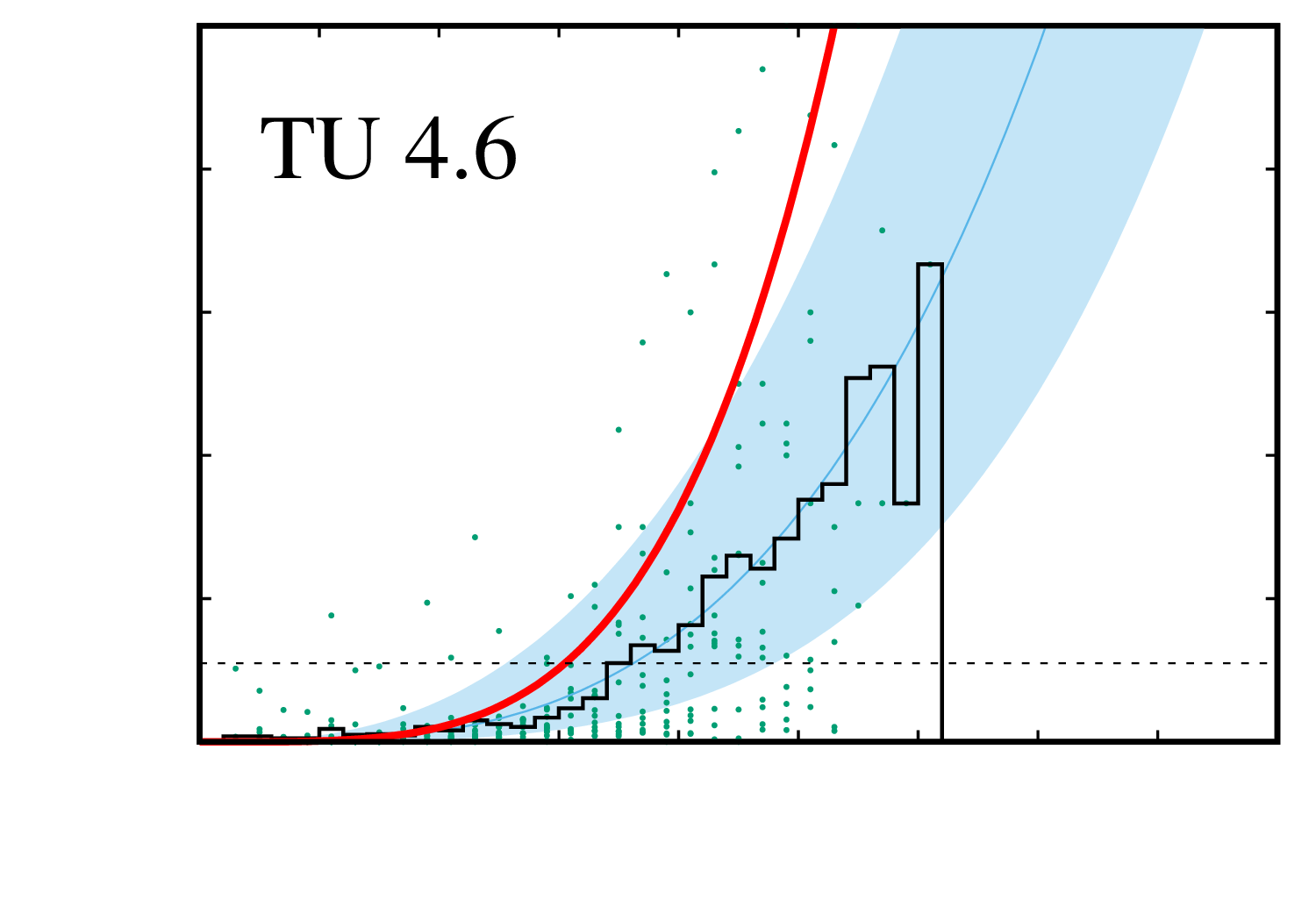}\hspace{-0.95cm}
    \includegraphics[width=0.28\textwidth]{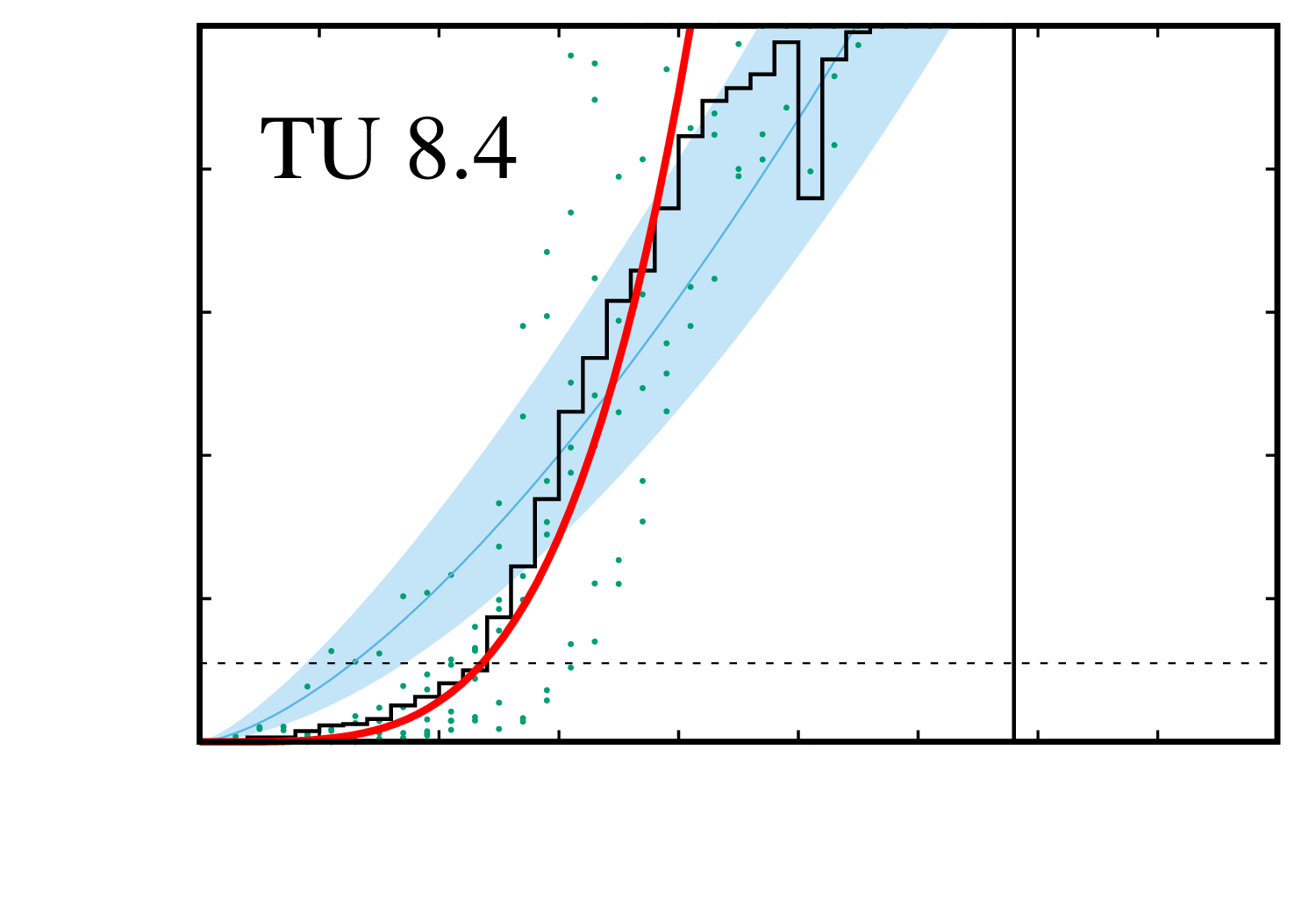}\hspace{-0.95cm}
    \includegraphics[width=0.28\textwidth]{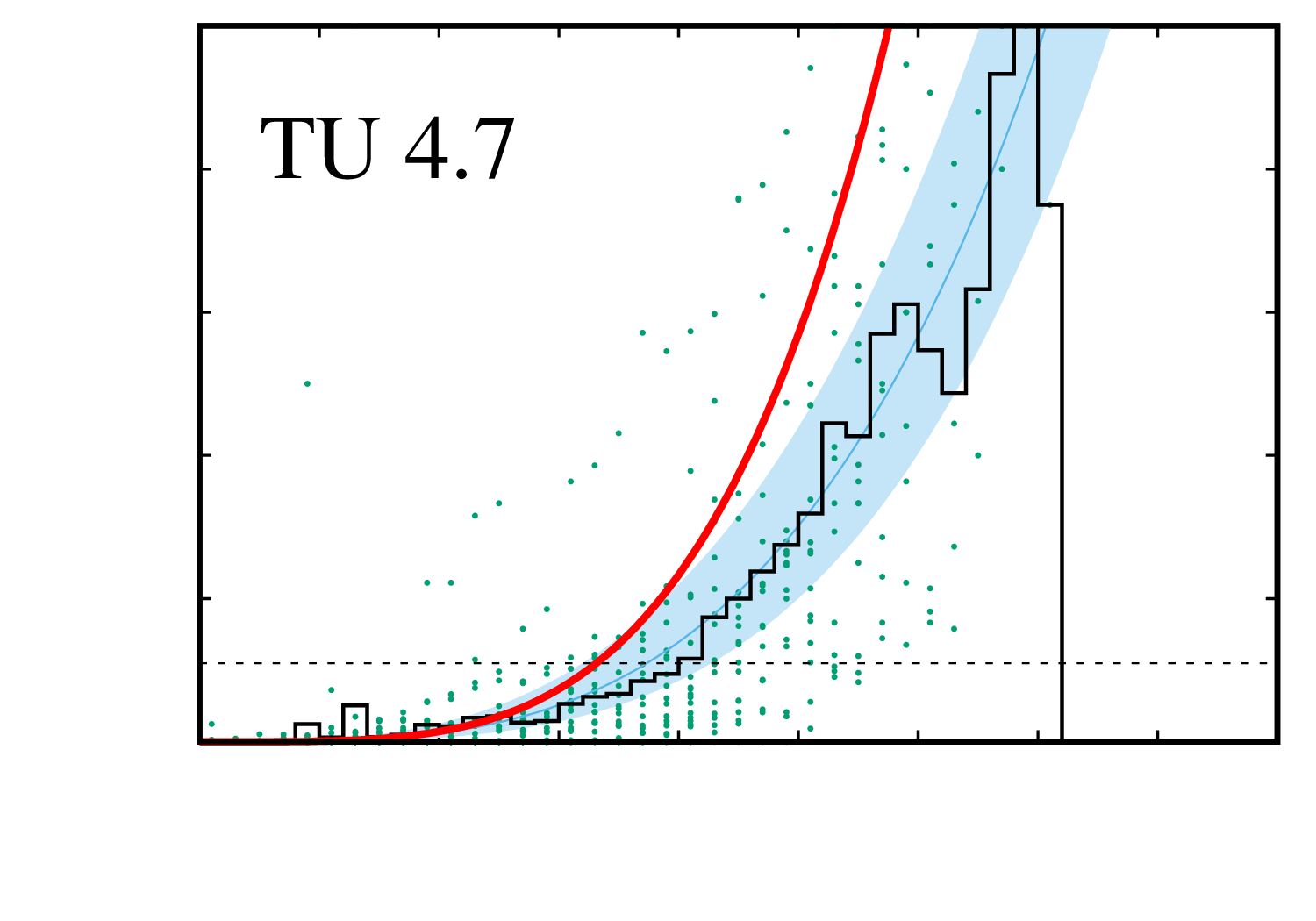}}
  \vskip -0.7cm
  \centerline{      
    \includegraphics[width=0.28\textwidth]{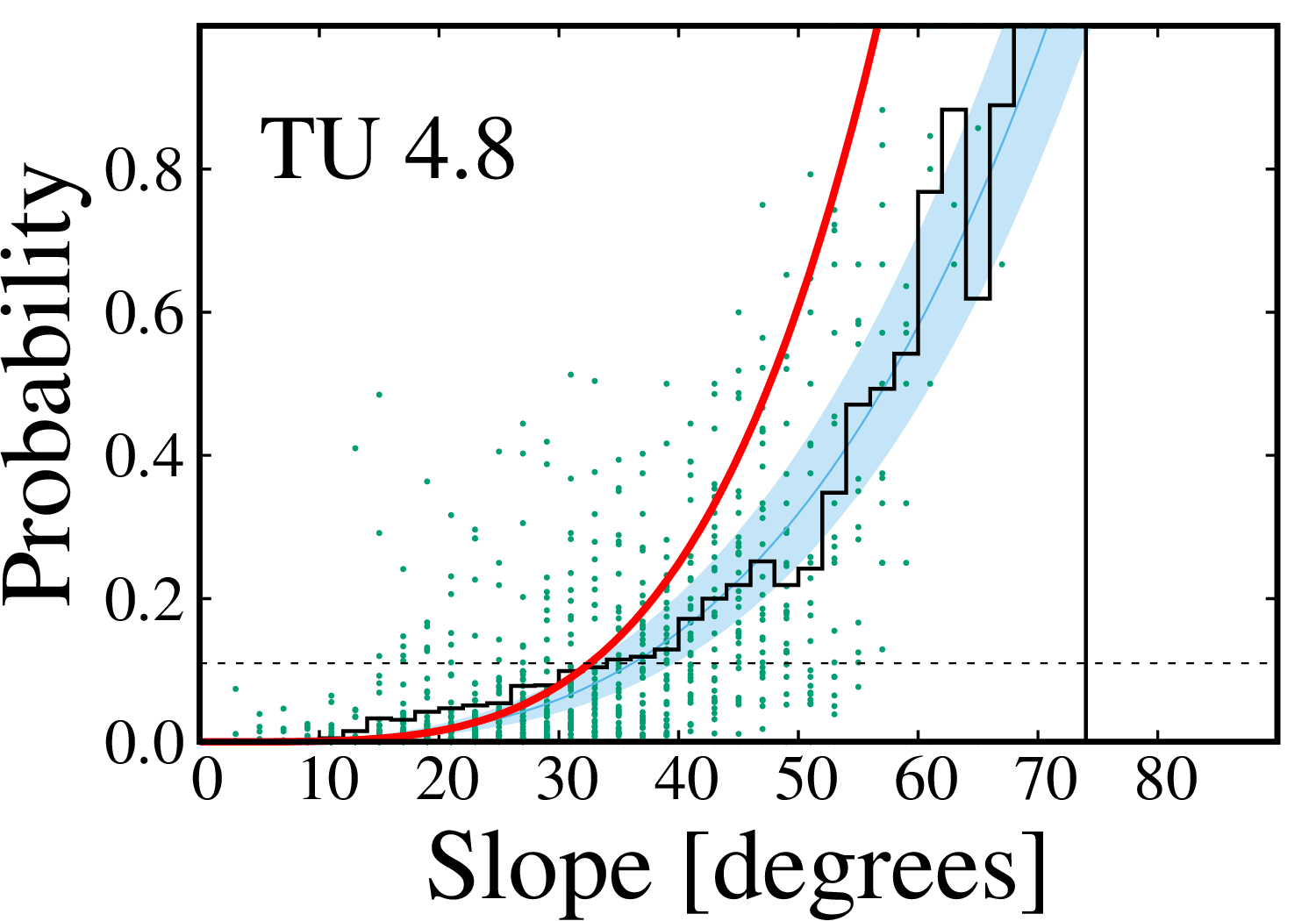}\hspace{-0.95cm}
    \includegraphics[width=0.28\textwidth]{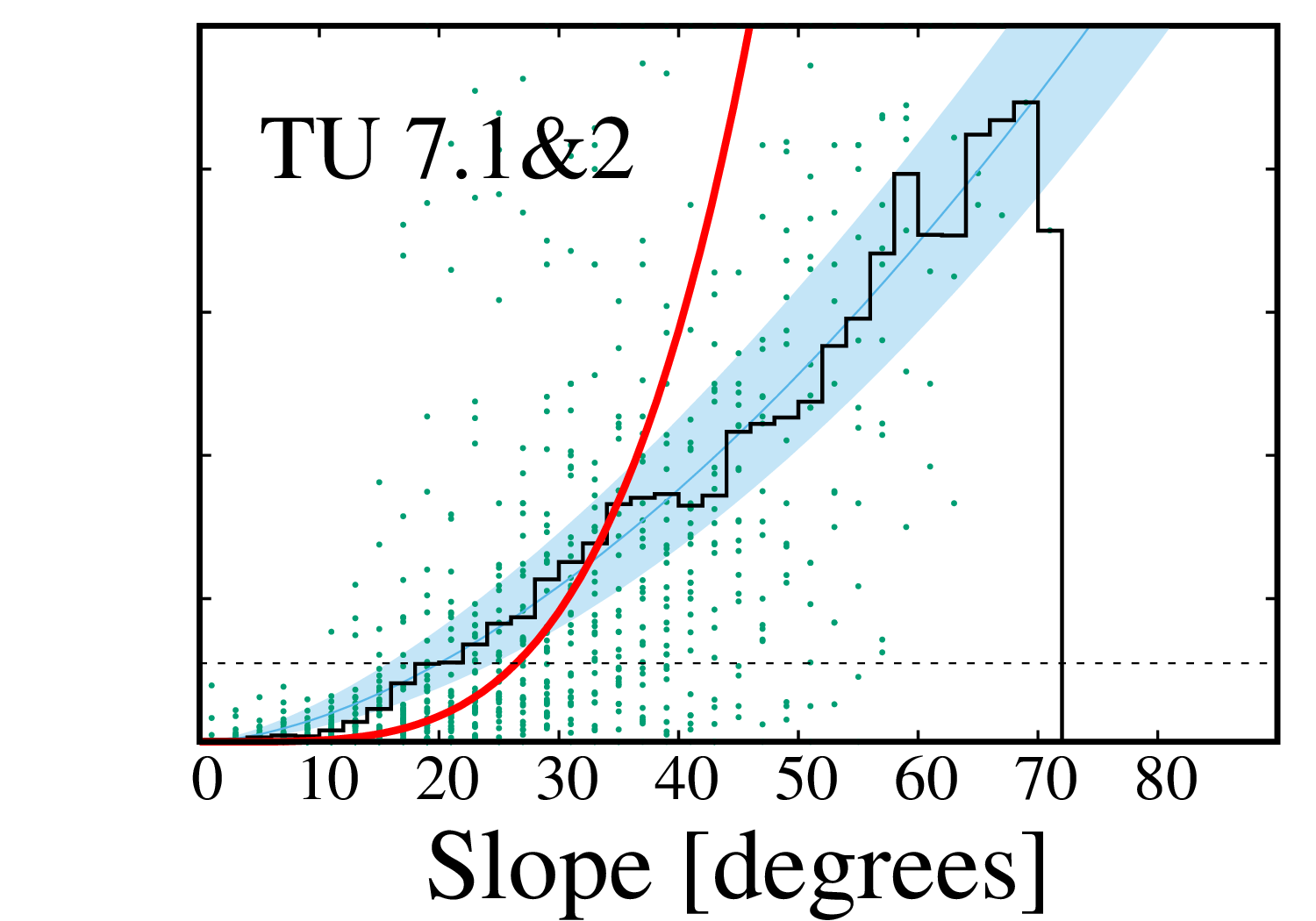}\hspace{-0.95cm}
    \includegraphics[width=0.28\textwidth]{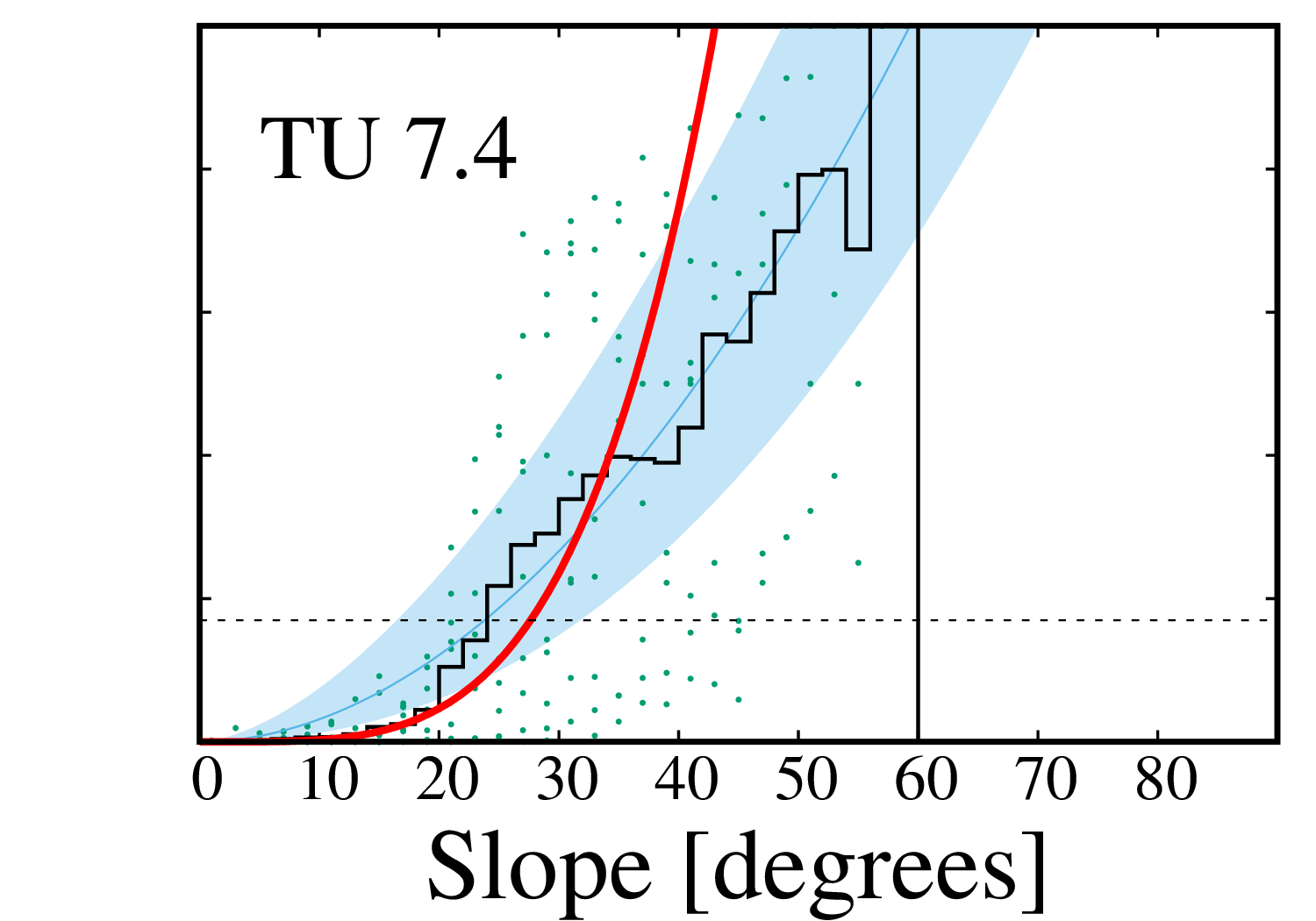}\hspace{-0.95cm}
    \includegraphics[width=0.28\textwidth]{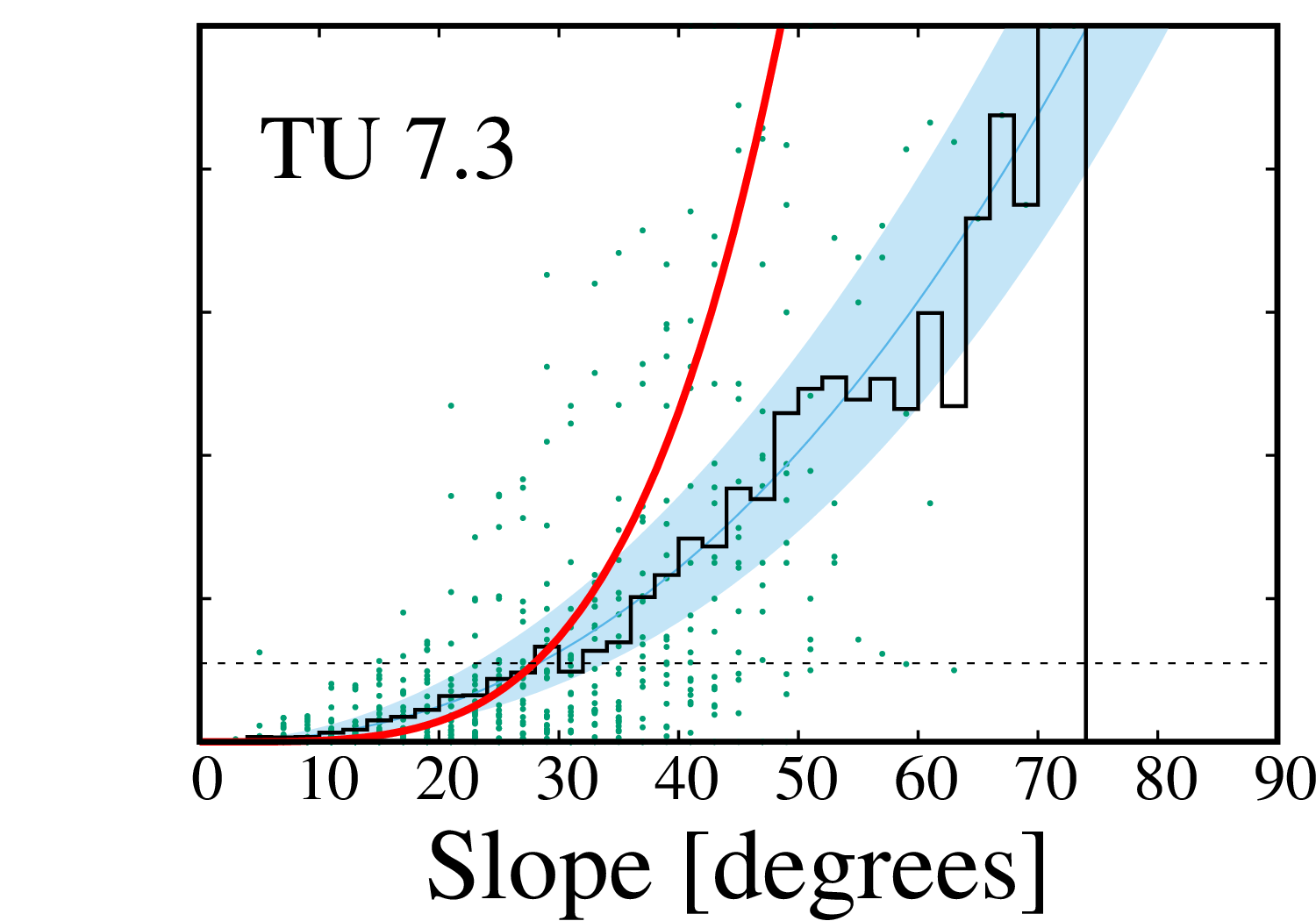}}
  \caption{Each panel is as in Fig. \ref{fig03}; topographic unit 1.2 \& 1.3 is shown in Fig.
    \ref{fig03}; 2.3a and 2.3b were grouped into one section, for what concerns modeling
    of source areas; section 3.1 does not contain mapped sources; sections 6.3 and
    8.2 contain no intersections between railway and slope units; this figure shows
    results for the remaining 24 sections.}
  \label{fig06}
\end{figure*}

The procedure described in \textbf{Section \ref{subs:methods_sources}} applies to each of the topographic
units adopted in this work as homogeneous domains for simulation with STONE.
\textbf{Figure \ref{fig06}} shows the probability--slope dependence in all of the TUs. Once source areas and the
associated probabilities were estimated, we visually checked that no evident errors existed along the entire railway
track. We systematically checked the results of the statistical classification of source areas along the entire
railroad network, through expert visual analysis. In a few selected locations, in which expert geomorphologists
considered that the infrastructure was potentially subject to manifest risk, additional source areas were delineated.
\textbf{Table \ref{tab03}} lists, for each topographic unit: the total area, the area of slope units considered
here (\textit{i.e.}, the extent of the study area, the number and total area of the expert--mapped
polygons of potential source grid cells.

\begin{figure}[!ht]
  \centerline{
    \includegraphics[width=0.45\textwidth]{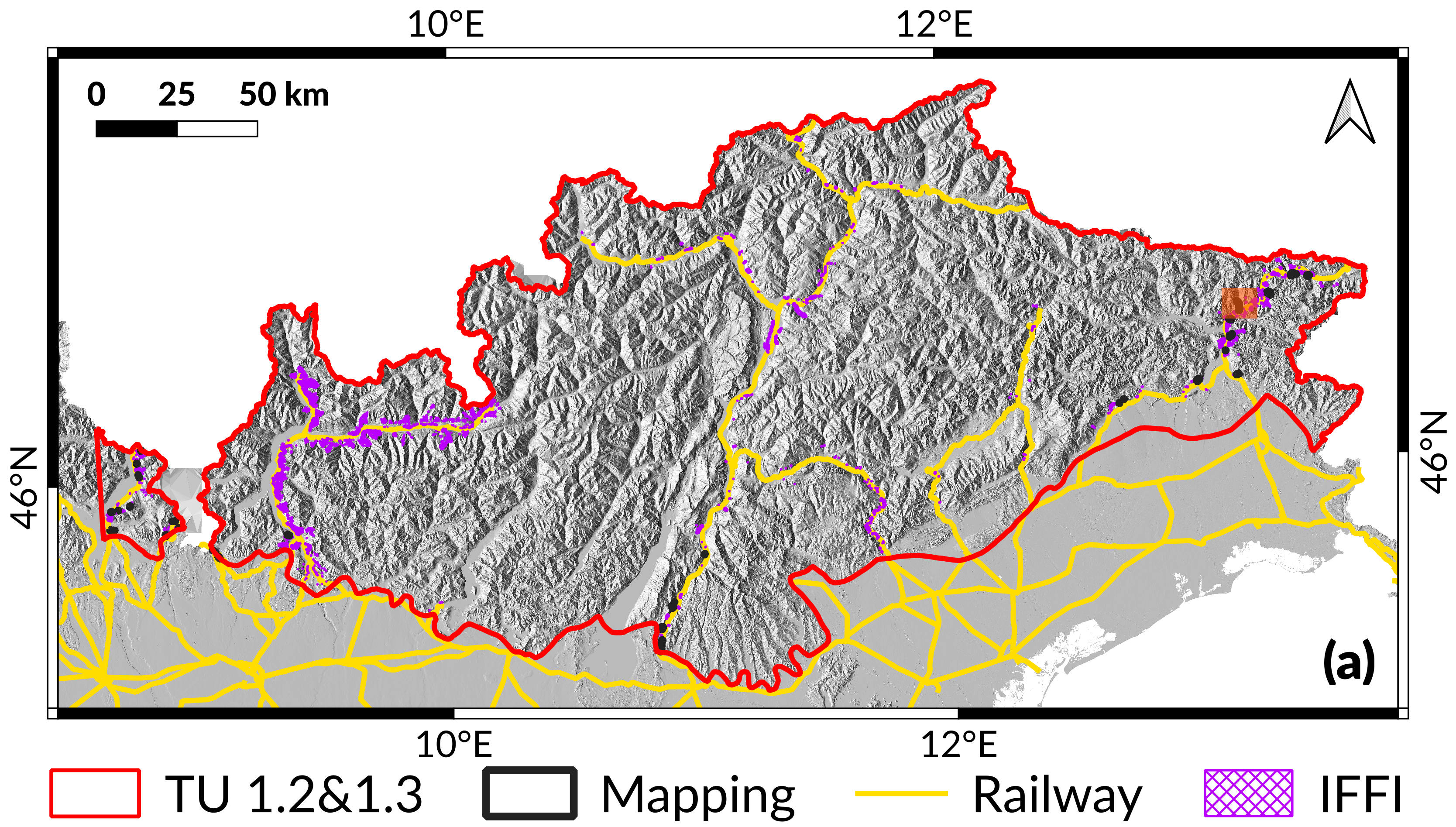}}        
  \centerline{
    \includegraphics[width=0.45\textwidth]{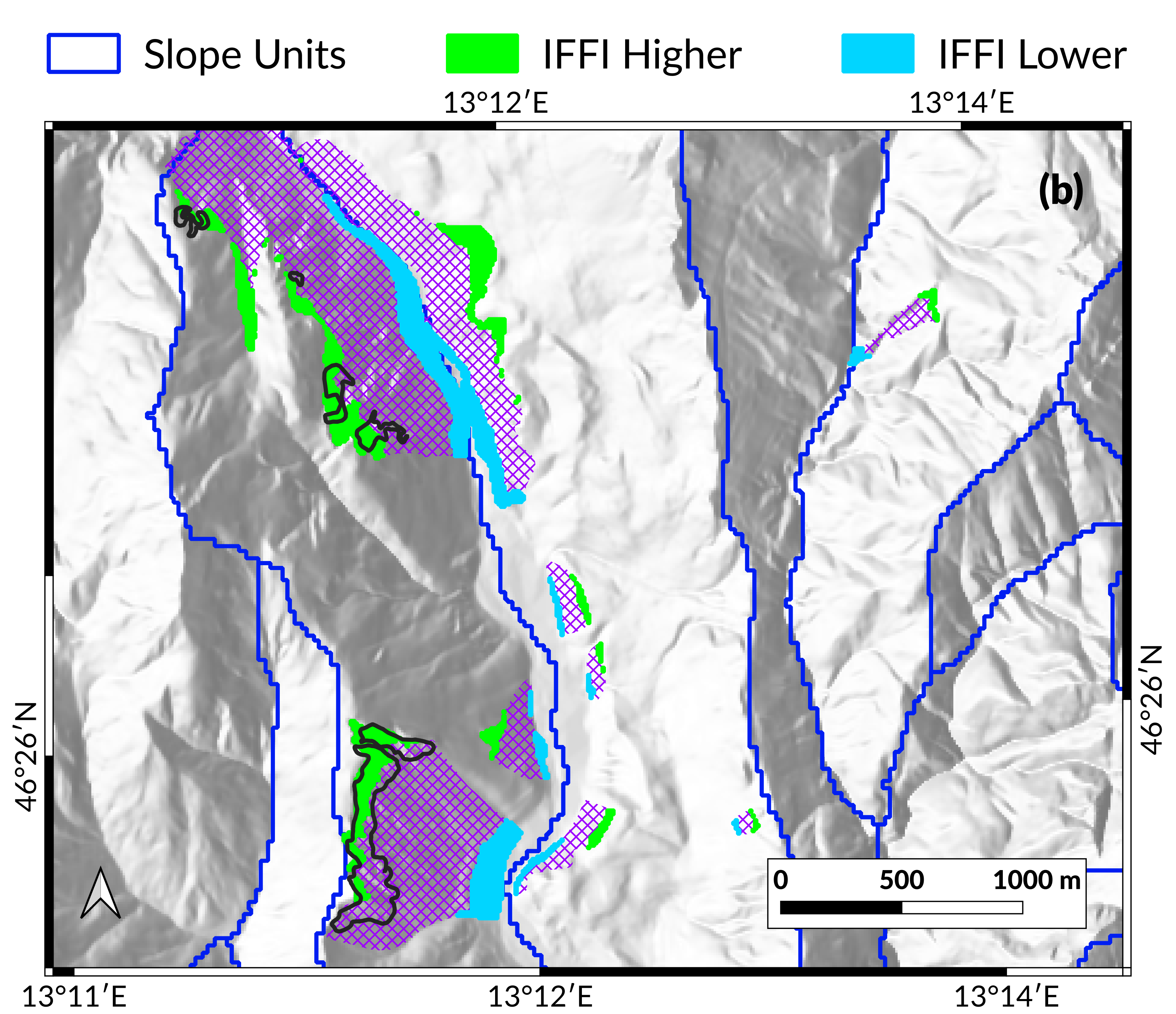}}        
  \caption{The relationship between expert--mapped source area and the IFFI inventory, limited to the
    buffer around the railroad track. (a) Overall view of one topographic unit. (b) A detail (small orange
    in the top figure) showing a few slope units in which both expert--mapped areas and IFFI polygons
    are present; one can see that, in this case, expert--mapped areas approximately agree with the 90$^{th}$
    elevation percentile of IFFI polygons, and that the 10$^{th}$ elevation percentile of the polygons extends
    down to the bottom of the valley.
  }
  \label{fig07}
\end{figure}

For illustrative purposes, due to the large extent of the study area, we show details of expert mapping and source
selection procedure, and their comparison, for one specific topographic unit, namely unit 1.2 \& 1.3 (Central-Eastern
Alps \& Carso) of \cite{Guzzetti1994}. \textbf{Figure \ref{fig07}(a)} shows an overall view of unit 1.2 \& 1.3, the
railway track, the polygons mapped by experts in this unit, along with the portion of IFFI rockfalls overlapping with
the set of SUs in this unit. Similar settings were found across the whole study area, about 25,000 km$^2$.

In \textbf{Fig \ref{fig07}(b)}, the detail shows a zoom of a few slope units in the north--east of the unit, in
which both expert-mapped and IFFI polygons are present (small orange box in the top figure). IFFI polygons are split
into main body (purple pattern), their portion below the 10$^{th}$ elevation percentile (``IFFI Lo'', cyan) and
the portion above 90$^{th}$ elevation percentile (``IFFI Hi'', green). Percentiles help showing that
most IFFI rockfall polygons actually contain both the source, runout and deposition zone. We infer that from
the fact that the upper percentile lies on top of slope units and approximately corresponds with expert--mapped
sources, while the lower percentile lies at the bottom of slopes and stretches down to the valley,
for the larger polygons. These characteristics were shared by all of the IFFI rockfall polygons that we explicitly
checked visually.  We use upper and lower percentiles to have a general overview of the performance of both source
selection and simulation results, in the following.

\begin{figure}[!ht]
  \centerline{
    \includegraphics[width=0.45\textwidth]{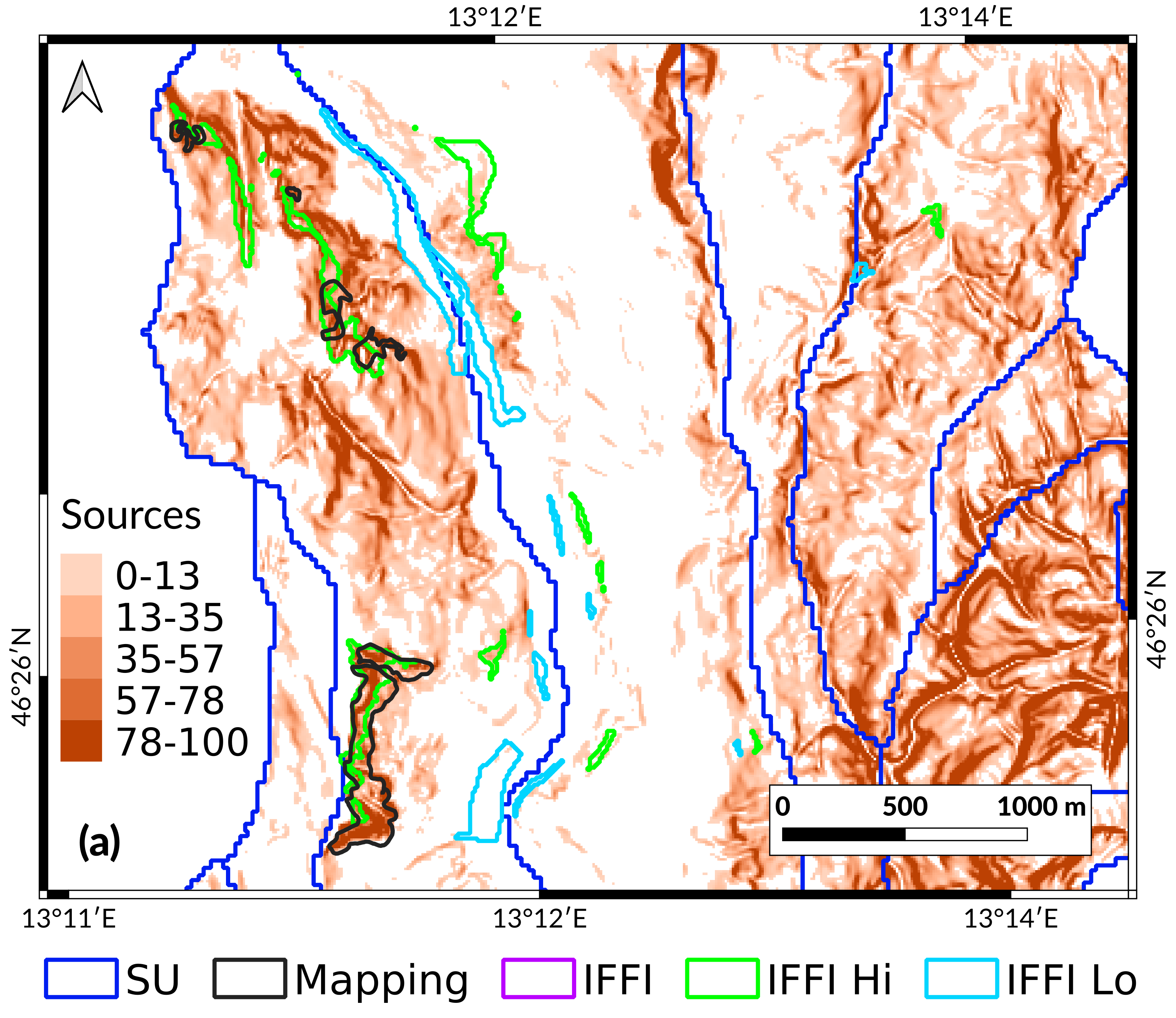}}    
  \centerline{
    \includegraphics[width=0.45\textwidth]{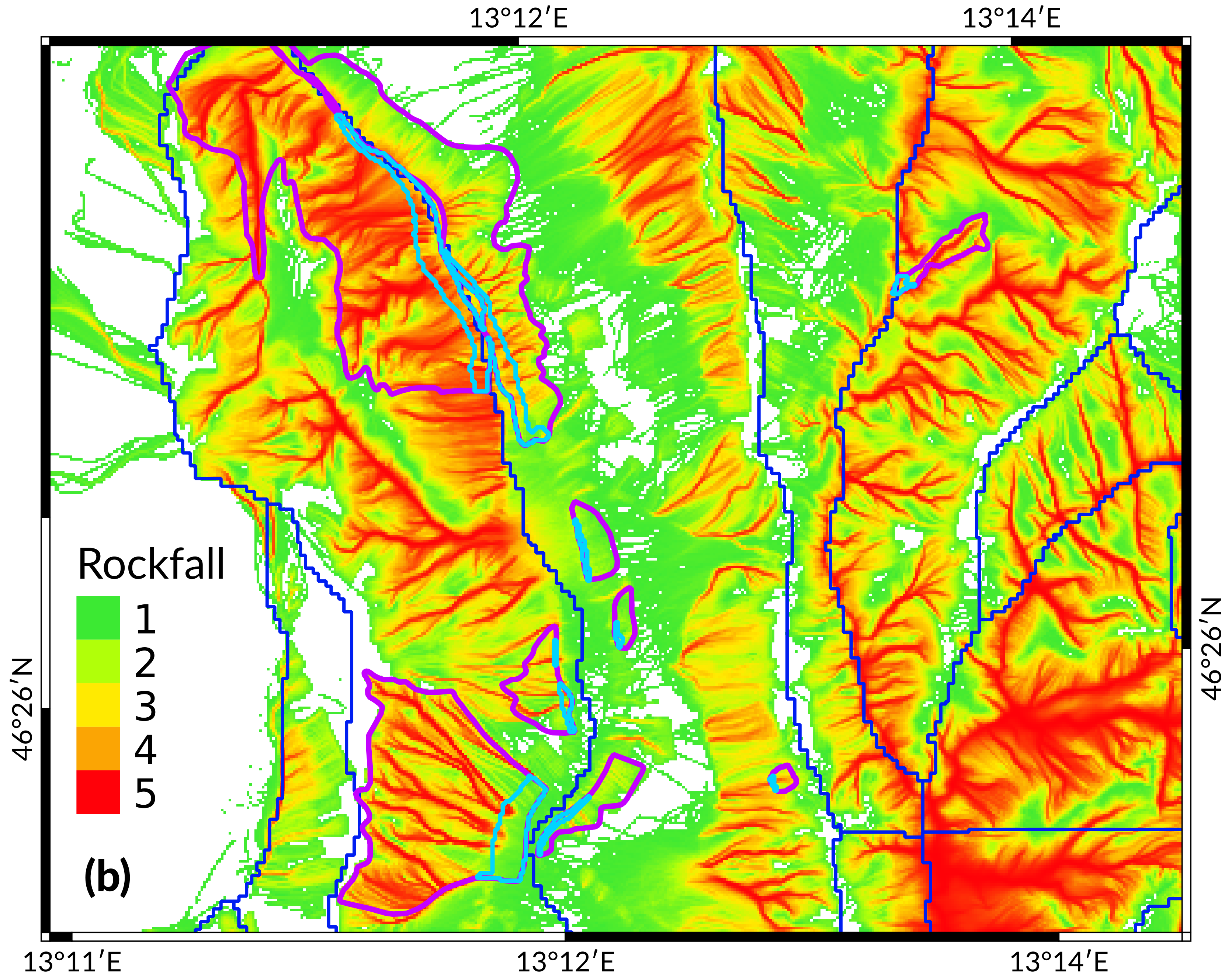}}
  \caption{(a) Modeled source areas (classified in shades of brown) and their relationship with
    expert--mapped source areas (black) and different elevation percentiles of IFFI polygons.
    (b) Modeled rockfall trajectory count (classified with a green--to--red color ramp) and
    its relationship with whole IFFI polygons (purple) and their 10$^{th}$ elevation percentile (cyan).
  }
  \label{fig08}
\end{figure}

The output of the generalization is a 10 m x 10 m grid map aligned with the TINITALY DEM, whose value
in each cell is the probability for a rockfall trajectory to originate from that specific location.
An example probabilistic source map selected within topographic unit 1.2 \& 1.3 is in \textbf{Fig. \ref{fig08}(a)}

A comparison between modeled probability and expert--mapped source areas represents an evaluation of
%
%
the performance of the calibration step of the source identification procedure. Performance is quantified
by hit rate of modeled probability versus ground truth, represented by mapped polygons. We considered
the modeled probability in two different ways. We either selected all of the grid cells for which the
statistical procedure assigned any non-null probability, and cells with values of probability larger
than 0.8, as an arbitrary threshold to distinguish values of ``high'' probability.

We devised a second comparison, using rockfall polygons in the IFFI inventory \citep{Trigila2010},
intended as a validation step of the procedure for source identification. We calculated, within each
polygon in the inventory, the portion corresponding to the upper 90\% in elevation. This was our best
(arbitrary) guess to what could have been the source area of each catalogued rockfall.

%
%
Results for the comparisons in each of the topographic units, in terms of HR, are listed in
\textbf{Table \ref{tab03}}.
We did not evaluate true negative rate, the counterpart of hit rate,
or a full confusion matrix: true negatives are unknown, here, because expert mapping was performed only
in selected location, and we did not know if IFFI is a complete inventory. Hit rate was calculated within the
slope units selected as study area for this work.

\textbf{Table \ref{tab03}} shows HR corresponding to four combinations of HR calculated as:
(i) modeled sources, with any probability value, against the expert--mapped source areas (``Mapped Total'');
(ii) modeled sources, limited to cells with highest probability, against the expert--mapped source areas
(``Mapped $P >$ 80\%''); 
(iii) modeled sources, with any probability value, against the upper 90\% elevation of IFFI polygons (``IFFI Total'');
(iv) modeled sources, with highest probability, against the upper 90\% elevation of IFFI polygons (``IFFI $P >$ 80\%'').

%
%
Combination (i): HR between modeled sources (cells assigned with non--null probability of being a source)
and expert--mapped source areas is below 75\% in 12 topographic units, namely: 2.2 (Veneto Plain), 3.1
(Monferrato Hills), 4.6 (Sila), 4.7 (Aspromonte), 4.8 (Sicilian Apennines), 5.1 (Tyrrhenian Borderland),
7.1 \& 7.2 (Marsala Lowland \& Sicilian Hills), 7.3 (Iblei Plateau), 8.1 (Sardinian Hills) and 8.3 (Campidano
Plain). Out of the 12 sections, five are below 50\%, of which two (units 2.2 and 3.1) are below 25\%.
The corresponding sections have small or very small total SU area, and many are located in the vicinities
of plains or coasts, which are probably the most difficult to model on a regional or national level as
they differ the most from the typical settings in which we mapped sources in expert way.

%
\begin{table*}[ht!]
\footnotesize
\begin{center}
  \caption{Numerical evaluation of the statistical generalization for the probability of grid cells to
    initiate a rockfall trajectory. We list the total area of each section (Section ID, as in
    \cite{Guzzetti1994}), the area covered by slope units selected for this work \cite{Alvioli2020a},
    the number of expert--mapped source polygons and their total area.
    HR is hit rate (or true positive rate), the ratio $TP$/$P$ = $TP$/($TP$+$FN$). ``Mapped'' refers to
    the cells underlying the expert--mapped source areas (colored in green--yellow--orange--deep orange,
    from high to small HR), and ``IFFI'' (colored in light to deep blue from large to small HR)
    refers to cells in the 90$^{th}$ elevation percentile of the rockfall polygons in the national inventory
    \citep{Trigila2010}.}
  \vskip 0.5cm
    {\setlength\arrayrulewidth{1pt}  
  \label{tab03}{\renewcommand{\arraystretch}{1.1}
  \begin{tabular}{c|r|r|c|r|c|c|c|c}
\hline
\multirow{3}{*}{\begin{tabular}{c}\textbf{Section}\\\textbf{ID}\end{tabular}} &
\multicolumn{1}{c|}{\textbf{Total}} &
\multicolumn{1}{c|}{\textbf{SU}} &
\multicolumn{1}{c|}{\textbf{Mapped}} &
\multicolumn{1}{c|}{\textbf{Mapped}} &
\multicolumn{1}{c|}{\textbf{HR}} &
\multicolumn{1}{c|}{\textbf{HR}} &
\multicolumn{1}{c|}{\textbf{HR}} &
\multicolumn{1}{c}{\textbf{HR}}\\
\multicolumn{1}{c|}{\textbf{ }} &
\multicolumn{1}{c|}{\textbf{Area}} &
\multicolumn{1}{c|}{\textbf{Area}} &
\multicolumn{1}{c|}{\textbf{Polygons}} &
\multicolumn{1}{c|}{\textbf{Area}} &
\multicolumn{1}{c|}{\textbf{(Mapped)}} &
\multicolumn{1}{c|}{\textbf{(Mapped)}} &
\multicolumn{1}{c|}{\textbf{(IFFI)}} &
\multicolumn{1}{c}{\textbf{(IFFI)}}\\
\multicolumn{1}{c|}{\textbf{ }} &
\multicolumn{1}{c|}{[km$^2$]} &
\multicolumn{1}{c|}{[km$^2$]} &
\multicolumn{1}{c|}{(\#)} &
\multicolumn{1}{c|}{[km$^2$]} &
\multicolumn{1}{c|}{(Total)} &
\multicolumn{1}{c|}{(P $>$ 80\%)} &
\multicolumn{1}{c|}{(Total)} &
\multicolumn{1}{c}{(P $>$ 80\%)}\\
\hline
\textbf{1.1}     & 16,274	& 1,590	& 152 &   3.2 &\ccdue 77\% &\ccbldue 13\% &\ccdue 86\% &\ccbluno 23\%\\
\textbf{1.2\&1.3}& 35,735	& 2,620	&  78 &   2.2 &\ccdue 80\% &\ccbldue 19\% &\ccdue 76\% &\ccbldue 10\%\\
\textbf{2.1}     & 32,702	& 373	&  11 & 0.039 &\ccdue 80\% &\ccbluno 27\% &\ccqua 63\% &\ccbltre  8\%\\
\textbf{2.2}     & 9,426	& 164	&  38 &  0.58 &\ccsei 11\% &\ccblqua  6\% &\ccdue 72\% &\ccbldue 11\%\\
\textbf{2.3a}    & 3,103	& 458	&  58 &  0.34 &\ccdue 75\% &\ccbldue 16\% &\ccdue 87\% &\ccbldue 12\%\\
\textbf{2.3b}    & 1,298	& 88	&   2 & 0.001 & - & - & - & -\\
\textbf{3.1}     & 2,322	& 332	&  39 & 0.027 &\ccsei 16\% &\ccblqua  0\% &\ccdue100\% &\ccbluno100\%\\
\textbf{3.2}     & 3,991	& 1,778	& 225 &  0.84 &\ccqua 51\% &\ccblqua  4\% &\cccin 29\% &\ccblqua  1\%\\
\textbf{4.1}     & 22,393	& 2,067	& 151 &  0.84 &\ccdue 78\% &\ccbltre  7\% &\ccqua 60\% &\ccbldue 15\%\\
\textbf{4.2}     & 16,835	& 1,894	& 105 &   1.5 &\ccdue 79\% &\ccbltre  9\% &\ccqua 59\% & \ccbltre 5\%\\
\textbf{4.3}     & 4,920	& 457	&  89 &   1.9 &\ccdue 77\% &\ccbldue 13\% &\ccdue 77\% &\ccbldue 13\%\\
\textbf{4.4}     & 8,097	& 1,585	& 102 &   1.6 &\ccdue 75\% &\ccbldue 14\% &\cccin 46\% &\ccbltre  6\%\\
\textbf{4.5}     & 12,890	& 1,379	& 200 &   7.2 &\ccdue 79\% &\ccbldue 18\% &\ccqua 62\% &\ccbldue 11\%\\
\textbf{4.6}     & 6,203	& 383	&  62 &  0.34 &\ccqua 50\% &\ccblqua  3\% &\cccin 44\% &\ccblqua  1\%\\
\textbf{4.7}     & 5,337	& 598	&  70 &  0.57 &\ccqua 59\% &\ccbltre  9\% &\cccin 46\% &\ccblqua  3\%\\
\textbf{4.8}     & 4,262	& 511	& 127 &   1.6 &\cccin 48\% &\ccblqua  3\% &\cccin 30\% &\ccbltre  7\%\\
\textbf{5.1}     & 25,346	& 2,086	&  80 &  0.19 &\ccqua 58\% &\ccbldue 12\% &\cccin 44\% &\ccbltre  6\%\\
\textbf{5.2}     & 6,136	& 972	&  11 & 0.013 &\ccdue 79\% &\ccbltre  5\% &\ccsei 16\% &\ccblqua  1\%\\
\textbf{5.3}     & 6,375	& 859	&  81 & 0.081 &\ccqua 62\% &\ccblqua  4\% &\ccqua 59\% &\ccbltre  6\%\\
\textbf{6.1}     & 9,023	& 930	&  46 & 0.035 &\ccdue 78\% &\ccbldue 10\% &\cccin 42\% &\ccbluno100\%\\
\textbf{6.2}     & 20,236	& 706	& 107 &   1.2 &\cccin 44\% &\ccbldue 11\% &\cccin 38\% &\ccbldue 11\%\\
\textbf{6.3}     & 1,731	& -	&   0 &     - &          - &            - &          - &            -\\
\textbf{7.1\&7.2}& 14,285	& 2,195	& 121 &   3.5 &\ccqua 56\% &\ccbldue 12\% &\cccin 33\% &\ccbltre  6\%\\
\textbf{7.3}     & 5,321	& 691	& 561 &  0.97 &\cccin 46\% &\ccbltre  9\% &\cccin 34\% &\ccbltre  5\%\\
\textbf{7.4}     & 1,499	& 210	&  31 &  0.37 &\ccdue 80\% &\ccbluno 28\% &\ccdue100\% &\ccbluno100\%\\
\textbf{8.1}     & 16,404	& 428	&  80 &   1.4 &\ccqua 63\% &\ccbldue 14\% &\cccin 25\% &\ccblqua  2\%\\
\textbf{8.2}     & 258	        & -     & -   &     - &          - &          - &            - &            -\\
\textbf{8.3}     & 1,946	& 4	&   6 & 0.061 &\ccqua 58\% &\ccbluno 21\% &\cccin 41\% &\ccbldue 18\%\\
\textbf{8.4}     & 2,844	& 42	&  11 &  0.66 &\ccdue 76\% &\ccbluno 22\% &\ccqua 63\% &\ccbldue 13\%\\
 \hline
\end{tabular}}}
\end{center}
\end{table*}
%
Combination (ii): the validity of this comparison is probably more difficult to understand on general
grounds. Out of the 26 sections (three out of the original 29 were ruled out by absence of railway
track or absence of slopes prone to rockfalls), four have HR larger than 20\%, 11 sections show smaller
than 10\%, five of them smaller than 5\%.

Combination (iii): HR results of the comparison between modeled sources and polygons labeled as ``rockfall''
or ``extended areas containing rockfalls'' in the IFFI national inventory. In this case, compared with the
region above the 90$^{th}$ elevation percentile, within each polygon. Results reveal a lower degree of match
with respect to case (i): 20 sections assigned with HR less than 75\%, of which 13 smaller than 50\%, of
which one smaller than 25\%.

Combination (iv): as in case (ii), interpretation of the comparison is less obvious than the one with
any value of probability for mapped sources, case (iii). Out of the 26 relevant sections, one
has HR larger than 20\%, 13 sections show HR smaller than 10\%, five of them smaller than 5\%.
\subsection{Results for rockfall susceptibility within the model STONE}\label{subs:results_stone}
The final product of this work is a classification of the railway track in Italy, split into 1--km segments,
with susceptibility values for the occurrence of rockfalls. To this end, the most relevant points are as
follows (\textit{cf.} enumerated list in \textbf{Section \ref{subs:methods_stone}}). We prepared maps of
trajectory count per cell using the model STONE. Then, we classified all of the cells encompassed by the
selected SUs (\textit{cf.} step 2a of the enumerated list in \textbf{Section  \ref{subs:methods_sources}});
we used the map to calculate a unique value of susceptibility per 1-km segment of the track.

We converted the trajectory count map, containing wildly varying values, into the [0,1] interval -- consistent
with a (relative) probability, \textit{i.e.} susceptibility. We did so using ECDF functions calibrated in the
subset of grid cells in which rockfalls are known to have occurred, reported in the national IFFI inventory \citep{Trigila2010}.
%
\begin{table*}[ht!]
\footnotesize
\begin{center}
  \caption{Hit rate for the validation of rockfall runout areas, \textit{i.e.} obtained from the
    comparison of probability of occurrence $P_{QR}$ from reclassification of trajectory count map
    by means of ECDFs as described in Section \ref{sec:methods}. The ID in the first column
    corresponds to topographic units of \textbf{Fig. \ref{fig01}(a)}, \textbf{Table \ref{tab01}}
    and \textbf{Table \ref{tab03}}.
    ``Total'' $HR$ values correspond to values of any non--null probability value, and partial $HR$
    values to probability between the specified intervals. The largest value among the partial ones
    is in bold. This comparison involves only the 10$^{th}$ elevation percentile of each polygon in
    the IFFI inventory.}
  \vskip 0.5cm
  {\setlength\arrayrulewidth{1pt}
  \label{tab04}{\renewcommand{\arraystretch}{1.1}
    \begin{tabular} {C{2.2cm}|C{1.4cm}|C{1.4cm}|C{1.4cm}|C{1.4cm}|C{1.4cm}|C{1.4cm}}
\hline
\multirow{2}{*}{\textbf{Section ID}} & \textbf{HR} & \textbf{HR} & \textbf{HR} & \textbf{HR} & \textbf{HR} & \textbf{HR}\\
& (Total) & [0,0.2) & [0.2,0.4) & [0.4,0.6) & [0.6,0.8) & [0.8,1.0]\\
\hline
 \textbf{1.1}       &\ccdue  95\% & \textbf{33\%} & 19\% & 17\% & 13\% & 13\%\\
 \textbf{1.2 \& 1.3}&\ccdue  95\% & 19\% & 16\% & 19\% & 20\% & \textbf{22\%}\\
 \textbf{2.1}       &\ccdue 100\% &  3\% & 14\% & 22\% & 19\% & \textbf{41\%}\\
 \textbf{2.2}       &\ccdue  98\% &  3\% &  4\% & 18\% & 27\% & \textbf{46\%}\\
 \textbf{2.3a}      &\ccdue  96\% & 13\% & 12\% & 20\% & 23\% & \textbf{29\%}\\
 \textbf{2.3b}      &  --  &  --  &  --  &  --  &  --  &  --\\
 \textbf{3.1}       &\ccsei   2\% &  0\% &  0\% &  0\% &  1\% &  1\%\\
 \textbf{3.2}       &\cctre  74\% & \textbf{23\%} & 14\% & 12\% & 10\% & 15\%\\
 \textbf{4.1}       &\ccdue  91\% & 13\% & 11\% & 21\% & 22\% & \textbf{23\%}\\
 \textbf{4.2}       &\ccdue  81\% & 16\% & 13\% &  5\% & \textbf{25\%} & 22\%\\
 \textbf{4.3}       &\ccdue  96\% & \textbf{37\%} & 15\% &  9\% & 21\% & 14\%\\
 \textbf{4.4}       &\ccdue  85\% & 15\% & 10\% & 10\% & 17\% & \textbf{33\%}\\
 \textbf{4.5}       &\ccdue  97\% &  9\% &  8\% & 11\% & 19\% & \textbf{49\%}\\
 \textbf{4.6}       &\cctre  60\% & 13\% &  8\% & 12\% & 11\% & \textbf{16\%}\\
 \textbf{4.7}       &\cctre  75\% & 14\% & 15\% & \textbf{18\%} & 16\% & 13\%\\
 \textbf{4.8}       &\ccdue  93\% & 13\% & 14\% & 20\% & 22\% & \textbf{24\%}\\
 \textbf{5.1}       &\cctre  71\% & \textbf{33\%} & 20\% & 14\% &  3\% &  2\%\\
 \textbf{5.2}       &\ccqua  43\% &  8\% &  7\% &  5\% &  5\% & \textbf{18\%}\\
 \textbf{5.3}       &\cctre  79\% & 21\% & 15\% & 12\% & 15\% & \textbf{16\%}\\
 \textbf{6.1}       &\ccqua  59\% & \textbf{14\%} & 10\% & \textbf{14\%} & \textbf{14\%} &  6\%\\
 \textbf{6.2}       &\ccdue  85\% & \textbf{52\%} & 14\% &  7\% & 10\% &  3\%\\
 \textbf{6.3}       &  --  &  --  &  --  &  --  &  --  &  --\\
 \textbf{7.1 \& 7.2}&\cctre  76\% & 16\% & 13\% & 13\% & \textbf{19\%} & 15\%\\
 \textbf{7.3}       &\cctre  77\% &  9\% & 12\% &  9\% & 11\% & \textbf{35\%}\\
 \textbf{7.4}       &\ccsei   8\% &  \textbf{3\%} &  2\% &  1\% &  1\% &  1\%\\
 \textbf{8.1}       &\ccqua  52\% & \textbf{13\%} &  9\% &  9\% & 11\% & 11\%\\
 \textbf{8.2}       &  --  &  --  &  --  &  --  &  --  &  --\\
 \textbf{8.3}       &\ccdue  81\% & 17\% & 17\% &  9\% & \textbf{21\%} & 17\%\\
 \textbf{8.4}       &\ccdue  97\% & \textbf{38\%} & 28\% & 11\% &  8\% & 12\%\\
 \hline
\end{tabular}}}
\end{center}
\end{table*}
%
The inventory contains polygons mapped as ``rockfalls'' or as ``extended areas containing rockfalls''. We calculated
specific ECDFs in each topographic unit shown in \textbf{Fig. \ref{fig01}(a)}.
\textbf{Figure \ref{fig04}} shows a few sample ECDFs corresponding to the first four topographic units
in \textbf{Tables \ref{tab03}}, \textbf{\ref{tab04}} and \textbf{\ref{tab05}}.
The remaining ECDFs are very similar, and curves would overlap
with those in \textbf{Fig. \ref{fig04}}.

\textbf{Figure \ref{fig08}(b)} shows the final, classified trajectory count (susceptibility) in a sample location.
The raster map is colored with a green--to--red color ramp from low to high values of susceptibility.
In the figure, we compare the susceptibility map with the IFFI
polygons existing in the area (\textit{cf.} \textbf{Fig. \ref{fig07}}). We performed validation
of the classified output in the susceptibility map as follows. We selected all of the rockfall polygons
from the IFFI inventory within the slope units buffer, and further selected the DEM grid cells under
the polygons and falling in the lowest 10$^{th}$ elevation percentile -- also shown in \textbf{Fig.
  \ref{fig08}}. We are reasonably confident that these cells represent rockfall deposit areas.

Comparison of probability maps among different topographic units with deposit areas inferred from IFFI
polygons, and with the whole IFFI polygons, provided values of hit rate ($HR$ = $TP/(TP+FN)$) listed in
\textbf{Table \ref{tab04}} and labeled as ``Total'' HR. Similarly, we defined ``partial'' HR values,       
limiting the calculations within the individual classes delimited by (0, 0.2, 0.4, 0.6, 0.8, 1) values,
  whose results are also listed in \textbf{Table \ref{tab04}} in five additional columns.

\textbf{Table \ref{tab05}} has a similar structure as \textbf{Table \ref{tab04}}, but it lists results
corresponding to comparison of modeled source areas with the whole extent of rockfall polygons in the
IFFI inventory. Topographic unit 2.3b does not contain rockfalls
recorded in the IFFI inventory relevant to the railway track, while sections 6.3 and 8.2 do not contain
intersections of the track with slope units. Thus, we reported no values for such zones in \textbf{Tables
  \ref{tab04}} and \textbf{\ref{tab05}}.

The next step was the classification of the railway network split into 1-km segments and classified using
a simple strategy. For each segment, overlapping many susceptibility values, we selected the largest value.
%
\begin{table*}[ht!]
\footnotesize
\begin{center}
  \caption{
    As in Table \ref{tab04}, but the comparison involves whole polygons in the IFFI inventory.
  }
  \vskip 0.5cm
  {\setlength\arrayrulewidth{1pt}
  \label{tab05}{\renewcommand{\arraystretch}{1.1}
    \begin{tabular} {C{2.2cm}|C{1.4cm}|C{1.4cm}|C{1.4cm}|C{1.4cm}|C{1.4cm}|C{1.4cm}}
\hline
\multirow{2}{*}{\textbf{Section ID}} & \textbf{HR} & \textbf{HR} & \textbf{HR} & \textbf{HR} & \textbf{HR} & \textbf{HR}\\
& (Total) & [0,0.2) & [0.2,0.4) & [0.4,0.6) & [0.6,0.8) & [0.8,1.0]\\
\hline
 \textbf{1.1}       &\ccdue 97\% & 19\% & 22\% & 20\% & 18\% & 18\%\\
 \textbf{1.2 \& 1.3}&\ccdue 95\% & 17\% & 19\% & 19\% & 19\% & 20\%\\
 \textbf{2.1}       &\ccdue 95\% & 13\% & 13\% & 14\% & 22\% & 33\%\\
 \textbf{2.2}       &\ccdue 93\% & 16\% & 18\% & 21\% & 20\% & 18\%\\
 \textbf{2.3a}      &\ccdue 93\% & 18\% & 19\% & 19\% & 19\% & 19\%\\
 \textbf{2.3b}      &         -- &   -- &   -- &   -- &   -- &   --\\
 \textbf{3.1}       &\ccsei  6\% &  1\% &  1\% &  0\% &  2\% &  1\%\\
 \textbf{3.2}       &\cctre 60\% & 12\% & 13\% & 12\% & 12\% & 12\%\\
 \textbf{4.1}       &\ccdue 90\% & 12\% & 15\% & 17\% & 21\% & 24\%\\
 \textbf{4.2}       &\ccdue 84\% & 24\% & 16\% & 15\% & 18\% & 12\%\\
 \textbf{4.3}       &\ccdue 95\% & 16\% & 19\% & 20\% & 20\% & 20\%\\
 \textbf{4.4}       &\ccdue 81\% & 14\% & 16\% & 17\% & 17\% & 17\%\\
 \textbf{4.5}       &\ccdue 94\% & 17\% & 19\% & 19\% & 19\% & 19\%\\
 \textbf{4.6}       &\cctre 65\% & 12\% & 13\% & 13\% & 13\% & 13\%\\
 \textbf{4.7}       &\cctre 72\% & 13\% & 16\% & 16\% & 15\% & 13\%\\
 \textbf{4.8}       &\ccdue 80\% & 15\% & 16\% & 16\% & 17\% & 17\%\\
 \textbf{5.1}       &\cctre 58\% & 24\% & 19\% & 10\% &  4\% &  1\%\\
 \textbf{5.2}       &\ccqua 36\% &  9\% &  6\% &  5\% &  7\% &  9\%\\
 \textbf{5.3}       &\ccdue 89\% & 17\% & 18\% & 18\% & 18\% & 18\%\\
 \textbf{6.1}       &\cctre 65\% & 10\% & 11\% & 14\% & 15\% & 15\%\\
 \textbf{6.2}       &\ccdue 78\% & 16\% & 15\% & 16\% & 16\% & 16\%\\
 \textbf{6.3}       &         -- &   -- &   -- &   -- &   -- &   --\\
 \textbf{7.1 \& 7.2}&\cctre 55\% & 12\% & 12\% & 12\% & 11\% &  9\%\\
 \textbf{7.3}       &\ccdue 83\% & 10\% & 15\% & 19\% & 20\% & 20\%\\
 \textbf{7.4}       &\ccsei \% & \% & \% & \% & \% & \%\\
 \textbf{8.1}       &\ccqua 40\% &  8\% &  8\% &  8\% &  8\% &  8\%\\
 \textbf{8.2}       &         -- &   -- &   -- &   -- &   -- &   --\\
 \textbf{8.3}       &\ccdue 82\% & 17\% & 17\% & 17\% & 19\% & 12\%\\
 \textbf{8.4}       &\ccdue 87\% & 15\% & 17\% & 17\% & 19\% & 19\%\\
 \hline
\end{tabular}}}
\end{center}
\end{table*}
%
This is a conservative and reasonable choice, given that if a single point of
the segment would be hit by a rockfall, the whole segment would be unusable.
On the other hand, splitting the whole link connecting two nodes into small segments of the same length
guarantees spatial homogeneity.
\textbf{Figure \ref{fig09}(a)} shows the railway segments classified as outlined above. Out of the total
16,084 km of railway track (the track length is less than the total, above 17,000 km, due to tunnels and
a few areas outside of the TINITALY), 14,724 km were classified in [0,0.2), 239 km in [0.2,0.4), 170 km in
    [0.4,0.6), 163 km in [0.6,0.8) and 789 km in [0.8,1.0].

Unfortunately, available data was insufficient to perform a validation of the result of the
classification of segments. To perform an additional assessment of the results,
we performed graph analysis, and evaluated its implications for the operativity of the railway network.
\subsection{Impact of rockfalls on the railway network: network--ranked susceptibility}\label{subs:results_graph}
\begin{figure*}[!ht]
  \leftline{\hspace{0.5cm}\includegraphics[width=0.45\textwidth]{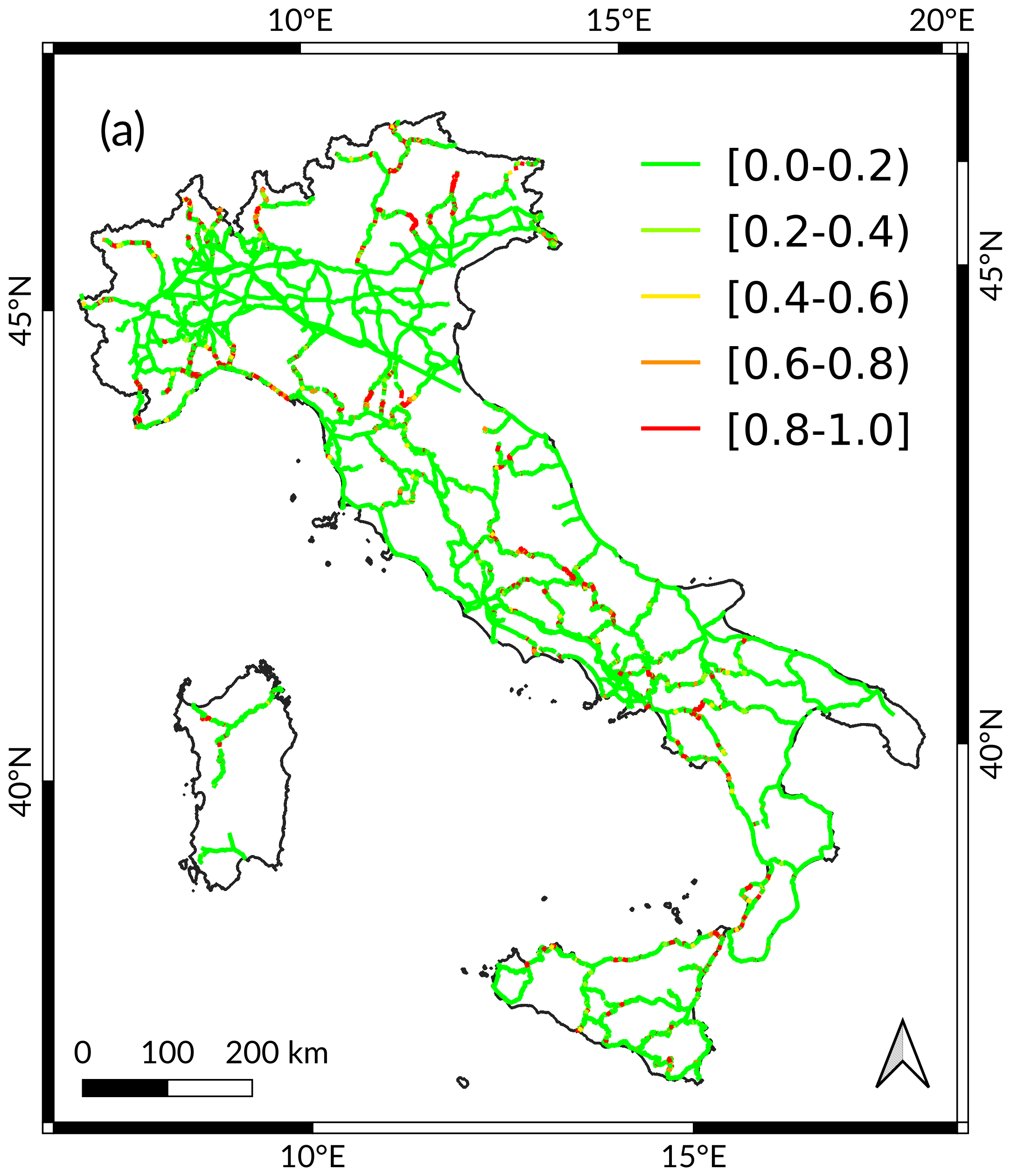}}
  \vskip -9.6cm
  \rightline{\includegraphics[width=0.45\textwidth]{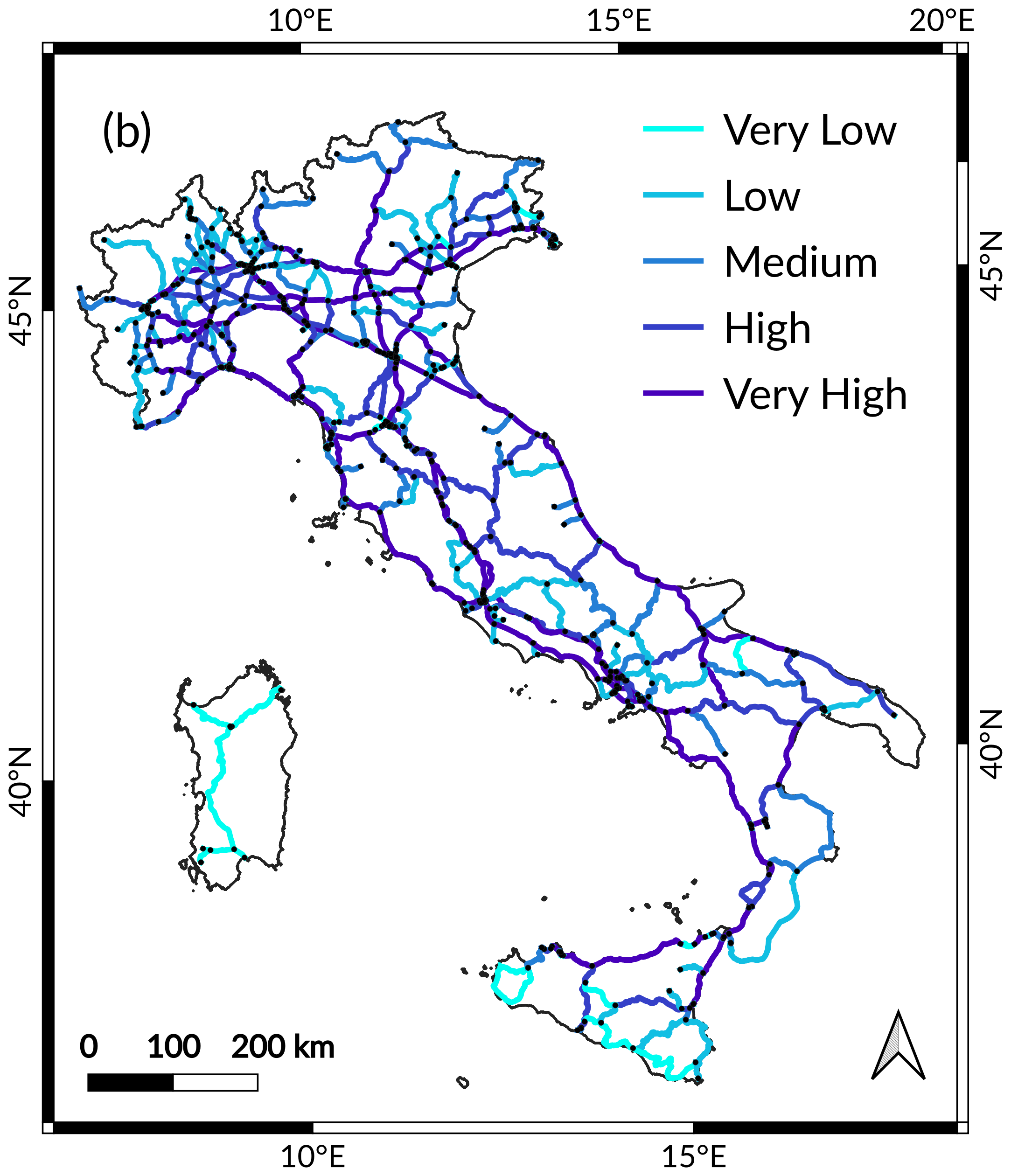}\hspace{0.5cm}}
  \vskip -0cm
  \leftline{\hspace{0.5cm}\includegraphics[width=0.45\textwidth]{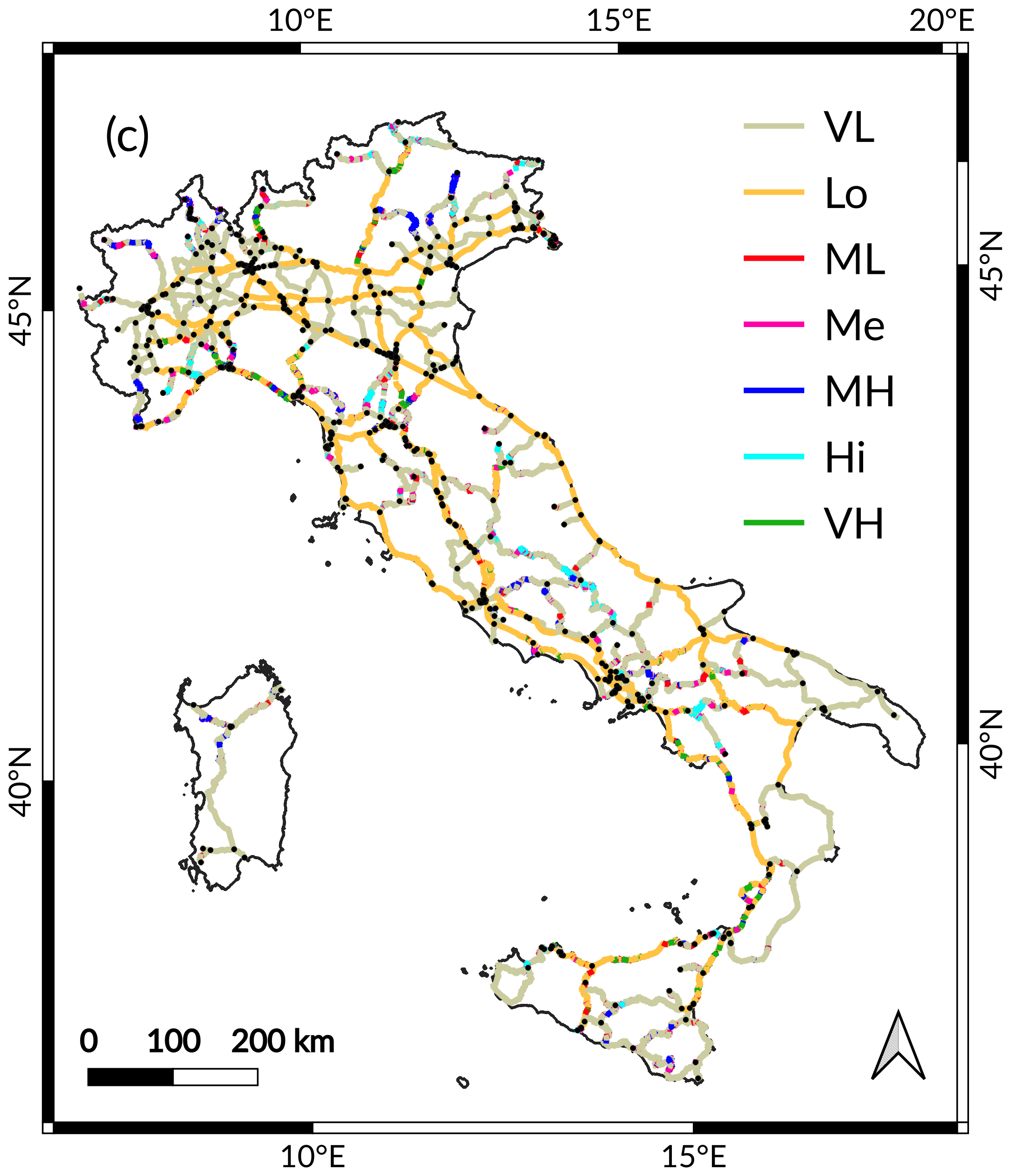}}
  \vskip -8cm
  \hspace{0.54\textwidth}\begin{minipage}[t]{0.42\textwidth} 
    \caption{(a) Classification of the railway network, split into segments of 1 km length,
      with the procedure developed in this work. Segments are colored with a red--to--green
      palette, corresponding to low--to--high susceptibility for rockfalls.
      (b) A measure of relevance of railway links in the Italian railway network, alternative
      to betweenness, calculated in \textbf{Section \ref{subs:methods_graph}}. Relevance is shown with
      a blue color palette; blue denotes more relevance. Relevance is high if total variation
      of vertex betweenness is high, when the corresponding edge is removed from the graph (\textit{cf.}
      \textbf{Fig. \ref{fig05}}).
             (c) Network--ranked susceptibility, defined as a combination of segment--wise rockfall susceptibility,
             modeled in this work with STONE and shown in (a), and relevance of the corresponding railway links, show in (b).
             \textbf{Table \ref{tab06}} lists a breakdown of the total railway length, in kilometers, within the seven classes
             shown in this figure in different colors.
    }\label{fig09}\end{minipage}
  \vskip 1.9cm  
\end{figure*}

In \textbf{Section \ref{subs:methods_graph}} we described an analysis of the railway network
in terms of graph theory. We defined a new index for each railway segment, obtained as the
variation of total edge betweenness on the network if the edge corresponding to the segment
is removed from the graph. \textbf{Figure \ref{fig09}(b)} shows the result of the analysis; edges
are colored with a blue color palette, for increasing value of the attribute calculated
by removing one edge at a time. The figure clearly highlights the most relevant edges, topology--wise,
and one can easily spot the difference between this analysis and edge betweenness, \textbf{Fig. \ref{fig05}}.

The result of \textbf{Fig. \ref{fig09}(b)} is an intermediate step towards the definition of a combined
quantity, defined considering simultaneously the variation of betweenness, thus strictly related to
network properties, and classification of the railroad based on rockfall susceptibility. The combination
was performed in the simplest
possible way, by crossing the classes of both quantities to obtain a new classification, the last
result of this work.

Classes based on rockfall susceptibility of railway links, split in 1-km segments, was defined in
\textbf{Section \ref{subs:methods_stone}}, and \textbf{Fig. \ref{fig09}(a)} shows the results.
Variation of betweenness on removal of one graph edge was defined in this section, and it is shown
in \textbf{Fig. \ref{fig09}(b)}. The latter quantity was categorized into five categories, using natural
breaks.

%
%
We refer to the joint classification of susceptibility and segment relevance as network--ranked
susceptibility. Results for network--ranked susceptibility, expressed in terms of kilometers or
railway in each class, are in \textbf{Table \ref{tab06}}. The table clearly shows the classification
obtained by crossing five classes in the two considered attributes, \textit{i.e.} segment relevance
(score) and rockfall susceptibility. The 25 classes are further colored (\textit{i.e.}, further
classified) using seven colors, in order of increasing ranked susceptibility: light brown, yellow,
red, magenta, blue, cyan, green. We defined the coloring scheme in a totally arbitrary
way, except that: (i) the lowest rockfall susceptibility class, corresponding to the largest portion
of railroad, was split into two classes of network--ranked susceptibility, and (ii) the remaining
portion of the railroad was (almost) evenly distributed into additional five classes.

We applied the same coloring scheme of \textbf{Table \ref{tab06}} to the graphical representation of
the classification of segments, \textbf{Fig. \ref{fig09}(c)}. Comparison with the results of classification
based on the only susceptibility highlights manifest differences. In fact, while reddish segments are
%
\begin{table*}[ht!]
  \footnotesize
  \begin{center}
    \caption{Table shows the cross--comparison of rockfall susceptibility (increasing from left to right)
      \textit{vs.} segment relevance (score, increasing form top down; classes are as follow: S1 (very low), 0--687;
      S2 (low), 688--3006; S3 (medium), 3007--9014; S4 (high), 9015--32914; S5 (very high), 329515--512079),
      as described in Section \ref{sec:results}. The different colors define a joint classification, named
      network--ranked susceptibility: light brown, very low (VL); yellow, low (Lo);
      red, medium-low (ML);
      magenta, medium (Me); blue, medium-high (MH); cyan,
      high (Hi); green, very high (VH).
      Values are in kilometers; total length (16,084 km) is
      shorter than the whole railway network because we removed tunnels and segments (few kilometers)
      laying outside the TINITALY DEM. Segments on plain topography (\textit{i.e.}, non overlapping with
      slope units) are included, here.}
    \vskip 0.5cm
    {\setlength\arrayrulewidth{1pt}
    \label{tab06}{\renewcommand{\arraystretch}{2}
      \begin{tabular}{C{1.0cm}|C{0.5cm}|R{1.1cm}|R{1.1cm}|R{1.1cm}|R{1.1cm}|R{1.1cm}|R{1.1cm}}
        \multicolumn{2}{c|}{\multirow{2}{*}{}} & \multicolumn{5}{c|}{\textbf{Rockfall Susceptibility}} & \multicolumn{1}{c}{\multirow{2}{*}{\textbf{Sum}}}\\\hhline{~~-----~}
        \multicolumn{2}{c|}{}& \multicolumn{1}{c|}{0.0--0.2} & \multicolumn{1}{c|}{0.2--0.4} & \multicolumn{1}{c|}{0.4--0.6} & \multicolumn{1}{c|}{0.6--0.8} & \multicolumn{1}{c|}{0.8--1.0} &\\\hhline{--------}                
        \multicolumn{1}{c|}{\multirow{5}{*}{\STAB{\rotatebox[origin=c]{90}{\textbf{Railway Segment Score\hspace{0.3cm}}}}}}
        &\rotatebox[origin=c]{90}{S1}& \funo 1,432 & \ftre 16 & \ftre 11 & \fqua 10 & \fcin  42 & 1,511\rule{0pt}{20pt}\\\hhline{~-------}
        &\rotatebox[origin=c]{90}{S2}& \funo   782 & \ftre 11 & \ftre  5 & \fqua  7 & \fcin  21 &   826\rule{0pt}{20pt}\\\hhline{~-------}
        &\rotatebox[origin=c]{90}{S3}& \funo 2,506 & \ftre 52 & \fqua 33 & \fqua 33 & \fcin 188 & 2,811\rule{0pt}{20pt}\\\hhline{~-------}
        &\rotatebox[origin=c]{90}{S4}& \funo 4,655 & \ftre 86 & \fqua 67 & \fqua 64 & \fsei 323 & 5,195\rule{0pt}{20pt}\\\hhline{~-------}
        &\rotatebox[origin=c]{90}{S5}& \fdue 5,348 & \ftre 73 & \fqua 54 & \fcin 50 & \fset 215 & 5,740\rule{0pt}{20pt}\\\hhline{--------}
        \multicolumn{2}{c|}{\textbf{Sum}} & 14,724 & 239 & 170 & 163 & 789 & 16,084\rule{0pt}{20pt}\\
        %
        %
    \end{tabular}}}
  \end{center}
\end{table*}
%
consistently located in hilly and mountainous areas, plain areas now contains the ``very low'' and ``low''
classes of network--ranked susceptibility, while they just had the lowest class of susceptibility; this
was a totally arbitrary choice. We discuss the possibility of optimizing the new network--ranked
susceptibility in the next section.
%
\section{Discussion}\label{sec:discussion}
As for the Methods and Results sections, we discuss separately the source areas identification
procedure, rockfall susceptibility assessment and implications from network analysis combined
with susceptibility.

We stress here that we used the national landslide inventory IFFI \citep{Trigila2010,IFFI2018}
for validation in several ways. Nevertheless, we did not actually use the inventory for building or calibrating
the rockfall model adopted here. In fact, we calculated hit rates of the statistical generalization
of expert--mapped source areas and hit rates of runout areas against proper subsets and sub--areas of the polygons in
the IFFI inventory.
\subsection{Identification of rockfall source areas}\label{subs:discussion_sources}
An illustration of the comparison between modeled source probability, expert--mapping sources
and the upper elevation percentile of IFFI polygons is in \textbf{Fig. \ref{fig08}}, for one
particular region in topographic unit 1.2 \& 1.3. We can see that in this region non--null pixels
of the probabilistic source map have a wider extent than both expert--mapping and IFFI upper
elevation percentile. This is by construction, because the statistical procedure described in
\textbf{Section \ref{subs:methods_sources}} was devised to conservatively assign non--null
probabilities to any grid cell, above a rather low threshold, based on their slope. This
is consistent with the observation that expert--mapped sources actually contains low values of slopes
(\textit{cf.} \textbf{Figs. \ref{fig03}} and \textbf{\ref{fig06}}). \textbf{Figure \ref{fig08}(a)}
shows a nice agreement between dark--shaded probabilistic map and both IFFI upper elevation percentile
and expert--mapped area, though not all areas match perfectly.

Nevertheless, values of model match against the mapped and IFFI polygons, in \textbf{Table \ref{tab03}}
are often rather low. We attribute that to the following reasons. To build a model of source areas, we
hypothesized a relationship
between the probability of a grid cell of initiating a rockfall and local slope. This represents
a compromise between an acceptable overall time needed for the procedure over a large study area
and a realistic product, but it certainly does not embed all of the local terrain properties that
influence the expert criteria applied for mapping potential source areas. Moreover, we made
  specific choices and assumptions, often arbitrary, in the modeling process.

Errors may also arise
from the discrepancy between the DEM used in the analysis and the apparent resolution of Google
Earth\texttrademark{ }imagery used for expert mapping, especially in the locations with steepest
relief, the ones in which we are mostly interested in. In the analysis, we
used a DEM generated from a triangulated irregular network (TIN); visual inspection of a shaded
relief generated form the DEM highlighted locations in which the triangulation used to prepare the
DEM is manifest, which surely affects the slope map, nation-wide.
%
\subsection{The model STONE and rockfall susceptibility}\label{subs:discussion_stone}
The output of the model STONE is a raster map: the classified trajectory count per grid cell.
We used it in several ways, to get to the final results of this work:
(i) we compared the classified output with the relevant subset of the IFFI inventory; (ii)
we split the railway network into 1-km segments and classified them on the basis of rockfall
susceptibility; (iii) we studied the properties of the railway network within graph theory,
considering the network in relation with the susceptibility results, which helped
defining a joint classification.

\textbf{Tables \ref{tab04}} and \textbf{\ref{tab05}} list results of the validation against
IFFI polygons. Values for the total hit rate in this case are rather large, which means that
most of the runout areas in the inventory were actually overlapping with at least one
trajectory. Considering the ``partial'' HR in \textbf{Table \ref{tab04}}, we find that in most
cases runout areas (maxima thereof, in bold) fall within the [0.8,1.0] susceptibility class, many
fall within the [0,0.2), a few within the [0.4,0.6) and [0.6,0.8) classes, while there are none
in the [0.2,0.4) class. This information does not call for straightforward conclusions, though it
might be interesting for future reference if an update of the methods presented in this work will
be implemented. This could most probably be in the procedure for the selection of source areas
(for example as suggested at the end of \textbf{Section \ref{subs:discussion_stone}}, or as in
\cite{Rossi2021}), or in the strategy for classification of the trajectory count.

Results in \textbf{Table \ref{tab05}} correspond to validation against the whole extent of polygons
in the IFFI inventory, at variance with \textbf{Table \ref{tab04}}. Values of ``Total'' hit rate are
rather similar to the case where we compared to the 10$^{th}$ elevation percentile of each polygon,
in a few cases a slightly higher. Interestingly, in this case, values of hit rates are almost always
equally distributed among the five sub--classes of probability. One reason for that could be
due to spatial inaccuracies of the polygons in the inventory.

The trajectory count obtained here, \textbf{Fig. \ref{fig08}(b)}, is a raster map with resolution of 10 m.
Nevertheless, we consider as a final product of this work the vector map of the railway network, split into
1--km segments, classified with five levels of susceptibility. The reason for that is twofold. First, the original
aim of this work was to prepare a map which could be used for ranking, or prioritizing, safety countermeasures
at the national level. Second, we acknowledge that we necessarily made assumptions and approximations to be
able to work homogeneously at the national scale; for this reason, a raster map with 10 m resolution could
contain local inaccuracies, which are mitigated by the classification of 1--km segments. This implies that
application of the classified map is intended to locate the segments with highest susceptibility at the national
scale; conclusions at smaller (local) scales can still be obtained within the model STONE, but should be supported
by higher resolution data.
%
\subsection{Network--ranked rockfall susceptibility}\label{subs:discussion_graph}
Network--ranked susceptibility for rockfalls, shown in \textbf{Table \ref{tab06}} and \textbf{Fig. \ref{fig09}(c)},
was defined in this work as a straightforward combination of five classes in edge ranking and rockfall
susceptibility. We already noted that the proposed
classes produce somewhat striking results, \textbf{Fig. \ref{fig09}(c)}, in that segments with negligible
rockfall susceptibility (\textit{cf.} \textbf{Fig \ref{fig09}(a)}) do not fall in the lowest class
of network--ranked susceptibility. We stress that the seven classes (colors) shown in \textbf{Table
  \ref{tab06}} were chosen almost arbitrarily -- hence the apparent mismatch.

We consider the
classification proposed here as a methodological proposal, and we did not attempt a higher--level
optimization of final classes, which we could actually perform with suitable validation data.
Validation of susceptibility requires accurate data about observed rockfall occurrences,
possibly including information about whether rockfalls did or did not cross the railway track.
Calibration and validation of network--ranked susceptibility, in addition, would require
information about the actual operation of the network.

Knowledge of actual traffic on the railroad would imply different network properties.
In fact, betweenness of nodes and edges was calculated here considering all--to--all
shortest paths. If we replaced them with actual routes we would obtain a different
network and corresponding graph \citep{Kurant2006a}. As another example, knowledge
of the number of trains running on each railroad link per day could be used to define relevance
of edges in a different way, beyond the simple topology of the network considered here. Additional
information on the average number of passengers per link/day could help in a similar way.
On the other hand, knowledge of real time presence of trains on each link of the network,
though not difficult to consider in technical terms, is not immediately relevant for this
work. That would require inclusion of some dynamical mechanism for the actual detachment of
rocks uphill form each railway link, along with real time monitoring of possible triggering
events.
%
\section{Conclusions}\label{sec:conclusions}
We described a comprehensive modeling chain for the assessment of rockfall susceptibility along
the railway network in Italy. To cope with issues related with the large size of the study area,
extensive mapping and computational challenges, we necessarily adopted assumptions and approximations.
These were discussed in details throughout this work, in which we described the performance of the
different steps of the procedure and we highlighted weak points and possible improvements. An
additional point to stress here is that, due to the large size of the study area and unavailability
of specific data, rockfalls on anthropic slopes were not considered.
On the other hand, we highlighted the merits of this work as an effective procedure for rockfall
susceptibility assessment over a very large and geomorphologically diverse study area.

We described three main steps: (i) localization of rockfall source areas on a digital landscape,
with different probabilities, (ii) simulation of rockfall trajectories stemming from any possible
source area, and (iii) assessment of the impact on the operativity on the railway network,
using a joint classification of network segments taking into account both rockfall susceptibility
and network properties. The relevant findings were as follows:
\begin{itemize}
\item[$\bullet$] The procedure for localizing source areas is a
  statistical generalization of the distribution of slope angle values within expert--mapped polygons.
  In general, values of hit rate were seldom in the higher
  quartile, and mostly in the third and second quartile. Results for the validation were similar.
  The procedure is suitable for identification of source areas in large areas (here, about 25,000
  km$^2$), though improvement of calibration performance would require including additional
  morphometric variables besides slope.
\item[$\bullet$] Rockfall trajectories were simulated using the
  program STONE. Validation of the classified runout in terms of hit rate, were
  rather satisfactory, in that the majority of values were in the higher quartile of the probability
  distribution. We further cast the susceptibility map onto the railway network, by splitting the railway
  link in 1--km segments and assigning a class of susceptibility to each segment. Validation of the
  susceptibility values for segments of the railway requires additional data, not available to us.
\item[$\bullet$] The analysis of possible rockfall impact on the railway considered the effect
  of disruption of a railway segment on the entire network. That was quantified by a
  new quantity which we called network--ranked susceptibility, considering the joint
  classification of rockfall susceptibility and graph properties.
  We consider this step as a methodological proposal, whose optimization and validation would
  require dedicated data and additional knowledge of the actual traffic over the railway network.
\end{itemize}
We stress that, in this work, the national landslide inventory IFFI \citep{Trigila2010} helped
validating results, but did not enter neither in the runout modeling, nor in the identification of
sources areas, which required expert source areas mapping. This step of the procedure leaves room for
improvement, in that we only used distributions of slope angle values to characterize expert--mapped
polygons. 

Both steps (identification of source areas and simulation of runout) can,
in principle, be extended outside of the buffer considered in this work, because most of the
information and methods developed here are still valid. The buffer itself, consisting of slope
units overlapping with the railway, can be easily enlarged, given that slope units are available
for the whole of Italy \citep{Alvioli2020a}.

Eventually, we stress that a natural evolution of the analysis performed on the topological
properties of the railway network is a real--time monitoring (still static, no traffic involved)
of the network as a function of the real--time evolution of possible triggering events. This was
actually the original motivation of this work. The segment--wise susceptibility map developed
here, along with analogous map obtained for debris flows susceptibility and soil slides susceptibility,
are key inputs for the prototype early warning system that was funded by the National railway company
(RFI) within the framework for the implementation of the SANF (an acronym for national early warning
system for rainfall-induced landslides) system \citep{Rossi2012,Guzzetti2020}.
%
\section{Acknowledgments}
This work was partially supported by RFI gruppo Ferrovie dello Stato Italiane, contract number 488, 01/02/2018.
%
\section{Declaration of competing interests}
The authors declare that they have no known competing financial interests or personal relationships that could
have appeared to influence the work reported in this manuscript.
%
%
%
\bibliography{main}

\end{document}